\documentclass[a4paper,11pt]{article}
\usepackage{jheppub} 
\usepackage{lineno}

\usepackage[utf8]{inputenc}
\usepackage{amsmath}
\usepackage{amssymb}
\usepackage{amsfonts}
\usepackage{amsthm}
\usepackage{xcolor}
\usepackage{hyperref}
\usepackage{rotating}
\usepackage{array}
\usepackage{appendix}
\usepackage{bbold}
\usepackage{textcomp}
\usepackage{float}
\usepackage{comment}
\usepackage{upgreek}

\def\sc{\textit{sc}}
\def\tr{\text{tr}}

\def\fZ{f^{\mathcal{Z}}}
\def\IZ{I^{\mathcal{Z}}}
\def\fp{f^{\langle\phi^2\rangle}}
\def\Measure{\mathcal{W}}

\def\fp{f^{\langle\phi^2\rangle}}

\def\gcZ{\mathcal{Z}}
\def\cZ{Z}
\def\bb{t}
\def\oneptcub{b_{\upphi}}

\title{Thermal One-point Functions and Asymptotic CFT Data: QFT in AdS}

\author[a]{Ilija Buri\'c,}
\author[b]{Francesco Mangialardi,}
\author[c]{Francesco Russo,}
\author[b]{Volker Schomerus}
\author[d]{and Alessandro Vichi}

\affiliation[a]{School of Mathematics and Hamilton Mathematics Institute, Trinity College, Dublin 2, Ireland}
\affiliation[b]{Deutsches Elektronen-Synchroton DESY, Notkestr. 85, 22607 Hamburg, Germany}
\affiliation[c]{CPHT, CNRS, \'Ecole polytechnique, Institut Polytechnique de Paris, 91120 Palaiseau, France}
\affiliation[d]{Department of Physics, University of Pisa and INFN, Largo Pontecorvo 3, I-56127 Pisa, Italy}

\emailAdd{burici@tcd.ie}
\emailAdd{francesco.mangialardi@desy.de}
\emailAdd{francesco.russo@polytechnique.edu}
\emailAdd{volker.schomerus@desy.de}
\emailAdd{alessandro.vichi@unipi.it}

\abstract{We investigate the thermal partition and one-point functions of the three-dimensional conformal field theory dual to a massive interacting scalar field in AdS$_4$. Using thermal inversion formulas, we determine the asymptotic behaviour of the spectral density and OPE coefficients involving heavy operators at fixed spin. We first analyse these CFT data for the generalised free field, corresponding to the non-interacting bulk theory. Then we compute the first-order perturbative corrections induced by the cubic and quartic bulk interactions. The thermal observables considered here probe a sector associated with operators of large dimension and, in the bulk description, a regime dominated by states with large particle number. This regime remains comparatively unexplored even in
generalised free field theory. Remarkably, the asymptotic formulas obtained from thermal inversion remain quantitatively accurate far from the asymptotic regime, describing CFT data reliably already at intermediate conformal weights. Our results show that this feature survives the inclusion of bulk interactions and provide new
analytic control over heavy-state data in conformal field theories.}

\begin{document}

\maketitle

\section{Introduction}

Correlation functions at finite temperature are observables of great interest 
in conformal field theory (CFT). In addition to their experimental relevance, 
these correlators are also important for a number of theoretical reasons. Most 
importantly, they provide direct access to heavy operators, i.e. operators of 
large weight/twist that involve a large number of constituent fields. This  
regime of CFTs is complementary to the low weight/twist sectors that have been 
well explored using modern numerical and analytical bootstrap methods. Through 
the holographic correspondence, thermal CFT opens a new window into multiparticle 
states of the theory in anti-de Sitter (AdS) spaces. For sufficiently high temperature, black holes may form so that finite-temperature CFT correlation functions can probe quantum aspects of black holes. Recently, thermal correlators, most notably the two-point functions, have also been studied using the {\it thermal bootstrap}, \cite{Iliesiu:2018fao,Iliesiu:2018zlz,Marchetto:2023xap,Barrat:2024aoa,Barrat:2024fwq,Buric:2025anb,Barrat:2025nvu,Buric:2025fye,Barrat:2025twb,Simmons-Duffin:2025qox}. 
\vskip0.05cm
 
Among the various {\it thermal manifolds} of the form $S^1 \times \mathcal{M}_{d-1}$, 
where $S^1$ denotes the Euclidean-time thermal circle of length $\beta$ and 
$\mathcal{M}_{d-1}$ is some spatial manifold, the geometry $S^1 \times S^{d-1}$ is
distinguished by the fact that correlation functions on this manifold are entirely
determined by {\it flat-space} CFT data. In particular, the spectrum of a CFT is encoded 
in its partition function on $S^1 \times S^{d-1}$. In this paper, we continue a systematic
study, initiated in \cite{Buric:2024kxo,Buric:2025uqt}, of the next simplest observable on
$S^1\times S^{d-1}$, namely the one-point function of a conformal primary field $\phi$.
\smallskip

Correlators on $S^1 \times S^{d-1}$ are particularly well understood in the high-temperature regime $\beta \to 0$. For the partition function, this limit is governed by a {\it thermal effective field theory} (EFT) \cite{Bhattacharyya:2007vs,Banerjee:2012iz,Jensen:2012jh,Benjamin:2023qsc,Benjamin:2024kdg}. The thermal EFT expansion of the partition function, in turn, yields an asymptotic expansion for the density of primary states $\rho(\Delta,\ell)$ in the limit $\Delta \to \infty$, at fixed spin $\ell$ \cite{Benjamin:2023qsc}. This nicely complements the regime 
accessible through the lightcone bootstrap \cite{Komargodski:2012ek,
Fitzpatrick:2012yx}, see also \cite{Anand:2025mfh,Komargodski:2026ain} for recent studies interpolating between these two regimes. Similarly, high-temperature one-point functions on $S^1 \times S^{d-1}$ are constrained by the fact that symmetry fixes their form on $S^1 \times 
\mathbb{R}^{d-1}$ \cite{Iliesiu:2018fao,Gobeil:2018fzy,Buric:2024kxo,Buric:2025uqt}. Combined with a set of additional mild assumptions, this observation was used in \cite{Buric:2025uqt} to derive an asymptotic expansion for the OPE coefficients $\lambda_{\phi \mathcal{O}_{\Delta,\ell}\mathcal{O}_{\Delta,\ell}}$ in the same large-$\Delta$, fixed-$\ell$ regime (these findings are consistent with and extend those coming from the eigenstate thermalisation hypothesis (ETH) \cite{Lashkari:2016vgj} and hydrodynamics \cite{Delacretaz:2020nit}). The resulting formulas, which are expected to hold for any local three-dimensional CFT \footnote{The asymptotic expansion takes a slightly different form depending on whether the CFT is gapped upon compactification on $S^1$ or not.}, were tested in the simplest example of the free scalar field. In this special case, it was found that the first few orders of the expansion at large $\Delta$ were sufficient to approximate the exact OPE coefficients of the free field at intermediate weights of $\Delta\sim 30$ 
extremely well, regardless of the choice of spin $\ell$ and tensor structure. We also re-examined the spectral densities of the free scalar field and found 
a similar behaviour: while the leading order in the large $\Delta$ expansion provides a very poor estimate, with an error of $40\%$ at $\Delta \sim 200$, higher order corrections can give extremely accurate results. This certainly raises some hope that high temperature expansions provide very accurate 
estimates for fixed spin CFT data even down to values of $\Delta$ that can be reached by the numerical bootstrap. 
\smallskip 

In the present work, we take a further step in this study by using thermal partition and one-point functions to extract fixed spin CFT data in the CFT dual of weakly interacting QFTs on AdS$_4$.
\smallskip

We begin by considering the dual of a free scalar $\Phi$ in AdS$_4$ with mass $m^2$, which is 
the well-known generalised free field (GFF) theory of a field $\phi$ of weight $\Delta_\phi$ 
satisfying \mbox{$m^2=\Delta_\phi(\Delta_\phi-3)$}. The high temperature behaviour of the GFF partition function differs from that of a local CFT, reflecting the non-local nature of the former. Therefore, deriving the asymptotic spectral densities and OPE coefficients in this theory forms an important part of our analysis. The resulting expressions are quite different from those in the free theory that we studied earlier. Nevertheless, the asymptotic expansions again approximate the exact CFT data very well down to intermediate weights. 
\vskip0.05cm

Next, we turn on perturbative interactions in the bulk. Weakly interacting quantum field theories 
on AdS provide a particularly useful laboratory, since thermal observables can be computed explicitly 
while still exhibiting nontrivial dynamical effects. Here, we shall look at a scalar field with cubic 
$\Phi^3$ and quartic $\Phi^4$ interactions in the bulk of AdS$_4$. For this theory, we analyse the 
partition function and the one-point function of the fundamental scalar $\phi$ dual to $\Phi$ in 
the boundary CFT, to first order in perturbation theory in the two couplings. Let us note that, 
at leading order, the partition function only receives corrections from the quartic interaction, 
while the one-point function of $\phi$ is only sensitive to the cubic interaction.
\vskip0.05cm

As in our previous work, we approach the problem using two complementary methods. On the one hand, we show how to derive exact anomalous dimensions (at leading order in perturbation theory) for low-lying operators that arise from the quartic interaction in AdS$_4$. This recovers known results for double-trace operators, \cite{Fitzpatrick:2010zm}, and extends them to some multi-trace cases. On the other hand, we will obtain the asymptotic behaviour of the anomalous dimensions as $\Delta\to\infty$, and show how these asymptotic formulas provide an accurate approximation even for the low-lying spectrum. The same strategy is also used for the one-point function, only that the corrections now come from the cubic interaction. Once again, we will compute exact OPE coefficients $\lambda_{\phi\mathcal{OO}}$ (at leading order in perturbation theory) and compare these with the asymptotic behaviour derived at large $\Delta_\mathcal{O}$.  
\smallskip


As has been pointed out before, thermal observables in general, and consequently also the asymptotic spectral density and the heavy-heavy-light (HHL) OPE coefficients we study here, receive dominant contributions from states containing increasingly many 
constituent fields as $\Delta$ increases. In order to gain some more insight into why the asymptotic expansions at large $\Delta$ seem to capture even low weight OPE data surprisingly well, we shall study thermal observables also in canonical ensembles in which we keep the particle number fixed. Note that the concept of particle number is indeed well defined in GFF, and it remains a useful perturbative family label for states continuously connected to free $n$-particle states, even when we turn on the interactions. In the remainder of this introduction we now summarise our main results. 

\subsection{Summary of results}

Let us now state the main new results of this work in more detail. We begin by studying the generalised free theory of a scalar field. This is a non-local CFT and its density of primary states as $\Delta\to\infty$ does not follow the behaviour of a local theory. In particular, we will show that the density reads
\begin{align}\label{rho-GFF-intro}
    \rho_0(\Delta,\ell) = &\frac{5^2 3^\frac72}{2^\frac52 \pi^9} (2\ell+1)
\left(\frac{30\Delta}{\pi^4}\right)^{-\frac{101}{32}-\frac{\Delta_\phi(12-9\Delta_\phi+2\Delta_\phi^{2})}{48}} e^{\left(\frac{128}{1215}\right)^\frac14 \pi\,\Delta^\frac34 + C_1^{(\Delta_\phi)}\Delta^\frac12 + C_2^{(\Delta_\phi)}\Delta^\frac14 + C_3^{(\Delta_\phi)}} \nonumber\\[1ex]
&\left(1 -\frac{P_0(\Delta_\phi) + P_1(\Delta_\phi)\zeta(3) + P_2(\Delta_\phi)\zeta(3)^2 + P_4(\Delta_\phi) \zeta(3)^4}{960 \sqrt[4]{30} \pi ^{11}} \Delta^{-\frac14} +\dots\right)\,,
\end{align}
where the coefficients $C_i^{(\Delta_\phi)}$ are given in equation \eqref{Ci-coefficients} and polynomials $P_i(\Delta_\phi)$ in equation \eqref{polys-Pi}. Further subleading corrections in equation \eqref{rho-GFF-intro} go in powers of $\Delta^{-\frac14}$ and are straightforwardly obtained - several of them are given in the attached Mathematica notebook. We see that the leading term in the exponent is proportional to $\Delta^{3/4}$, as is the case for local CFTs in four spacetime dimensions. We compare the asymptotic formula \eqref{rho-GFF-intro} to exact multiplicities of operators in GFF and find excellent agreement. These comparisons are shown in Figure \ref{fig:GFF_spectrum}.
\smallskip

In the generalised free theory, the particle number is a well-defined quantum number and can be used to refine the spectral density. By focusing on sectors with a fixed number of particles, $n$, we obtain the asymptotic multiplicities of primaries
\begin{equation}\label{GFF-U1-asymptotics-spectrum-intro}
    N(n,\ell,k)\sim \frac{(2\ell+1)\Gamma\left(n-\frac52\right)}{8\sqrt{\pi}\,n!(n-2)!(3n-7)!}\;k^{3n-7} \quad \text{as} \quad k\to\infty\,, \qquad \text{for }n\geq4\ .
\end{equation}
Here, the scaling dimension of the operator is $\Delta = n\Delta_\phi + k$ and it corresponds schematically to an operator with $n$ copies of the field $\phi$ and $k$ derivatives. For two- and three-particle states, the formula \eqref{GFF-U1-asymptotics-spectrum-intro} is appropriately modified, see equation \eqref{triple-trace-asymptotics} and the discussion around it. While we derive equation \eqref{GFF-U1-asymptotics-spectrum-intro} by analysing the thermal partition function, the same result may be obtained in a more combinatorial manner, similar to \cite{deMelloKoch:2018klm}. We compare with this alternative approach in Appendix \ref{app:deMelloKoch}. In addition to the regime of fixed $n$ and $\Delta\to\infty$, there are several interesting limits in which both $n$ and $k$ are sent to infinity at different rates. Indeed, the latter are ‘responsible' for the total spectral density \eqref{rho-GFF-intro}, as we briefly discuss at the end of Section \ref{subsect-GFF-U1-spectrum}.
\smallskip

Having studied the spectrum, we turn to the OPE coefficients. Our main result is the asymptotic density of averaged HHL OPE coefficients
\begin{equation}\label{GFF-OPE-asymptotics-intro}
    \overline{\lambda_{\phi^2\mathcal{OO}}^a}(\Delta,\ell) \sim \sqrt{2}\, \binom{-\Delta_\phi}{a}\, \zeta(2\Delta_\phi)\, \mathcal{N}_{a,\ell}\, \left(\frac{30\Delta}{\pi^4}\right)^{\frac{\Delta_\phi}{2}}\Delta^{-2a} \quad\text{as}\quad \Delta\to\infty\ .
\end{equation}
Here, $(\Delta,\ell)$ are the scaling dimension and spin of the operator $\mathcal{O}$ and $a$ labels a $\langle\phi^2\mathcal{OO}\rangle$ tensor structure in a particularly chosen basis, see the discussion around equation \eqref{GFF-OPE-asymptotics} for details. The average is taken among all operators $\mathcal{O}$ sharing the same quantum numbers $(\Delta,\ell)$. Finally, $\mathcal{N}_{a,\ell}$ is a normalisation constant given in equation \eqref{coefficient_N_a_l}. Similarly to the findings of \cite{Buric:2025uqt} for local CFTs, the tensor structure label $a$ establishes a hierarchy of OPE coefficients as $\Delta\to\infty$, the suppression being by the power $\Delta^{-2a}$. However, the dominant OPE coefficients scale differently compared to a local CFT. Indeed the scaling is the same as for a local CFT in four spacetime dimensions. In addition to the leading asymptotics \eqref{GFF-OPE-asymptotics-intro}, we derive several subleading corrections, see equation \eqref{GFF-OPE-asymptotics-a0} and the attached Mathematica notebook. The comparison of asymptotic formulas to exact OPE coefficients, obtained through the low-temperature expansion of $\langle\phi^2\rangle$, are shown in Figures \ref{OPE-GFF-plot-scalars} and \ref{OPE-GFF-plot-spin-1}. Again, we observe excellent agreement.
\smallskip

\begin{figure}[h!]
    \centering
    \begin{minipage}[b]{0.49\textwidth}
        \includegraphics[width=\textwidth]{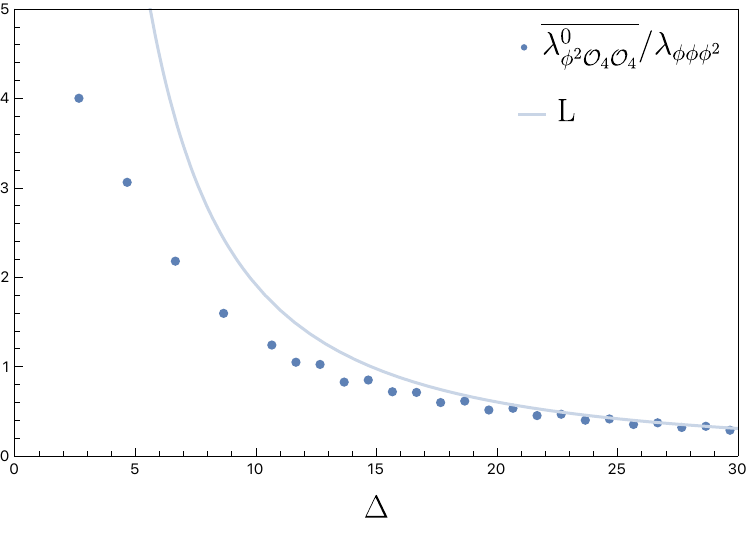}
    \end{minipage}
    \hfill
    \begin{minipage}[b]{0.49\textwidth}
        \includegraphics[width=\textwidth]{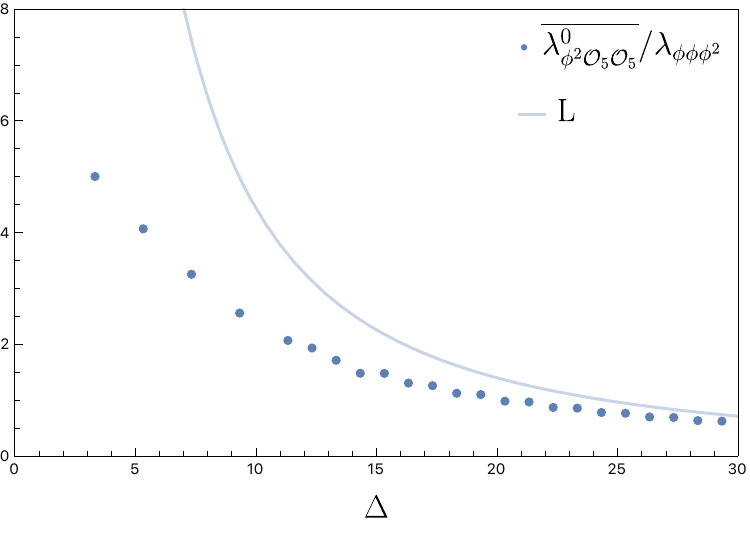}
    \end{minipage}
    \caption{OPE coefficients with four- and five-particle scalars internal operators, $\Delta_\phi = \frac23$. We compare normalised exact OPE coefficients with the leading asymptotic formula \eqref{eq:GFF_fixFamilyAsymp}.}
    \label{fig:OPE_fixed_family_GFF}
\end{figure}

Similarly as for the spectrum, we extend the analysis of OPE coefficients to sectors with a fixed particle number. This leads to the asymptotic formula
\begin{equation}\label{OPE-fixed-n-asymptotics-GFF-intro}
    \overline{\lambda_{\phi^2\mathcal{OO}}^0}(\Delta,\ell,n)\sim\frac{2\sqrt{2} n \,(3n-6)!}{3(-7+2n+2\Delta_\phi)\Gamma(3n+2\Delta_\phi-9)}\;\Delta^{2\Delta_\phi-3}\,, \quad\text{as}\quad \Delta\to\infty\ .
\end{equation}
We have restricted our attention to the case of the {\it dominant} tensor structure, $a=0$, and to the very leading order of asymptotic behaviour, although extensions are possible. The formula \eqref{OPE-fixed-n-asymptotics-GFF-intro} applies to $n\geq4$. It is compared to exact OPE coefficients in Figure \ref{fig:OPE_fixed_family_GFF}.
\medskip

Our study of weakly interacting theories in AdS$_4$ begins with anomalous dimensions which, to leading order in perturbation theory, 
are induced by a $\lambda_4\Phi^4$ interaction term in the bulk. Anomalous dimensions of double-trace (i.e. two-particle) operators in this theory are known since \cite{Fitzpatrick:2010zm}. Our new results include anomalous dimensions of low-lying operators with any number of particles, see e.g. Figure \ref{plot-low-lying-anom-dim}, and certain infinite families of operators, such as those given in equations \eqref{anom-dim-nDelta}, \eqref{anom-dim-nDelta-plus-2} and the Appendix~\ref{AA:Anomalous dimensions}. The main new result of this part is the asymptotic form of anomalous dimensions for $n$-particle primary states
\begin{equation}\label{finite-n-anom-dim-asymptotics-intro}
    \bar{\gamma}(\Delta,\ell,n)\sim \frac{3(3n-7)(3n-8)n(n-1)(n-2)}{2\pi^2(2n-7)}\Delta^{-2}\,, \quad\text{as}\quad \Delta\to\infty \ .
\end{equation}
Here, $n\geq4$ and $\ell$ are kept fixed and the bar signifies the fact that anomalous dimensions are averaged over all operators with the same quantum numbers. Both the exact and asymptotic results are derived using the integral representation \eqref{anomalous-dimensions-density} for the anomalous dimensions which can be controlled in either small or large $\Delta$ limits. Even the very leading asymptotic formula \eqref{finite-n-anom-dim-asymptotics-intro} approximates the exact anomalous dimensions well starting from \mbox{$\Delta- n\Delta_\phi\sim5$}, see Figure \ref{fig::anomal_dim_comparison_fixedN}.
\smallskip

Finally, we obtain various results for OPE coefficients of the form $\lambda_{\phi\mathcal{OO}}$. To leading order, these are induced by 
a cubic interaction term $\lambda_3\Phi^3$ in the bulk. Our starting observation, already pointed out in \cite{Gobeil:2018fzy}, is that the thermal one-point function of $\phi$ in this theory is closely related to thermal conformal blocks, see equation \eqref{cubic-one-point-function}. This allows to obtain all low-lying OPE coefficients just from the low-temperature expansion of the blocks without doing any AdS integrals. For a selection of results, see Figure \ref{OPE-exact-plot}, equations \eqref{leading-OPE-family-cubic}, \eqref{OPE-cubic-nDelta-plus-2} and Appendix~\ref{AA:OPE coefficients}. In addition, we derive asymptotic expansions of HHL OPE coefficients. To this end, we develop a new high-temperature expansion of scalar-external scalar-exchange conformal blocks. The latter are also of interest in their own right. To lowest orders,
\begin{equation}
    g^{\Delta_\phi,0}_{\Delta,0}(\beta,\Omega,s) = \frac{b_{0,0,0}}{\beta^{\Delta_\phi}
   \left(1+s\Omega^2\right)^{\frac{\Delta\phi}{2}}} \left(1+\beta^2\,\frac{c_1(1-s \Omega^4)+c_2\,\Omega^2(1-s)}{(1+s\,\Omega^{2})}+O(\beta^4)\right)\ .
\end{equation}
Here, $\beta$ is the inverse temperature, variables $\Omega,s$ are defined in Section \ref{S:Background} and coefficients $b_{0,0,0}$, $c_1$, $c_2$ are given in equations \eqref{coefficient-b000} and \eqref{coefficients-c1-c2}, respectively. Along the way, we also obtain several new exact expressions for thermal conformal blocks at zero chemical potential for spinning exchanged fields. For instance, the blocks at vanishing chemical potential and with extremal tensor structure label $a=\ell$ are given in terms of the generalised $_3F_2$ hypergeometric function
\begin{align}\label{exact-blocks-subdominant-intro}
    & g^{\Delta_\phi,a=\ell}_{\Delta,\ell}(\beta,\Omega=0,s) = \,c_{\ell,\ell}\; 
    \frac{16^{\ell}\,\left(\tfrac12\right)_{\ell}\,\left(\tfrac32-\tfrac{\Delta_{\phi}}{2}\right)_{\ell}^{2}
    }{\left(2-\Delta\right)_\ell\,\left(2\Delta-2\right)_{\ell}\,\left(\Delta+\ell\right)_\ell} \left(\frac{1}{4\sinh^2\frac{\beta}{2}}\right)^{\Delta+\ell}
    \\[1ex]
    & \hskip1cm
    _3F_2\left(\Delta+\ell-1,\Delta+\ell-\frac{\Delta_\phi}{2},\Delta+\ell+\frac{\Delta_\phi-3}{2};\Delta+2\ell,2 \Delta+\ell-2;\;\frac{1}{4\sinh^2\frac{\beta}{2}}\right)\ .\nonumber
\end{align}
At non-zero $\Omega$, the blocks are much more complicated. However, for the the exchange of scalar operators, it is possible to write a closed form expansion as a three-variable hypergeometric function, see equation \eqref{exact-full-scalar-block}. Some other exact results are collected in Appendix~\ref{A:New results on blocks with external scalars}.
\smallskip

Returning to OPE coefficients induced by the cubic interaction, we use the above results for conformal blocks to derive the asymptotic behaviour
\begin{align}
    & \overline{\lambda_{\phi\mathcal{OO}}^a}(\Delta,\ell) \sim C_a^{(\Delta_\phi>2)}\, \mathcal{N}_{a,\ell}\, \left(\frac{30\Delta}{\pi^4}\right)^{\frac{\Delta_\phi}{4}}\Delta^{-2a} \hskip0.75cm\text{for}\quad \Delta_\phi>2\,,\\
    & \overline{\lambda_{\phi\mathcal{OO}}^a}(\Delta,\ell) \sim C_a^{(\Delta_\phi<2)}\, \mathcal{N}_{a,\ell}\, \left(\frac{30\Delta}{\pi^4}\right)^{\frac12}\Delta^{-2a} \hskip1cm\text{for}\quad 1<\Delta_\phi<2\,,
\end{align}
as $\Delta\to\infty$. The coefficients $C_a^{(\Delta_\phi)}$ are given in equations~\eqref{C>}-\eqref{C<}. As explained in Appendix \ref{A:Flat space limit}, the first case admits a well-defined (convergent) flat space limit. Similarly as in GFF, we derive many subleading orders to the above formula. These are required in order to approximate the exact OPE coefficients well down to $\Delta\sim15$, as shown in Figures \ref{cubic-OPE-plot-spin-1} and \ref{cubic-OPE-plot-spin-2}. The finite particle number asymptotic OPE coefficients are also derived and read
\begin{equation}\label{cubic-finite-particle-number-OPE-intro}
    \overline{\lambda_{\phi\mathcal{OO}}^0}(\Delta,\ell,n)\sim
    \begin{cases}
        K_n^{(\Delta_\phi>2)}\Delta^{\Delta_\phi-3}, \quad &\text{if }\Delta_\phi>2\\
        K_n^{(\Delta_\phi<2)}\Delta^{-1}, \quad &\text{if }1<\Delta_\phi<2
    \end{cases}
\end{equation}
with coefficients $K_n^{(\Delta_\phi)}$ given in equations \eqref{K>}-\eqref{K<}. These are compared to the low lying exact data in Figure \ref{fig:OPE_fixed_family_hol}.
\medskip

The paper is organised as follows. In Section \ref{S:Background}, we review some facts about thermal partition and one-point functions, their character/conformal block decompositions and associated inversion formulas. Section \ref{S:Generalised free field} applies these techniques to the generalised free theory, while Section \ref{S:Interacting Scalar Fields in AdS$_4$} studies weakly interacting theories on AdS$_4$. In the final Section \ref{S:Summary and discussion} we discuss some open questions and future directions. In this section, we also show how, and to what extent, thermal one-point functions obtained in this paper can be put in a thermal EFT form.

\section{Background}
\label{S:Background}

In this section, we review some definitions and techniques for studying conformal field theories at finite temperature, introduced in \cite{Buric:2024kxo,Buric:2025uqt}, that will be used throughout this work. In Section \ref{sec:Review_definitions}, we introduce the notation for finite-temperature correlation functions and conformal blocks. We present the character/block decomposition of the partition and one-point functions, together with the corresponding inversion formulas that express CFT data in terms of these quantities. Section \ref{sec:Review_blocks} is dedicated to two expansions of conformal blocks for thermal one-point functions, valid respectively at low temperatures and at large exchanged scaling dimensions.

\subsection{Thermal correlators and inversion formulas}
\label{sec:Review_definitions}

We consider three-dimensional conformal field theories at finite temperature, or equivalently CFTs on the manifold $S^1 \times S^2$. Indeed, starting from a CFT on $\mathbb{R}^3$, one can use the plane-cylinder map to pass to the geometry $\mathbb{R} \times S^2$ and then compactify the Euclidean time direction by turning on a finite temperature. Altogether, correlators on $S^1\times S^2$ can be expressed as traces over the Hilbert space of products of local operators and (exponentials of) conformal symmetry generators.

\paragraph{Partition function.} The simplest correlator to consider is the partition function
\begin{equation}\label{partition-function}
    \gcZ(\beta,\mu) = \tr_\mathcal{H}\left(e^{-\beta D} e^{i\mu M}\right) =  \tr_\mathcal{H}\left(q^D y^M\right)\ .
\end{equation}
Here, $\mathcal{H}$ is the Hilbert space of the theory, $D$ the generator of dilations on $\mathbb{R}^3$ and $M$ any one of the rotation generators. We shall choose $M$ as the generator of rotations around the $x^1$-axis. Thus, we regard the partition function as a function of two variables, the inverse temperature $\beta$ and the {\it chemical}, or {\it angular}, potential $\mu$. For various purposes in this work, it will be convenient to use different sets of variables rather than $(\beta,\mu)$. We shall write
\begin{equation}\label{sets-of-variables}
    q = e^{-\beta} \,, \qquad y = e^{i\mu} = e^{i\beta\Omega} \ .
\end{equation}
The partition function contains all the information about the spectrum of the theory. This information is encoded in the decomposition into conformal characters
\begin{equation}\label{eq:character_dec}
    \gcZ(q,y) = \sum_{\Delta,\ell} n_{\Delta,\ell} \, \chi_{\Delta,\ell} (q,y)\ .
\end{equation}
Here, $n_{\Delta,\ell}$ denote multiplicities of primary operators and $\chi_{\Delta,\ell}$ are the characters of $SO(3,2)$ representations, see equation \eqref{characters}. We define the density of primary states $\rho(\Delta,\ell)$, also referred to as the spectral density, by
\begin{equation}\label{function-character-decomposition}
    \gcZ(q,y) = \int_0^\infty d\Delta \sum_{\ell=0}^\infty \rho(\Delta,\ell)\chi_{\Delta,\ell}(q,y) \ .
\end{equation}
In general, CFTs are expected to have discrete spectra, so $\rho(\Delta,\ell)$ is a sum of delta-functions. Using the orthogonality of characters \eqref{orthogonality-characters}, it is possible to invert the decomposition \eqref{function-character-decomposition} and write the spectral density in terms of the partition function,
\begin{equation}\label{inverse-character-transform}
    \rho(\Delta,\ell) = \frac{1}{2\pi i} \int\limits_{\gamma-i\infty}^{\gamma+i\infty} 
    d\beta \int\limits_{-\pi}^{\pi}d\mu\ \omega(\beta,\mu)\, \chi_{3-\Delta,\ell}(\beta,\mu)\, 
    \gcZ(\beta,\mu)\ .
\end{equation}
Here, $\gamma$ is some positive real number and the measure $\omega(\beta,\mu)$ is given in equation \eqref{Haar-measure}. Finally, note that one may put $\mu=0$ and consider the partition function as a function of a single variable $\beta$. From this quantity, one can extract the spectral density summed over spin.

\paragraph{One-point functions.}

The next simplest correlator after the partition function is the one-point function
\begin{equation}\label{one-point-function}
    \langle \phi(x) \rangle_{q,y} = \frac{1}{\gcZ(q,y)} \tr_\mathcal{H} \left(\phi(x) q^D y^M\right) \ .
\end{equation}
Here, $\phi$ is a local primary field inserted at the point $x\in\mathbb{R}^3$. In this work, we shall restrict to scalar operators $\phi$ and denote its scaling dimension by $\Delta_\phi$. See \cite{Buric:2024kxo} for general properties of spinning thermal one-point functions.
\smallskip

Unlike flat-space one-point functions, the one-point functions \eqref{one-point-function} neither vanish generically nor are they fixed by conformal symmetry. Rather, they admit a nontrivial conformal block decomposition,
\begin{equation}\label{eq:block_decomposition}
    \gcZ(q,y)\langle\phi(x)\rangle_{q,y} = r^{-\Delta_\phi} \sum_{\mathcal{O},a} \lambda_{\phi\mathcal{O}\mathcal{O}}^a\, g_{\Delta,\ell}^{\Delta_\phi,a}(q,y,s)\ .
\end{equation}
Here, $(r,\theta,\varphi)$ are spherical polar coordinates on $\mathbb{R}^3$ and $s = \sin^2\theta$. The sum runs over all primary operators $\mathcal{O}$ of the theory and over all possible $\langle \phi \mathcal{O} \mathcal{O} \rangle$ three-point tensor structures. The latter are labelled by $a$. For example, if $\mathcal{O}$ has spin $\ell$, $a$ runs over $(\ell+1)$ values $0,1,\dots,\ell$. The $\lambda_{\phi\mathcal{O}\mathcal{O}}^a$ are corresponding OPE coefficients. Finally, $g_{\Delta,\ell}^{\Delta_\phi,a}$ are conformal blocks -- in addition to $\Delta_\phi$, they are labelled by quantum numbers $(\Delta,\ell)$ of the internal operator $\mathcal{O}$ and the tensor structure label $a$.
\smallskip

For theories with operator degeneracies, we define the averaged OPE coefficients by
\begin{equation}\label{averaged-OPE-coefficients}
    \overline{\lambda_{\phi \mathcal{O}\mathcal{O}}^a} = 
    \left( \sum_{i=1}^{n_{\Delta,\ell}} 
    \lambda_{\phi \mathcal{O}_i \mathcal{O}_i}^a \right) / n_{\Delta,\ell}\ .
\end{equation}
Similarly to the spectral density, we may pass to the continuum OPE density by writing
\begin{equation}\label{1pt-function-decomposition}
   \gcZ(\beta,\mu) \langle\phi(x)\rangle_{\beta,\mu} = 
   r^{-\Delta_\phi}\int_0^\infty d\Delta \sum_{\ell=0}^\infty \,  \rho(\Delta,\ell)\sum_{a=0}^\ell \overline{\lambda_{\phi\mathcal{O}\mathcal{O}}^a}
   (\Delta,\ell)\,g_{\Delta,\ell}^{\Delta_\phi,a}(\beta,\mu,\theta)\ .
\end{equation}
Again, it is possible to invert the decomposition \eqref{1pt-function-decomposition} and write the OPE density in terms of the thermal one-point function. We have
\begin{equation}\label{inversion-formula}
   \rho(\Delta,\ell) \overline{\lambda_{\phi\mathcal{O}\mathcal{O}}^a}(\Delta,\ell)=
   \frac{r^{\Delta_\phi}}{2\pi i} \int \Measure\  
   g^{3-\Delta_\phi,a}_{3-\Delta,\ell}(\beta,\mu,\theta)\, \gcZ(\beta,\mu) 
   \langle\phi(x)\rangle_{\beta,\mu}\,,
\end{equation}
where the measure $\Measure$ is given by
\begin{equation}\label{measure-blocks}
     \Measure(\beta,\mu,\theta) = \frac12 \omega(\beta,\mu)\sin\theta\  .
\end{equation}
The integration region for $\theta$ is $[0,\pi]$, while over $\beta$ and $\mu$ it is the same as in equation \eqref{inverse-character-transform}. For a derivation of equation \eqref{inversion-formula}, based on Laplace-Casimir differential equations, see \cite{Buric:2025uqt}. The main intermediate step leading to the inversion formula is the orthogonality relation satisfied by conformal blocks,
\begin{equation}\label{orthogonality_blocks_integrated}
    \int \Measure\ g^{\Delta_\phi,a}_{\Delta_1,\ell_1}(\beta,\mu,\theta)\, g^{3-\Delta_\phi,b}_{3-\Delta_2,\ell_2}(\beta,\mu,\theta) = 2\pi i\, \delta(\Delta_1-\Delta_2)\delta_{\ell_1,\ell_2} \delta_{a,b}\ .
\end{equation}

\paragraph{Limit to $S^1 \times \mathbb{R}^2$.} In the literature, it is very common to refer to CFTs at finite temperature as those on the geometry $S^1 \times \mathbb{R}^{d-1}$. The latter arises as a limit of our geometry in which the radius of $S^{d-1}$ is sent to infinity. Due to conformal invariance, one can equivalently think of taking the limit $\beta\to0$. In other words, the limit is that of high temperatures. One-point functions on $S^1 \times \mathbb{R}^{d-1}$ are fixed by conformal invariance up to a dynamical coefficient. These coefficients are in principle fully determined by the flat space CFT data - however, they are related to the latter in a very complicated way. Therefore, when working on $S^1 \times \mathbb{R}^{d-1}$ it is often useful to work directly with one-point coefficients and not try to express them in terms of the flat space OPE coefficients, see \cite{Iliesiu:2018fao,Buric:2024kxo} for more details.

\subsection{Expansions of conformal blocks}
\label{sec:Review_blocks}

The conformal block decomposition \eqref{eq:block_decomposition} and the inversion formula \eqref{inversion-formula} are rather abstract and only become useful once sufficient control over conformal blocks is available. In this subsection, we describe, following \cite{Buric:2024kxo,Buric:2025uqt}, two expansions of conformal blocks that allow for efficient extraction of CFT data. For other works on one-point blocks on $S^1\times S^{d-1}$, see \cite{Alkalaev:2024jxh,Ammon:2025cdz,Alkalaev:2026sha}.
\smallskip

All the results that we shall discuss follow from the Casimir differential equation
\begin{equation}\label{Casimir-eqn}
    \mathcal{C} _{\Delta_\phi} \, g^{\Delta_\phi,a}_{\Delta,\ell}(q,y,s) = 
    -2\Big(\Delta (\Delta-3) + \ell
    (\ell+1)\Big) g^{\Delta_\phi,a}_{\Delta,\ell}(q,y,s)\,,
\end{equation}
satisfied by thermal conformal blocks. The explicit expression for the Laplace-Casimir operator $\mathcal{C} _{\Delta_\phi}$ is given in equations \eqref{Laplace-Casimir-op}-\eqref{D2-s-variable}.

\paragraph{Small $q$-expansion.} At low temperatures, i.e. small $q$, we expand the blocks as
\begin{align}\label{eq:gqpower}
    g^{\Delta_\phi,a}_{\Delta ,\ell }(q,u,s) & = 
    q^{\Delta }\sum_{n_1=0}^{\infty} q^{n_1}\, f_{n_1}(u,s) \,, \\
    f_{n_1}(u,s)=\sum_{n_2,n_3} A_{(n_1,n_2,n_3)} u^{n_2} s^{n_3}\,, \qquad & \text{with} \qquad  
    0\leq n_2\leq n_1+ \ell  \,, \ \ 0\leq n_3 \leq n_2\,,        
\end{align}
where
\begin{equation}
    u = y + y^{-1} -2 \ .
\end{equation}
The coefficients $A_{(n_1,n_2,n_3)}$ are uniquely fixed by solving the Casimir equation order by order in $q$, with the appropriate initial conditions for $n_1=0$, \cite{Buric:2024kxo}. In practice, the Casimir equation is first re-written as a set of recursion relations for the $A_{(n_1,n_2,n_3)}$-s, which proves to be much more efficient, see Section 3 of \cite{Buric:2025uqt} for full details. The recursion relation method is inspired by similar treatments of four-point function conformal blocks in flat space, \cite{Hogervorst:2013sma,Costa:2016xah}. A Mathematica code implementing the low temperature expansion is available at \href{https://gitlab.com/russofrancesco1995/thermal_blocks}{gitlab.com/russofrancesco1995/thermalblocks}.

\paragraph{Large $\Delta$-expansion.} Another expansion of blocks that is useful for the extraction of CFT data is that in large internal dimension $\Delta$. We can strip off a prefactor from the block
\begin{equation}\label{g-f-large-Delta}
    g^{\Delta_\phi,a}_{\Delta,\ell}(q,y,s) = 
    q^{\Delta} F^{\Delta_\phi,a}_{\Delta,\ell}(q,y,s)\,,
\end{equation}
and expand the remaining function $F^{\Delta_\phi,a}_{\Delta,\ell}$ as
\begin{equation}\label{large-Delta-expansion}
    F(q,y,s) = \sum_{n=0}^\infty F^{(n)}(q,y,s) \Delta^{-n} \ .
\end{equation}
Similarly to the discussion above, solving the Casimir differential equation order by order in $\Delta^{-1}$ uniquely determines all terms once the initial function $F^{(0)}(q,y,s)$ has been specified. In turn, the latter is fixed by the leading term in the low-temperature expansion. For more details, see \cite{Buric:2025uqt}.

\section{The Generalised Free Field}
\label{S:Generalised free field}

In this section, we study the generalised free theory (GFF) of a real scalar field $\phi$ in three dimensions. This is equivalent to a free massive scalar in AdS$_4$ with mass $m^2 = \Delta_\phi(\Delta_\phi-3)$. Using the finite temperature partition function and the one-point function of $\phi^2$ we derive the asymptotic spectral density $\rho(\Delta,\ell)$ and $\lambda_{\phi^2\mathcal{OO}}^a$ OPE coefficients in the limit $\Delta\to\infty$, with fixed $\ell$. We also derive spectral densities and OPE coefficients for families of operators $\mathcal{O}$ with fixed particle number $n$.

\subsection{Partition function and spectral density}

This first subsection is devoted to spectral densities of the GFF. Below we shall 
first recall the form of the partition function before studying its high temperature expansion. Then we apply the inversion formula \eqref{inverse-character-transform} to compute the asymptotic expansion of the (spin resolved) density of primaries in GFF. As we explained in the introduction, this asymptotics is dominated by states with large particle numbers. Fixing the particle number leads to 
a different behaviour that we determine in the last 
part of this subsection.

\subsubsection{The partition function} 

The Hilbert space $\mathcal{H}$ of the generalised free theory is the Fock space 
built upon the space $\mathcal{H}_1$ of one-particle states,
\begin{equation}
    \mathcal{H} = S^0\mathcal{H}_1 \oplus S^1\mathcal{H}_1 \oplus 
    S^2\mathcal{H}_1 \oplus\dots\ .
\end{equation}
The one-particle space is that of the state representation $V_{\Delta_\phi} = 
V_{\Delta_\phi,\ell_\phi = 0}$ of the fundamental scalar field $\phi$, i.e.\ it is given 
by $\mathcal{H}_1 \cong V_{\Delta_\phi}$. The character of this state 
representation can be found in equation  \eqref{characters}. Consider some state 
in $V_{\Delta_\phi}$ of conformal weight $\Delta$ and $M$-eigenvalue $m$. 
Multi-particle states constructed out of this single excitation contribute 
a factor $(1-q^\Delta y^m)^{-1}$ to the partition function, as is seen by 
summing a geometric series. Since each single-particle mode behaves 
independently, the full partition function is given by  
\begin{equation}\label{partition-function-1}
    \gcZ_0(q,y) = \prod_{\Delta,m} \frac{1}{1 - q^\Delta y^m}\ .
\end{equation}
The product runs over a basis of eigenstates of $D$ and $M$ in $V_{\Delta_
\phi}$. We may manipulate this product into an alternative expression 
for $\gcZ_0$ that will be useful for the analysis below
\begin{equation}\label{argument-Z-char}
    \log \gcZ_0(q,y) = -\sum_{\Delta,m}\log(1 - q^\Delta y^m) = 
    \sum_{\Delta,m} \sum_{n=1}^\infty \frac{(q^\Delta y^m)^n}{n} 
    = \sum_{n=1}^\infty\frac{1}{n} \sum_{\Delta,m} q^{n\Delta} y^{nm}\ .
\end{equation}
In the derivation we used the Taylor expansion of $\log(1-x)$ and exchanged the order of two 
sums. Notice that the second sum is nothing but the character of $V_{\Delta_\phi}$
(see equation  \eqref{characters}), evaluated at shifted values of fugacities $q,y$, 
i.e. it is simply $\chi_{\Delta_\phi,0}(q^n,y^n)$. With this observation equation  
\eqref{argument-Z-char} becomes 
\begin{equation}\label{partition-function-2}
    \log \gcZ_0(q,y) = \sum_{n=1}^\infty\frac{1}{n}  
    \frac{q^{n\Delta_\phi}}{(1-q^n)(1-q^n y^n)\left(1-\frac{q^n}{y^n}\right)}\ .
\end{equation}
According to the holographic dictionary, this partition function counts 
multi-particle states of the free massive scalar field in AdS$_4$ with 
any number $n$ of particles. It is therefore quite natural to introduce 
another fugacity $g$ that keeps track of the particle number, i.e. we 
introduce the following refined partition function
\begin{equation}\label{eq:Z_U(1)}
    \log \gcZ_0(q,y,g) = \sum_{n=1}^\infty\frac{1}{n}  
    \frac{q^{n\Delta_\phi} g^n}{(1-q^n)(1-q^n y^n)\left(1-\frac{q^n}{y^n}
    \right)} \,.
\end{equation}
We shall often refer to $\gcZ_0(q,y,g)$ as the grand canonical partition 
function. We can expand this function in a power series 
\begin{equation} \label{eq:canonicalZ}
  \gcZ_0(q,y,g) = \sum_{n=0}^\infty g^n 
  \cZ^{(n)}_0(q,y)\ , \quad \textrm{ with} \quad 
  \cZ^{(n)}_0(q,y) = \tr_{\mathcal{H}^{(n)}}\left( q^D y^M\right) 
\end{equation}
The coefficient functions $\cZ^{(n)}_0(q,y)$ count the 
number of fields that involve $n$ constituents $\phi$. In the dual 
theory, $\cZ^{(n)}_0$ is the contribution of $n$-particle 
states to the partition function.  We shall also refer to the 
functions $\cZ^{(n)}_0(q,y)$ as canonical partition 
functions.

\subsubsection{High-temperature expansion of the partition function}

The goal of this subsection is to derive an asymptotic expansion of the 
(logarithm of the) partition function \eqref{partition-function-2} near 
$\beta=0$. To this end, we shall apply the method reviewed in \cite{Zagier}.\footnote{The relevant equation (45) of \cite{Zagier} 
contains a typo. Terms $b_\lambda \zeta(-\lambda)(-t)^\lambda$ should 
instead read $b_\lambda \zeta(-\lambda)t^\lambda$.} The partition 
function\footnote{For the moment we set the fugacity $g$ we introduced at the end of the previous subsection to $g=1$.} takes the form of an ‘image sum',
\begin{equation}
    \beta^{-1} \log\gcZ_0 = \sum_{n=1}^\infty \fZ(n\beta,\Omega)\,, 
    \qquad \fZ(\beta,\Omega) = \frac{e^{\left(\frac32-\Delta_\phi\right)\beta}}{4\beta\sinh\frac{\beta}{2}\, (\cosh\beta-\cos(\beta\Omega))}\ .
\end{equation}
The asymptotic behaviour of $\gcZ_0$ near $\beta=0$ is determined 
by the behaviour of the simpler function $\fZ$ in the same limit. We expand 
$\fZ$ in a power series of the form
\begin{align}
     &\fZ(\beta,\Omega) = \sum_{m=-4}^\infty \fZ_m(\Omega) \beta^m 
     = \frac{1}{(1+\Omega^2)\beta^4} + \\[2mm]
    & + \frac{3-2\Delta_\phi}{2(1+\Omega^2)\beta^3} 
    + \frac{12 - 18\Delta_\phi + 6\Delta_\phi^2 + \Omega^2}{12(1+\Omega^2)\beta^2} 
    + \frac{(3-2\Delta_\phi)(3-6\Delta_\phi+2\Delta_\phi^2+\Omega^2)}{24(1+\Omega^2)\beta} + \dots \ . \nonumber
\end{align}
This implies the following asymptotic expansion of $\beta^{-1} \log\gcZ_0$,
\begin{equation*}
    \beta^{-1} \log\gcZ_0 = 
    \sum_{m=-4}^{-2} \zeta(-m) \fZ_m(\Omega)\beta^m 
    - \fZ_{-1}(\Omega)\frac{\log\beta}{\beta} + \frac{\IZ(\Omega)}{\beta} 
    + \sum_{m\geq0} \fZ_m(\Omega) \zeta(-m) \beta^m\,,
\end{equation*}
where
\begin{equation}
   \IZ(\Omega)= \int_0^\infty d\beta \left( \;\fZ(\beta,\Omega)-\sum_{m=-4}^{-2} 
   \fZ_m(\Omega) \beta^m - \frac{\fZ_{-1}(\Omega)e^{-\beta}}{\beta} \right)\ .
\end{equation}
The function $\IZ(\Omega)$ may be evaluated order by order in $\Omega$ 
by expanding the integrand and then integrating term by term. Denoting
\begin{equation}
  \IZ(\Omega)=\sum_{k=0}^{\infty}\IZ_{2k}\;\Omega^{2k}\,,
\end{equation}
we have, for example,
\begin{equation}\label{GFF-I0Z}
    \IZ_0 = \frac12 \left(\zeta_H'\!\left(-2,\Delta_{\phi}\right)+
    (3-2\Delta_{\phi})\,\zeta_H'\!\left(-1,\Delta_\phi\right)
+(\Delta_\phi-2)(\Delta_\phi-1)\,\zeta_H'\!\left(0,\Delta_\phi\right)
\right)\,,
\end{equation}
where $\zeta_H'(s,a)=\partial_s\zeta_H(s,a)$ is a derivative of the Hurwitz zeta function. Due to trivial zeros of the Riemann $\zeta$-function, among the positive powers of $\beta$ in the expansion of $\beta^{-1} \log\gcZ_0$, only the odd ones come with non-vanishing coefficients, leading to the final expression
\begin{align}\label{GFF-highT-logZ}
    \log\gcZ_0 &= \sum_{m=-4}^{-2} \zeta(-m) \fZ_m(\Omega)\beta^{m+1}\\[2mm]
    & \quad \quad - \fZ_{-1}(\Omega)\log\beta + \IZ(\Omega) -\frac{\fZ_0(\Omega)}
    {2}\beta+\sum_{m=1}^\infty \fZ_{2m-1}(\Omega) \zeta(1-2m) \beta^{2m} \ .\nonumber
\end{align}
We note that this high-temperature behaviour is not of the form expected for a 
generic CFT in three dimensions. Indeed, for a generic (gapped) CFT$_d$, the 
thermal EFT, \cite{Benjamin:2023qsc}, predicts that $\log \mathcal{Z}$ has a leading 
singularity of the type $\beta^{1-d}$, followed by corrections in even powers 
of $\beta$ (compared to the leading term). Thus, we see that the partition 
function \eqref{GFF-highT-logZ} resembles that of a theory in \textit{four}
dimensions, with, however, both even and odd powers of $\beta$, as well as 
$\log \beta$ appearing in the expansion. This is in line with the fact that 
the GFF theory is a non-local CFT$_3$, but can be viewed as a local QFT in four 
dimensions, namely the free massive scalar on AdS$_4$.

\subsubsection{Asymptotics of the density of primaries}
\label{subsect-GFF-asymptotics-spectrum}

\begin{figure}[h!]
    \centering
    \begin{minipage}[b]{0.49\textwidth}
        \includegraphics[width=\textwidth]{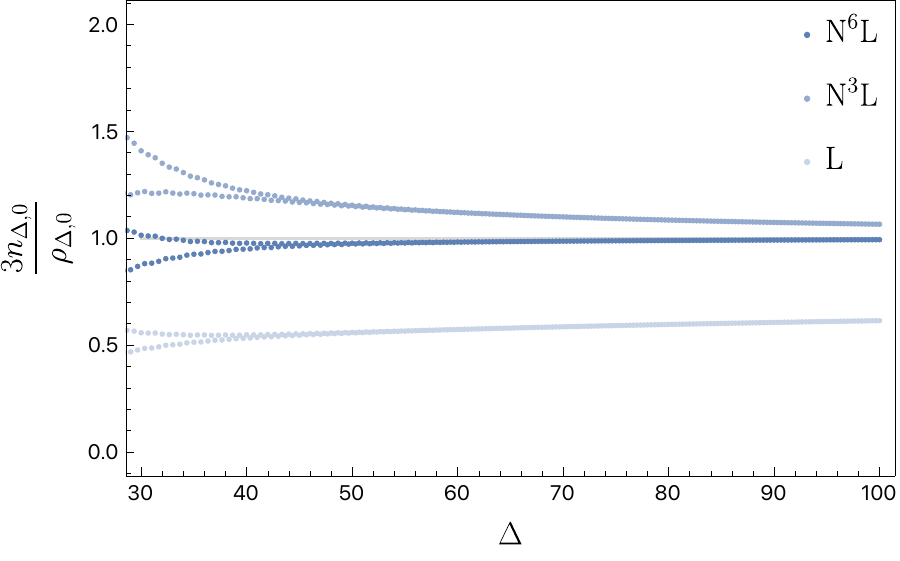}
    \end{minipage}
    \hfill
    \begin{minipage}[b]{0.49\textwidth}
        \includegraphics[width=\textwidth]{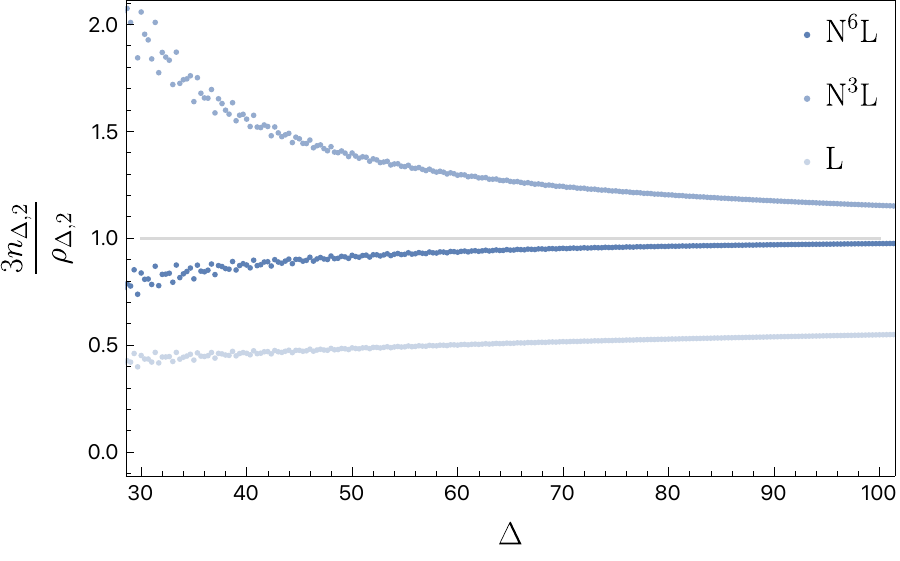}
    \end{minipage}
    \caption{GFF spectral density of scalars operators with $\Delta_\phi = \frac23$ (left) and spin two operators with $\Delta_\phi = \frac43$ (right). Here $\rho_{\Delta,\ell}$ stands for the asymptotic formula in \eqref{rho-GFF}, provided with more subleading terms. Instead, $n_{\Delta,\ell}$ are multiplicities of primaries operators as in \eqref{eq:character_dec}. Different curves correspond to keeping a different number of terms in the asymptotic expansion. In particular, the notation N$^n$L stands for $n$ terms in the asymptotic series after the leading order.}
    \label{fig:GFF_spectrum}
\end{figure}

Using the inversion formula \eqref{inverse-character-transform}, together with the high-temperature behaviour \eqref{GFF-highT-logZ} of the (logarithm of the) partition function, we can extract the asymptotics of the density $\rho_0(\Delta,\ell)$ of primary states in GFF for $\Delta\to\infty$ and finite $\ell$. In \cite{Benjamin:2023qsc,Buric:2025uqt}, it was described how to perform this analysis in the case of a generic gapped CFT in three dimensions, as well as for the free bosonic theory. However, as we mentioned at the end of the previous subsection, the GFF partition function displays a high temperature behaviour which is of neither of these two types. This implies that our analysis requires some minor adjustments. We start by extending the $\mu$ domain of integration in equation \eqref{inverse-character-transform} to the whole real line. This introduces an error which is non-perturbative (i.e. exponentially supressed) for $\Delta\rightarrow\infty$, and can thus be safely neglected in our asymptotic analysis,
\begin{equation*}
    \rho_0(\Delta,\ell) = \frac{1}{2\pi i} \int\limits_{\gamma-i\infty}^{\gamma+i\infty} 
    d\beta \int\limits_{-\infty}^{\infty}d\mu\ \omega(\beta,\mu)\, \chi_{3-\Delta,\ell}(\beta,\mu)\, 
    \gcZ_0(\beta,\mu) + \text{non-pert}\ .
\end{equation*}
For large $\Delta$, the integral develops a saddle\footnote{This is not the only saddle of the integral, but is the one that contributes the most to the overall result. In particular, the other saddles are exponentially suppressed compared to this dominant one.} on the positive real $\beta$ line and at $\mu\sim0$, the location of which is mainly controlled by the competing effects of the exponential growth of the shadow character $\chi_{3-\Delta,\ell}$ at low temperature with the high-temperature singular behaviour \eqref{GFF-highT-logZ} of $\gcZ_0$. To capture the behaviour of the integrand around this saddle, we introduce a new set of variables $\beta_N,\mu_N$ as
\begin{equation}\label{saddle-point-variables}
\begin{aligned}
    \beta = \, & \frac{\pi}{30^\frac14}\Delta^{-\frac14}+\frac{15^\frac12(2\Delta_\phi-3)\zeta(3)}{2^\frac32\pi^2}\Delta^{-\frac12}+i\beta_N\frac{\pi^\frac12}{2^\frac58 15^\frac18}\Delta^{-\frac58}\,,\\[2mm]
    \mu = \, & \mu_N\frac{3^\frac38 \pi^\frac12}{10^\frac18}\Delta^{-\frac58}\ .
\end{aligned}
\end{equation}
In these new variables, the integral expression for the density of primaries can be written as
\begin{align}
    \rho_0(\Delta,\ell) &= \left(\frac{3}{640}\right)^\frac14 \Delta^{-\frac54}\,e^{\left(\frac{128}{1215}\right)^{1/4}\pi\,\Delta^\frac34+C_1^{(\Delta_\phi)}\Delta^\frac12 + C_2^{(\Delta_\phi)}\Delta^\frac14 + C_3^{(\Delta_\phi)}}\\
    & \hskip4cm\int\limits_{-\infty}^\infty  d\beta_N d\mu_N\ h(\beta_N,\mu_N;\Delta)\,e^{-\beta_N^2-\mu_N^2} + \text{non-pert}\,,\nonumber
\end{align}
where the $C_i^{(\Delta_\phi)}$ are constants depending on $\Delta_\phi$, 
\begin{align}
    C_1^{(\Delta_\phi)} & = \frac{\sqrt{30}}{2\pi^2} (3-2\Delta_\phi)\,\zeta(3)\,,\nonumber\\[2mm]
    C_2^{(\Delta_\phi)} & = \left(\frac{5}{3456}\right)^{1/4}
\left(\pi(\Delta_\phi-1)(\Delta_\phi-2)-\frac{45(3-2\Delta_\phi)^2\zeta(3)^2}{\pi^5}\right)\,,\label{Ci-coefficients}\\[2mm]
    C_3^{(\Delta_\phi)} & = I_0^{\gcZ}-\frac{5(3-2\Delta_\phi)(\Delta_\phi-1)(\Delta_\phi-2)\,\zeta(3)}{8\pi^2}
+\frac{75}{2}\,\frac{(3-2\Delta_\phi)^3\zeta(3)^3}{\pi^8}\,.\nonumber
\end{align}
We recall that $I_0^{\mathcal{Z}}$ is the constant given in equation \eqref{GFF-I0Z}. Finally, the function $h$ has an expansion around $\Delta\rightarrow\infty$ of the form
\begin{equation}
h(\beta_N,\mu_N;\Delta)=\Delta^{-\frac{61}{32}-\frac{\Delta_\phi(12 - 9\Delta_\phi + 2\Delta_\phi^2)}{48}}\sum_{n=0}^\infty h_n(\beta_N,\mu_N)\Delta^{-\frac{n}{8}}\ .
\end{equation}
Exchanging the order of the $\Delta$-expansion and the integral, and then integrating term by term, we obtain our asymptotic formula for the density of primary states in GFF
\begin{align}\label{rho-GFF}
    & \rho_0(\Delta,\ell) = \frac{5^2 3^\frac72}{2^\frac52 \pi^9} (2\ell+1)
\left(\frac{30\Delta}{\pi^4}\right)^{-\frac{101}{32}-\frac{\Delta_\phi(12-9\Delta_\phi+2\Delta_\phi^2)}{48}} e^{\left(\frac{128}{1215}\right)^\frac14 \pi\,\Delta^\frac34 + C_1^{(\Delta_\phi)}\Delta^\frac12 + C_2^{(\Delta_\phi)}\Delta^\frac14 + C_3^{(\Delta_\phi)}} \nonumber\\[1ex]
&\hskip0.4cm\left(1 -\frac{P_0(\Delta_\phi) + P_1(\Delta_\phi)\zeta(3) + P_2(\Delta_\phi)\zeta(3)^2 + P_4(\Delta_\phi) \zeta(3)^4}{960 \sqrt[4]{30} \pi^{11}} \Delta^{-\frac14} + O\left(\Delta^{-\frac12}\right)\right)\ .
\end{align}
Here, the polynomials $P_i(\Delta_\phi)$ read
\begin{align}
    P_0(\Delta_\phi) & = \pi^{12} \left( 45 \Delta_\phi^4 - 270 \Delta_\phi^3  + 565\Delta_\phi^2 - 480\Delta_\phi + 1578\right)\,,\nonumber\\[1mm]
    P_1(\Delta_\phi) & = 300 \pi^8 \left(8 \Delta_\phi^4 - 48\Delta_\phi^3 + 102\Delta_\phi^2+30 \Delta_\phi-153\right)\,,\nonumber\\[1mm]
    P_2(\Delta_\phi) & = - 6750 \pi^6 (3-2 \Delta_\phi)^2 \left(\Delta_\phi^2 - 3\Delta_\phi+2\right)\,,\label{polys-Pi} \\[1mm]
    P_4(\Delta_\phi)&  = 354375 (2\Delta_\phi - 3)^4\ . \nonumber
\end{align}
Note that the exponents of the weight $\Delta$ in the asymptotic density of primaries once again put the 4-dimensional nature of the GFF on display. In any local $d$-dimensional CFT the leading exponent of $\Delta$ in the exponential, for example, is given by $(d-1)/d$. A further feature of equation \eqref{rho-GFF} is that the leading exponential growth is independent of the spin $\ell$. The spin dependence enters only through the prefactor $(2\ell+1)$ and subleading corrections, similarly to gapped and free theories, \cite{Benjamin:2023qsc}. It is a simple matter to generate any number of subleading terms in equation \eqref{rho-GFF}, whose powers go in steps of $\Delta^{-1/4}$. In Figure \ref{fig:GFF_spectrum} we compare the asymptotic formula \eqref{rho-GFF} and its extensions involving further subleading terms, with the exact multiplicities of operators in GFF\footnote{Recall that, when comparing the multiplicities $n_{\Delta,\ell}$ to our asymptotic formula \eqref{rho-GFF} for the density, we have to multiply the former by the spacing between neighbouring operators in the spectrum.}. The darker points correspond to keeping more subleading terms. The comparison illustrates the surprisingly rapid convergence of the asymptotic expansion, even for relatively low operator dimensions.

\subsubsection{Asymptotic density for fixed particle number}
\label{subsect-GFF-U1-spectrum}

As we stressed in the introduction, the high temperature regime of the 
GFF is dominated by fields with a large number $n$ of constituents. This 
explains why the formulas for the densities of primaries we found in the 
previous subsection are very different from those of multi trace operators
with a finite number $n$ of constituent fields. The contributions of the 
latter can be studied systematically. To this end we go back to the  
grand canonical partition function \eqref{eq:Z_U(1)} with the additional
insertion of the fugacity $g$ and analyse the spectrum with a fixed 
particle number separately. 

More precisely, we can can compute the density of $n$-particle primaries  
if we replace the partition function $\gcZ_0$ in the inversion formula
\eqref{inverse-character-transform} by the canonical partition function 
$\cZ^{(n)}_0$, 
\begin{equation}\label{density-inversion-U1}
\rho_0\left(\Delta, \ell,n\right)=
\int\limits_{\gamma-i \infty}^{\gamma+i \infty} \frac{d\beta}{2 \pi i}
\int\limits_{-\pi}^{\pi} d \mu\, \omega(\beta, \mu)\, \chi_{3-\Delta, \ell}
(\beta, \mu)\, \cZ^{(n)}_0(\beta,\mu)\,,
\end{equation}
Compared to the formula \eqref{inverse-character-transform} for the whole 
spectrum of primaries, the integrand in this case is much more tractable, 
since the contribution coming from the partition function involves only a
finite sum. This allows for a simple derivation of the asymptotic spectral
densities in different particle sectors, as well as various exact results 
that we shall discuss presently.
\smallskip

On the one hand, working case by case at small values of $n$, one can perform 
the integration over $\beta,\mu$ exactly and find the density of primaries 
as a sum of delta functions weighted by the multiplicities
\begin{equation}
  \rho_0\left(\Delta,\ell,n\right) = \sum_{k=0}^{\infty} 
  N(n,\ell,k)\delta(\Delta-n\Delta_\phi-k)\ .
\end{equation}
Here, the summation index $k$ counts the number of derivatives of the 
contributing field. This procedure is straightforward, but the obtained 
closed-form expressions for the multiplicities, $N(n,\ell,k)$, very quickly
become cumbersome to write. We explicitly spell out only the first 
nontrivial example, that of triple trace operators
\begin{equation}\label{triple-trace-multiplicity}
\begin{split}
    N(3,\ell,k)&=\frac{1}{288}\Bigg(3(2\ell+1)k^2
  + 6\left(3 + 3(-1)^\ell + 4\ell - 2\ell^2\right)k\\[2mm]
  &+ 33 + 15\ell - 27\ell^2 + 6\ell^3
  + (-1)^\ell\cdot9(5-2\ell)+\frac{32}{\sqrt{3}}
  \sin\frac{(2\ell+1)\pi}{3}\\[1ex]
  &+(-1)^{k+\ell}\left(
      69 + 9(-1)^\ell - 3\ell(7+\ell) - 32\cos\frac{2(\ell-1)\pi}{3} + 3k(12+k)
    \right)\\[2mm]
  &\hspace{-10mm} - \frac{32\left(\sin\frac{2k\pi}{3} + 
  2\sin\frac{2(k+\ell-1)\pi}{3}\right)}
  {\sqrt{3}}+ (-1)^{\lfloor \ell/2\rfloor} \cdot 36 
  \cos\frac{k\pi}{2}+(-1)^\ell \cdot 32 \cos\frac{k\pi}{3}
\Bigg)\ .
\end{split}
\end{equation}
Here, it is understood that the number $k$ of derivatives exceeds the spin
$\ell$, i.e. $k\geq\ell$, since otherwise the multiplicities are zero. 
We have chosen to display the different contributions by grouping them
according to their $k$-dependence. For a given value of spin, one can
distinguish a purely polynomial part in $k$, followed by a number of periodic
corrections with different periodicities (in this case $k\text{ mod}$ $2,3,4,6$). The same type of structure persists for multiplicities of operators with higher particle numbers.
\vskip0.05cm

The multiplicities $N(n,\ell,k)$ also possess a nice 
representation-theoretic interpretation that has been developed in \cite{deMelloKoch:2018klm}, see also Appendix~\ref{app:deMelloKoch} for a brief review. Even though the combinatorics that underlies the 
computation of the exact multiplicities is rather intricate, the behaviour often simplifies in appropriate limits. A particularly well-known case is the large spin limit of the leading triple twist 
trajectory,
\begin{equation*}
    N(3,\ell,\ell) \sim \frac{\ell}{6} \qquad\text{as}\qquad \ell\to\infty\ . 
\end{equation*}
Indeed, this is easily recovered from the exact expression 
\eqref{triple-trace-multiplicity} for the triple twist multiplicities if we specialize to the leading twist family $k = \ell$ and send the spin $\ell$ to infinity.%
\medskip 

In order to estimate the contributions of primaries with fixed number $n$
to the density $\rho_0(\Delta,\ell)$ in equation \eqref{rho-GFF} at fixed spin 
$\ell$ we are interested in a different limit in which the number of
derivatives $k$ is sent to infinity while keeping $\ell$ finite. This 
means that we consider contributions of all twist trajectories, not 
just the leading one. Starting from the inversion formula 
\eqref{density-inversion-U1} and performing a saddle-point analysis, we 
find the leading asymptotic behaviour of the multiplicities to be
\begin{equation}\label{GFF-U1-asymptotics-spectrum}
    N(n,\ell,k)\sim \frac{(2\ell+1)\Gamma(n-\frac52)}
    {8\sqrt{\pi}\,n!(n-2)!(3n-7)!}\;k^{3n-7} \quad \text{as} 
    \quad k\to\infty\,, \qquad \text{for }n\geq4\ .
\end{equation}
Note that the formula is only valid for $n \geq 4$. The case of 
triple trace operators needs to be considered separately, since 
there is a correction to the coefficient multiplying $k^2$,
\begin{equation}\label{triple-trace-asymptotics}
    N(3,\ell,k)\sim \frac{(2\ell+1)+(-1)^{k+\ell}}{96}\;k^2\,,
\end{equation}
as can be checked also from the exact formula 
\eqref{triple-trace-multiplicity}. An alternative derivation of 
these results using the representation-theoretic formulas for 
multiplicities from \cite{deMelloKoch:2018klm} can be found in Appendix~\ref{app:deMelloKoch}. 
\smallskip

To compare the asymptotic behaviour of the finite twist primaries 
in equation \ \eqref{GFF-U1-asymptotics-spectrum} with the density 
\eqref{rho-GFF}, we take into account that $k = 
\Delta - n \Delta_\phi \sim \Delta$. The differences between the 
two formulas are striking: while the growth with $\Delta$ is merely 
polynomial if we keep $n$ fixed, it becomes exponential for the full 
density of states. Such a transition was to be expected. In fact, a
similar behaviour is well understood for the simpler multiplicities 
$P(k,n)$ that count the number of unrestricted partitions of $k$ 
having $n$ or fewer positive parts. Extending earlier work by Hardy 
and Ramanujan, Szekeres obtained the following asymptotic formula 
\cite{Szekeres:1953}, see also \cite{Canfield_1996},  
\begin{equation}
P(k,n) \sim \frac{f(u)}{k} e^{k^{1/2}g(u) + O(k^{-1/6+\epsilon})}
\end{equation} 
with some known functions $g(u)$ and $f(u)$ of $u = n/\sqrt{k}$. This formula applies provided that $n$ is sufficiently large, which 
means that $n \geq k^{1/6}$. This is qualitatively similar to the behaviour of our formula \eqref{rho-GFF}. If, on the other hand, we keep $n$ fixed as $k$ goes to infinity, the numbers $P(k,n)$ behave as $P(k,n) \sim k^{n-1}$. We may compare this behaviour with the polynomial growth we found in equation \eqref{GFF-U1-asymptotics-spectrum} for the spectral density in the canonical ensemble. Even though the spin-resolved counting functions $N(n,\ell,k)$ for GFF are more complicated than the numbers $P(k,n)$ of partitions that were studied by Szekeres, we conclude that asymptotic behaviour \eqref{rho-GFF} of the spectral density is dominated by operators with a large number of constituents, where large now means $n \sim k^{3/4}$. 

\subsection{One-point function and OPE coefficients}

Let us now turn to a discussion of OPE coefficients in the GFF. In complete analogy with our discussion of the spectral density, we shall first review the well-known formula for the thermal one-point function of the field $\phi^2$ in GFF and analyse its high temperature expansion. We then use the inversion formula \eqref{inversion-formula} to compute the asymptotic expansion of the (spin resolved) OPE coefficients $\lambda_{\phi^2\mathcal{O}\mathcal{O}}$ in GFF. Once again, we compute the asymptotic behaviour of the OPE coefficients in the limit of large $\Delta_\mathcal{O}$ from the full thermal one-point function before we restrict to the contributions from fixed particle number $n$.  

\subsubsection{The one-point function of \texorpdfstring{$\phi^2$}{phi2}}

At zero temperature and chemical potential, the two-point function of the fundamental field reads
\begin{equation}
    \langle\phi(x_1)\phi(x_2)\rangle = \frac{1}{|x_{12}|^{2\Delta_\phi}}\ .
\end{equation}
The finite-temperature two-point function is obtained by the method of images. At finite chemical potential, the method instructs us to compute
\begin{equation}\label{two-point-function}
    \langle\phi(x_1)\phi(x_2)\rangle_{q,y} = \sum_{n=-\infty}^\infty \frac{q^{n\Delta_\phi}}{|x_{12}^{(n)}|^{2\Delta_\phi}}\,,
\end{equation}
where the points $x_1^{(n)}$ and $x_2^{(n)}$ are given by
\begin{equation}
    x_1^{(n)} = (r_1,\theta_1,\varphi_1)\,, \qquad x_2^{(n)} = (q^n r_2,\theta_2,\varphi_2+n\mu)\,,
\end{equation}
in radial polar coordinates on $\mathbb{R}^3$. We will evaluate this expression in the partial frame $r_1 = 1$, $\varphi_1 = 0$ and denote $r_2 \equiv r$, $\varphi_2\equiv\varphi$.
\smallskip

Using the operator product expansion, one can obtain from equation \eqref{two-point-function} the thermal one-point functions of all fields appearing in the $\phi\times\phi$ OPE. To extract the one-point function of $\phi^2$, we just need to perform the leading order OPE, which amounts to subtracting the identity contribution $\{n=0\}$ and setting $r=1$, $\theta_1 = \theta_2\equiv\theta$, $\varphi=0$ in equation \eqref{two-point-function}. Thus, we learn
\begin{align}
    \lambda_{\phi\phi\phi^2} \langle\phi^2(x)\rangle_{q,y} & = \sum_{n\neq0}\frac{q^{n\Delta_\phi}}{\left(1+q^{2n}-2q^n(\cos^2\theta+\sin^2\theta\cos(n\mu))\right)^{\Delta_\phi}} \nonumber \\
    & = \sum_{n=1}^\infty \frac{2}{\left(q^{n}+q^{-n}-2(\cos^2\theta+\sin^2\theta\, T_n\left(1+\frac{u}{2}\right))\right)^{\Delta_\phi}}\ . \label{phi2-one-pt-function}
\end{align}
Here, $T_n$ denote Chebyshev polynomials of the first kind. The last expression can readily be expanded in powers of $q$ and $u$. The expansion can be matched to that of thermal conformal blocks to read off the OPE coefficients $\lambda_{\phi^2\mathcal{OO}}^a$ for low-lying operators $\mathcal{O}$. 

As in the case of the partition function, we also want to introduce the corresponding one-point function in the grand canonical ensemble by adding the fugacity $g$ that counts the particle number. This refined one-point function can then be used to extract the contributions from a fixed particle number $n$ as follows. Put 
\begin{equation}
\label{eq:GFF_1pt_U(1)} 
\langle\phi^2(x)\rangle_{q,y,g}  = \sum_{n=1}^\infty \frac{\sqrt{2}\, g^n}{\left(q^n+q^{-n}-2(\cos^2\theta+\sin^2\theta\, T_n\left(1+\frac{u}{2}\right))\right)^{\Delta_\phi}} \,,
\end{equation}
where we have used that $\lambda_{\phi\phi\phi^2}=\sqrt{2}$. We can use this expression to define the one-point function 
in the canonical ensemble. To this end, we multiply the one point function \eqref{eq:GFF_1pt_U(1)} by the grand canonical partition function and expand the product in a power series in $g$, 
\begin{equation} \label{eq:Z0phi2expansion}
\gcZ_0(q,y,g) \langle\phi^2(x)\rangle_{q,y,g} = 
\sum_{n=0}^\infty g^n \tr_{\mathcal{H}^{(n)}}\left(
\phi^2 (x) q^D y^M\right)\ . 
\end{equation}
As indicated, the expansion coefficients correspond to 
partial traces were we trace only over fields with $n$ constituents $\phi$. In order to pass to the canonical 
one-point functions, we must divide by the corresponding 
canonical partition function, 
\begin{equation}
\langle\phi^2(x)\rangle^{(n)}_{q,y} = \frac{1}{\cZ_0^{(n)}(q,y)}\, \tr_{\mathcal{H}^{(n)}}
\left(\phi^2(x) q^D y^M\right)   
\end{equation}
We shall return to the analysis of these restricted one-point functions in the canonical ensemble in the final subsection. For now, let us analyse the full thermal one-point function \eqref{phi2-one-pt-function} of the grand 
canonical ensemble. 

\subsubsection{High-temperature expansion of the one-point function}

In order to compute the high temperature expansion of the one-point function \eqref{phi2-one-pt-function} we first observe that it may be rewritten in the form
\begin{equation}
\begin{split}    \langle\phi^2(x)\rangle_{\beta,\Omega}  & = \sqrt{2}\sum_{n=1}^\infty \fp(n\beta,\Omega,s)\, , \\[2mm]  \textrm{with} & \qquad \fp(\beta,\Omega,s) = \frac{2^{-\Delta_\phi}}{(\cosh\beta - s \cos(\beta\Omega) + s - 1)^{\Delta_\phi}}\ .
\end{split} 
\end{equation}
Therefore, using the method of \cite{Zagier} once again, we may find the asymptotic expansion of $\langle\phi^2(x)\rangle_{\beta,\Omega}$ from that of $\fp$. The latter function may be expanded in a power series of the form
\begin{equation}
     \fp(\beta,\Omega,s) = \beta^{-2\Delta_\phi}\sum_{m=0}^\infty \fp_{2m}(\Omega,s) \beta^{2m} = \frac{1}{\beta^{2\Delta_\phi} (1 + s \Omega^2)^{\Delta_\phi}} + \dots\,,
\end{equation}
where it is straightforward to compute any number of coefficient functions $\fp_{2m}(\Omega,s)$. The corresponding asymptotic expansion of $\langle\phi^2(x)\rangle_{\beta,\Omega}$ reads
\begin{align}\label{GFF-highT-1pt-full}
    \frac{1}{\sqrt{2}}\langle\phi^2(x)\rangle_{\beta,\Omega} & = \frac{I^{\langle\phi^2\rangle}}{\beta} + \beta^{-2\Delta_\phi}\sum_{m=0}^{\infty} \beta^{2m}\zeta(2\Delta_\phi-2m) \fp_{2m}(\Omega,s) \\
    & = \frac{\zeta(2\Delta_\phi)}{\beta^{2\Delta_\phi} \left(1 + s \Omega^2\right)^{\Delta_\phi}} \left(1 - \frac{\Delta_\phi\left(1-s\Omega^4\right)}{12\left(1+s\Omega^2\right)} \frac{\zeta(2\Delta_\phi-2)}{\zeta(2\Delta_\phi)} \beta^2 + O\left(\beta^4\right)\right) + \frac{I^{\langle\phi^2\rangle}}{\beta} \ . \nonumber
\end{align}
Here,
\begin{equation}\label{betaminusoneGFF}
    I^{\langle\phi^2\rangle} = I^{\langle\phi^2\rangle}(\Omega,s)=\int_0^\infty d\beta\, \left(\fp(\beta,\Omega)-\beta^{-2\Delta_\phi}\sum_{2\Delta_\phi - 2m < 1} \fp_{2m}(\Omega,s) \beta^{2m}\right)\ .
\end{equation}
We have assumed that $\Delta_\phi$ is not a half-integer. In such cases, there are minor modifications that can be dealt with following \cite{Zagier}. Let us note in passing that the leading term in equation \eqref{GFF-highT-1pt-full} allows to read off the coefficient $b_{\phi^2}$ of the one-point function on $S^1_\beta\times\mathbb{R}^2$, 
\begin{equation}\label{one-pt-coefficient-phi2}
    b_{\phi^2} = \sqrt{2}\, \zeta(2\Delta_\phi)\,,
\end{equation}
in agreement with \cite{Iliesiu:2018fao}. 

\subsubsection{Asymptotics of OPE coefficients}
\label{SSS:Asymptotics of OPE coefficients}

Using the inversion formula \eqref{inversion-formula}, the high-temperature expansion \eqref{GFF-highT-1pt-full} of the one-point function $\langle\phi^2\rangle$ and the high-temperature expansion \eqref{GFF-highT-logZ} of the partition function, we can obtain the asymptotics of the (averaged) OPE coefficients $\overline{\lambda_{\phi^2\mathcal{O}\mathcal{O}}^a}(\Delta,\ell)$ as $\Delta\to\infty$, keeping $\ell$ finite. As already mentioned, GFF does not belong to the class of gapped local CFTs for which the general results of \cite{Buric:2025uqt} apply. Nevertheless, it is possible to repeat a similar analysis with some minimal modifications.
\smallskip

Given the similarities with the analysis of \cite{Buric:2025uqt}, we shall be brief. The inversion integral \eqref{inversion-formula} is performed as an expansion around a saddle-point, using for the blocks their large $\Delta$ expansion \eqref{g-f-large-Delta}-\eqref{large-Delta-expansion}. This is achieved by making the change of variables \eqref{saddle-point-variables}. Performing a series of Gaussian-like integrals over $\beta_N$ and $\mu_N$, together with the integral over $s$, then produces the asymptotic expansion of $\rho_0(\Delta,\ell)\overline{\lambda_{\phi^2\mathcal{O}\mathcal{O}}^a}(\Delta,\ell)$ around $\Delta=\infty$. Dividing by the asymptotic expansion for $\rho_0(\Delta,\ell)$ obtained in the previous subsection, we arrive at the asymptotic expansion for $\overline{\lambda_{\phi^2\mathcal{O}\mathcal{O}}^a}(\Delta,\ell)$. Let us note that the integration is performed over $\beta_N,\mu_N\in(-\infty,\infty)$, which differs from the true integration domain that is read off from equation \eqref{saddle-point-variables}. This change of integration range affects the result through ``non-perturbative'' terms, i.e. those exponentially suppressed at large $\Delta$ - see \cite{Buric:2025uqt} for a detailed discussion. 
\smallskip

Performing the above calculation, we find the leading behaviour of OPE coefficients
\begin{equation}\label{GFF-OPE-asymptotics}
    \overline{\lambda_{\phi^2\mathcal{OO}}^a}(\Delta,\ell) \sim \sqrt{2}\, \binom{-\Delta_\phi}{a}\, \zeta(2\Delta_\phi)\, \mathcal{N}_{a,\ell}\, \left(\frac{30\Delta}{\pi^4}\right)^{\frac{\Delta_\phi}{2}}\Delta^{-2a} \quad\text{as}\quad \Delta\to\infty\,,
\end{equation}
where
\begin{equation}\label{coefficient_N_a_l}
    \mathcal{N}_{a,\ell} = \frac{3^{2a}\, \Gamma(1+2a)}{2^{6a+1}\,\Gamma(\frac32+2a)}
    \sqrt{\frac{\pi(4a+1)\Gamma(2(1+\ell+a))}{(2\ell+1)\Gamma(1+2(\ell-a))}}\ .
\end{equation}
The formula  \eqref{GFF-OPE-asymptotics} has been explicitly checked for the dominant ($a=0$) and the first subdominant ($a=1$) tensor structure, although it is expected to hold for all $a$. It is interesting to compare this result to that valid for gapped CFTs, obtained in \cite{Buric:2025uqt}. In the latter, the large-$\Delta$ scaling of OPE coefficients reads $\Delta^{\Delta_{\phi^2}/3 - 2a} = \Delta^{2\Delta_\phi/3 - 2a}$. Specifically, for the dominant tensor structure, the predicted scaling in any number of spacetime dimensions, related to the eigenstate thermalisation hypothesis, is $\Delta^{\Delta_{\phi^2}/d} = \Delta^{2\Delta_\phi/d}$, \cite{Lashkari:2016vgj,Gobeil:2018fzy}. Therefore, we see that, as was the case for the spectrum, the behaviour of OPE coefficients in GFF resembles that of a gapped theory defined in one dimension higher.
\smallskip

\begin{figure}[h]
        \centering
        \includegraphics[width=0.9\linewidth]{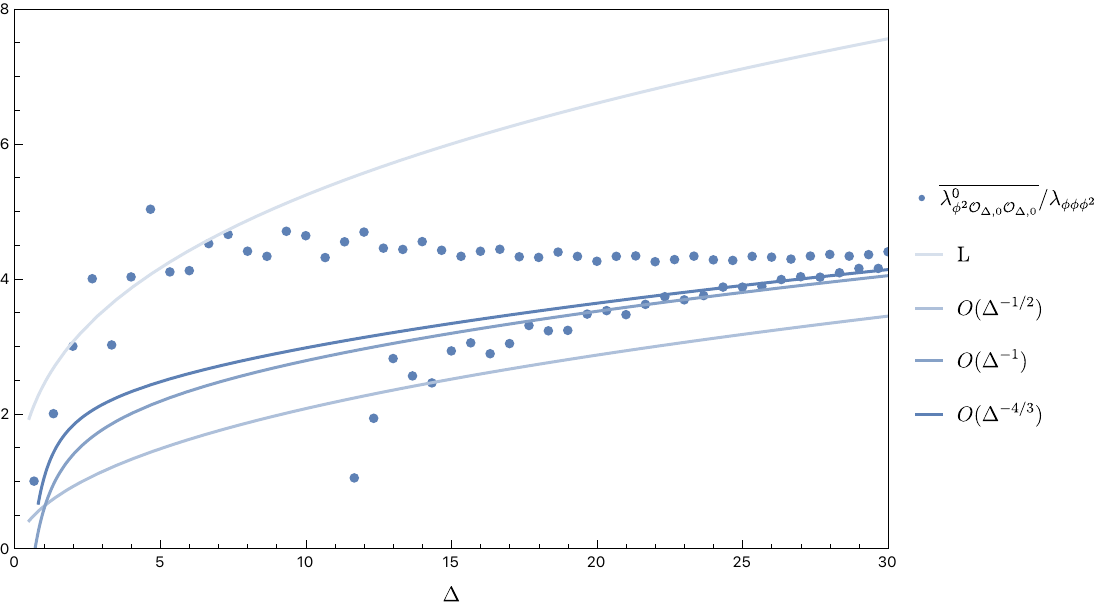}
        \caption{GFF OPE coefficients $\overline{\lambda^0_{\phi^2\mathcal{OO}}}$ for scalars, $\Delta_\phi = \frac23$. We compare normalised exact OPE coefficients for scalar exchanged operators with the asymptotic formula \eqref{GFF-OPE-asymptotics-a0} with additional subleading corrections.}
        \label{OPE-GFF-plot-scalars}
\end{figure}

\begin{figure}
        \centering
        \includegraphics[width=0.9\linewidth]{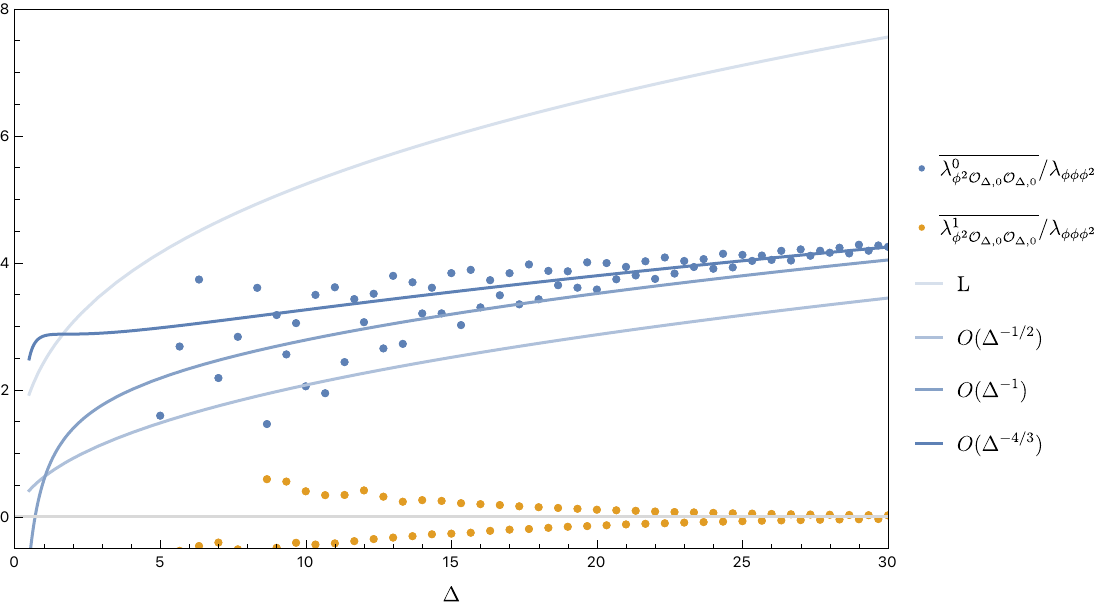}
        \caption{GFF OPE coefficients $\overline{\lambda^0_{\phi^2\mathcal{OO}}}$ for spin one, $\Delta_\phi = \frac23$. We compare normalised exact OPE coefficients for spin one exchanged operators with the asymptotic formula \eqref{GFF-OPE-asymptotics-a0} for the dominant tensor structure with additional subleading corrections.}
        \label{OPE-GFF-plot-spin-1}
\end{figure}

It is straightforward to obtain subleading corrections to the asymptotic formula \eqref{GFF-OPE-asymptotics}. We restrict our attention to OPE coefficients of the dominant tensor structure, $a=0$. They read
 \begin{align}\label{GFF-OPE-asymptotics-a0}
    &\overline{\lambda_{\phi^2\mathcal{OO}}^0}(\Delta,\ell)= \\
    & = \sqrt{2}\, \zeta(2\Delta_\phi) \left(\frac{30\Delta}{\pi^4}\right)^{\frac{\Delta_\phi}{2}} \left(1 + \frac{30\zeta(3)}{\pi^4}\Delta_\phi \left(\Delta_\phi-\frac32\right)\left(\frac{30\Delta}{\pi^4}\right)^{-\frac14} + O\left(\Delta^{-\frac12}\right) \right) \nonumber \\
    & + \sqrt{2}\, \frac{\Gamma(\Delta_\phi) \Gamma\left(\frac12-\Delta_\phi\right)}{4^{\Delta_\phi}\sqrt{\pi}} \left(\frac{30\Delta}{\pi^4}\right)^{\frac14} \left( 1 + \frac{15\zeta(3)}{\pi^4}\left(\Delta_\phi-\frac32\right) \left(\frac{30\Delta}{\pi^4}\right)^{-\frac14} + O\left(\Delta^{-\frac12}\right) \right) \nonumber \ .
\end{align}
For generic values of $\Delta_\phi$, the series contains two towers of terms, starting with powers $\Delta^{\Delta_\phi/2}$ and $\Delta^{1/4}$ and decreasing in integer powers of $\Delta^{1/4}$. By the unitarity bound $\Delta_\phi>1/2$, the first tower is leading in the $\Delta\to\infty$ limit. We have only displayed the first subleading term in each tower for reasons of space. The comparison of the asymptotic expansion \eqref{GFF-OPE-asymptotics-a0} containing further subleading terms with exact OPE coefficients is shown in Figures \ref{OPE-GFF-plot-scalars} and \ref{OPE-GFF-plot-spin-1}. Different curves correspond to different orders of the expansion, with darker shades indicating higher orders. The plot in Figure \ref{OPE-GFF-plot-spin-1} is for spin one exchanged operators and the OPE coefficients corresponding to $a=0$ and $a=1$ tensor structures are shown in blue and orange, respectively. The asymptotic and exact results show excellent agreement even for intermediate values $\Delta \sim 30$ of the weight $\Delta = \Delta_\mathcal{O}$
of the exchanged operator $\mathcal{O}$.

\subsubsection{Asymptotic OPE coefficients for fixed particle number}
\label{SSS:Asymptotics of OPE_GFF_fixed particles}

Similarly to what was discussed in Section \ref{subsect-GFF-U1-spectrum} for the spectrum of primaries, we can refine the asymptotic analysis of the OPE coefficients $\overline{\lambda_{\phi^2\mathcal{OO}}^a}$ just described by distinguishing the operators $\mathcal{O}$ according to their particle number $n$.\\
The refined version of the inversion formula \eqref{inversion-formula} reads
\begin{equation}\label{GFF-inversion-formula-OPE-fixedN}
    \overline{\lambda_{\phi^2\mathcal{O}\mathcal{O}}^a}(\Delta,\ell,n)=
   \frac{1}{\rho_0(\Delta,\ell,n)}\frac{r^{\Delta_{\phi^2}}}{2\pi i} \int \Measure\  
   g^{3-\Delta_{\phi^2},a}_{3-\Delta,\ell}(\beta,\mu,\theta)\,  \cZ^{(n)}_0(\beta,\mu) 
   \langle\phi^2(x)\rangle^{(n)}_{\beta,\mu}\ .
\end{equation}
where the product $\cZ^{(n)}_0\langle\phi^2(x)\rangle^{(n)}$ is given by 
the coefficients in the series expansion \eqref{eq:Z0phi2expansion}. For simplicity, we will now focus on the case $a=0$ corresponding to the dominant tensor structure, but the whole discussion that follows can be adapted to the other cases with minor modifications. In order to study the large $\Delta$ asymptotics of the inversion integral, we follow a strategy analogous to that described in the previous section when discussing OPE asymptotics without the additional refinement of the particle number. In particular, we expand the blocks in a large $\Delta$ expansion and for $\cZ_0^{(n)}$ and $\langle\phi^2(x)\rangle^{(n)}$ use their high temperature expansions. The integral is approximated using a saddle-point analysis. The main difference in the present case lies in the type of high temperature singularity coming from the partition function and the one-point function. Due to the projection onto a specific particle sector, we now have
\begin{equation}
   \cZ^{(n)}_0(\beta,\Omega) 
 \langle\phi^2(x)\rangle^{(n)}_{\beta,\Omega}\sim \frac{\sqrt{2}}{(n-1)!(1+s\Omega^2)^{\Delta_\phi}(1+\Omega^2)^{n-1}\beta^{2\Delta_\phi+3n-3}}, \quad \text{as}\quad\beta\to0\,,
\end{equation}
to be contrasted with the much stronger singularity \eqref{GFF-highT-logZ} present in $\gcZ_0$. In this case, working in the $\beta,\Omega,s$ variables, the expansion around the dominant saddle of the integral can be captured by simply rescaling $\beta\to\beta/\Delta$, and then expanding the integrand around $\Delta\to\infty$. Exchanging the series expansion with the integration, we may perform the integrals exactly, leading to the following prediction for the asymptotics of the OPE coefficients
\begin{equation}\label{eq:GFF_fixFamilyAsymp}
    \overline{\lambda_{\phi^2\mathcal{OO}}^0}(\Delta,\ell,n)\sim\frac{2\sqrt{2} n \,(3n-6)!}{3(2n+2\Delta_\phi-7)\Gamma(3n+2\Delta_\phi-9)}\;\Delta^{2\Delta_\phi-3}\ .
\end{equation}
In performing the integrals we have assumed $n\geq3$ and we also excluded the cases $\Delta_\phi \in \mathbb{N}/2$ which must be treated 
separately. As for the case of the spectrum (recall the discussion around \eqref{triple-trace-multiplicity}), on top of subleading power law corrections in \eqref{eq:GFF_fixFamilyAsymp}, we expect also oscillatory terms that become less and less important the higher the value of $n$. These corrections should be captured by additional saddles in the $\beta,\mu$ plane of the inversion formula integral \eqref{GFF-inversion-formula-OPE-fixedN}.

It is interesting to note that, for $\Delta_\phi > 2$, 
the leading averaged OPE coefficients grow much faster with $\Delta$ when we restrict to some finite particle number than within the full unrestricted ensemble, see equation \eqref{GFF-OPE-asymptotics-a0}. But since heavy operators are dominated by those with 
large particle number $n > \Delta^{3/4}$, the faster growth of the averaged OPE coefficient for finite particle number $n$ is washed 
out in the unrestricted grand canonical ensemble.  

\section{Interacting Scalar Fields in \texorpdfstring{AdS$_4$}{AdS4}}
\label{S:Interacting Scalar Fields in AdS$_4$}

The generalised free theory that we studied above can be thought of as the theory of a dual free massive scalar $\Phi$ on AdS$_4$. In this section we will go a step further and consider weakly interacting fields on AdS
with cubic and quartic interaction terms. 
After reviewing elements of the thermal AdS geometry in the first subsection, we will compute anomalous dimensions of multi trace operators in the second. To first order in perturbation theory these are induced by the $\Phi^4$ interaction. The final subsection 
is devoted to the computation of OPE coefficients of the form $\lambda_{\phi\mathcal{OO}}^a$. Once again, we restrict to the leading order, which arises from the $\Phi^3$ interaction.

\subsection{Scalar fields in thermal AdS}

The thermal AdS and the AdS-Kerr black hole are two solutions to Einstein's equations that dominate the gravitational path integral in regimes of low and high temperatures, respectively. Therefore, in a theory whose dual has dynamical gravity, high temperature correlation functions are obtained using wave propagation on the black hole background. By contrast, in this work we focus on quantum field theories on the fixed AdS background. In this case, correlators at all temperatures are computed using the thermal AdS space. Throughout this section, we restrict ourselves to the simplest case of a weakly-interacting single scalar field. In addition, for simplicity, we will assume that the scaling dimension associated to the field is not an integer or a half-integer, $2\Delta_\phi\notin\mathbb{N}$. In the latter case, the formulas derived below include additional terms involving logarithms.
\medskip

We parametrise the Euclidean AdS$_4$, i.e. the hyperbolic space $\mathbb{H}^4$, by coordinates $(\tau,\rho,\theta,\varphi)$, often referred to as {\it global coordinates}, in which the metric reads
\begin{equation}
    ds^2 = \cosh^2\rho\, d\tau^2 + d\rho^2 +  \sinh^2\rho \left( d\theta^2 + \sin^2\theta d\varphi^2 \right)\ .
\end{equation}
In what follows, we shall slightly abuse terminology and omit the adjective ‘Euclidean' when talking about this space. In terms of global coordinates, the embedding coordinates are given by
\begin{align}
    & X_0 = \cosh\tau \cosh\rho\,, \quad X_4 = -\sinh\tau \cosh\rho\,,\\
    & X_3 = \sinh\rho\cos\theta\,, \quad\ \ X_1 = \sinh\rho\sin\theta\cos\varphi\,, \quad X_2 = \sinh\rho\sin\theta\sin\varphi\ .
\end{align}
They lie on the hyperboloid
\begin{equation}
    - X_0^2 + X_1^2 + X_2^2 + X_3^2 + X_4^2 = - 1\,,
\end{equation}
viewed as a hypersurface in the Minkowski space $\mathbb{R}^{1,4}$. We denote the (squared) chordal distance between two points $X,Y$ on the hyperboloid by $v=(X-Y)^2$. The action of dilation and rotation generators in the above coordinates reads
\begin{align}
    & D = - \partial_\tau\,,\\
    & M_{12} = \partial_\varphi\,, \quad M_{13} = \sin\varphi\partial_\theta + \cot\theta\cos\varphi\partial_\varphi\,, \quad M_{23} =  \cos\varphi\partial_\theta  - \cot\theta \sin\varphi \partial_\varphi\,,
\end{align}
and the volume measure is given by
\begin{equation}\label{measure-AdS}
    \sqrt{g} = \cosh\rho \sinh^2\rho \sin\theta\ .
\end{equation}
The conformal boundary $\mathbb{R}\times S^2$ is located at $\{\rho=\infty\}$ and may be parametrised by coordinates $(\tau,\theta,\varphi)$. The AdS propagator for a scalar field of mass $m$ depends only on the chordal distance $v$ between the two points and takes the form 
\begin{equation}\label{AdS-propagator}
    G_{\Delta_\phi}(x,y) = C_{\Delta_\phi} v^{-\Delta_\phi}\ _2F_1\left(\Delta_\phi,\Delta_\phi-1,2\Delta_\phi-2,\frac{-4}{v}\right)\,,
\end{equation}
where
\begin{equation}\label{C-Delta}
    C_{\Delta_\phi} = \frac{\Gamma(\Delta_\phi)}{2\pi^{3/2} \Gamma(\Delta_\phi-1/2)}\,, \qquad \quad m^2 = \Delta_\phi(\Delta_\phi-3)\ .
\end{equation}
The thermal AdS is obtained as the quotient tAdS$_4=\mathbb{H}^4/\mathbb{Z}$ by the group generated by $\exp({-\beta D + \mu M})$, i.e. by making identifications  $(\tau,\rho,\theta,\varphi)\sim(\tau-\beta,\rho,\theta,\varphi+\mu)$. Its propagator is given by the sum of thermal images of the propagator \eqref{AdS-propagator},
\begin{equation}\label{thermal-propagator}
    G_{\Delta_\phi}^{\beta,\mu}(x,x') = \sum_{n=-\infty}^\infty G_{\Delta_\phi}(x,x'_n)\,, \quad x'_n = e^{n(-\beta D + \mu M)}\cdot x' = (\tau'-n\beta,\rho',\theta',\varphi'+n\mu)\ .
\end{equation}
On the other hand, we shall denote the bulk-to-boundary thermal propagator by $\Pi^{\beta,\mu}_{\Delta_\phi}(x,\hat x)$, where $x$ is the bulk and $\hat x$ the boundary point. In ordinary AdS, the bulk-to-boundary propagator is given by
\begin{equation}\label{bulk-bdy-propagator-ads}
    \Pi_{\Delta_\phi}(x,\hat x) = C_{\Delta_\phi} \left(2 \left(\cosh(\tau-\hat\tau)\cosh\rho - \Omega\sinh\rho\right) \right)^{-\Delta_\phi}\,,
\end{equation}
where
\begin{equation}
    \Omega = \cos\theta \cos\hat\theta + \cos(\varphi-\hat\varphi)\sin\theta\sin\hat\theta\ .
\end{equation}
The thermal propagator is again obtained as the sum of images, see e.g. \cite{Alday:2020eua},
\begin{equation}
    \Pi^{\beta,\mu}_{\Delta_\phi}(x,\hat x) = \sum_{n=-\infty}^\infty \Pi_{\Delta_\phi}(x,\hat x_n)\,, \qquad \hat x_n = e^{n(-\beta D + \mu M)}\cdot \hat x = (\hat\tau-n\beta,\hat\theta,\hat\varphi+n\mu)\ .
\end{equation}
Notice that we have summed over images of the boundary point. However, from equation \eqref{bulk-bdy-propagator-ads} it is clear that one may equivalently sum over images of the bulk point. For future reference, let us denote the chordal distance between a point $x$ and its $n$-th image by $v_n = v(x,x_n)$. Then
\begin{equation}\label{chordal-distance-expression}
    v_n 
    = q^{-n}(1-q^n)^2 \cosh^2\rho - y^{-n} (1-y^n)^2\sinh^2\rho \sin^2\theta \ .
\end{equation}
We note that $v_n$ does not depend on coordinates $\tau,\varphi$ of the point.
\smallskip

We end this subsection by reviewing the result of \cite{Gobeil:2018fzy} that gives an AdS integral representation of thermal conformal blocks. Using the notation for the thermal blocks
we introduced above, the representation reads
\begin{equation}\label{block-bulk-integral}
    g^{\Delta_\phi,0}_{\Delta,0}\left(q,y,\hat s\right) = A(\Delta_\phi,\Delta)\, \hat r^{\Delta_\phi}\int\limits_{\text{AdS}_4} d^4x\sqrt{g}\, G_{\Delta}(x,x_1) \Pi_{\Delta_\phi}(x,\hat x)\ .
\end{equation}
Recall, $x_1$ denotes the first thermal image of the bulk point $x$. The constant $A$ is given by
\begin{equation}
    A(\Delta_\phi,\Delta) = \frac{1}{C_{\Delta_\phi} C_\Delta} \frac{2\Gamma(\Delta_\phi) \Gamma(\Delta)^2}{\pi^{\frac32}\Gamma\left(\frac{\Delta_\phi}{2}\right)^2\Gamma\left(\Delta-\frac{\Delta_\phi}{2}\right)\Gamma\left(\Delta-\frac{3-\Delta_\phi}{2}\right)}\,,
\end{equation}
where $C_{\Delta_\phi},C_\Delta$ are as in equation \eqref{C-Delta}. The representation \eqref{block-bulk-integral} is valid under the conditions
\begin{equation}\label{conditions-bulk-integral}
    2 \Delta > \Delta_\phi\,, \qquad \Delta_\phi + 2\Delta>3\,,
\end{equation}
which ensure that the integral on the right hand side converges. An alternative way to write equation \eqref{block-bulk-integral} is as an integral over the thermal AdS, but with the bulk-to-boundary propagator replaced by its thermal version,
\begin{equation}\label{block-bulk-integral-2}
    g^{\Delta_\phi,0}_{\Delta,0}\left(q,y,\hat s\right) = A(\Delta_\phi,\Delta)\, \hat r^{\Delta_\phi} \int\limits_{\text{tAdS}_4} d^4x\sqrt{g}\, G_\Delta(x,x_1) \Pi^{q,y}_{\Delta_\phi}(x,\hat x)\ .
\end{equation}
Indeed, below we will make use of a slightly more general result
\begin{equation}\label{propagator-integral-identity}
    \int\limits_{\text{tAdS}_4} d^4x\sqrt{g}\ \Pi^{q,y}_{\Delta_\phi}(x,\hat x)\, G_\Delta(x,x_k) = \int\limits_{\text{AdS}_4} d^4x\sqrt{g}\ \Pi_{\Delta_\phi}(x,\hat x)\, G_\Delta(x,x_k)\,,
\end{equation}
for any integer $k\neq0$. To prove equation \eqref{propagator-integral-identity}, we expand the definition of the left hand side and use the fact that the propagator between a point $x$ and its image $x_k$ does not depend on coordinates $\tau,\varphi$,
\begin{align*}
    & \text{LHS} = \int \sqrt{g}\, d\rho d\theta\ G_\Delta(x,x_k) \int\limits_0^\beta d\tau \int\limits_0^{2\pi} d\varphi \sum_{n=-\infty}^\infty \Pi_{\Delta_\phi}(x_n,\hat x)\\
    & \hskip3cm = \int \sqrt{g}\, d\rho d\theta\ G_\Delta(x,x_k) \int\limits_{-\infty}^\infty d\tau \int\limits_0^{2\pi} d\varphi\ \Pi_{\Delta_\phi}(x,\hat x) = \text{RHS}\ .
\end{align*}
To get to the second line, we noticed that the sum over images $x_n$ effectively extends the range of integration over $\tau$ from $(0,\beta)$ to $(-\infty,\infty)$. The resulting integral coincides with the right hand side of equation \eqref{propagator-integral-identity}. Also, we clearly have
\begin{equation}
    G_\Delta(x,x_k) = G_\Delta(x,x_{-k})\,,
\end{equation}
a fact that will be used below. This concludes our review of the thermal AdS geometry.

\subsection{Partition function and spectral density}

We now turn to the study of interacting massive scalar fields in AdS$_4$ with cubic and quartic interaction, i.e. we consider the 
theory with the following action
\begin{equation}
    S = \int\limits_{\text{AdS}_4} d^4x \sqrt{g}\, \left(\frac12\Phi(\nabla^2+m^2)\Phi + \lambda_3 \Phi^3 + \lambda_4\Phi^4\right)\ .
\end{equation}
Due to the interaction terms, operators acquire anomalous dimensions and hence they modify the spectral density. In first order perturbation theory, only the quartic interaction term can contribute
to the partition function. Let $\gcZ_0$ denote the partition function of the free theory with $\lambda_3=0=\lambda_4$. This is the GFF partition function \eqref{partition-function-1}. To first order 
in the couplings, the partition function of the interacting theory reads, see e.g. \cite{Kraus:2020nga}, 
\begin{equation}\label{interacting-partition-function}
    \log \gcZ = \log \gcZ_0 - 3\lambda_4 \int\limits_{\text{tAdS}_4} d^4x\sqrt{g}\ G_{\Delta_\phi}^{\beta,\mu}(x,x)^2  + O(\lambda^2)\ .
\end{equation}
Here and in the following, $O(\lambda^2)$ denotes all higher order corrections terms which depend on both $\lambda_3$ and $\lambda_4$. The terms coming from the $n=0$ image in the sum-over-images expression for the thermal propagator \eqref{thermal-propagator} diverge and are discarded. This is the standard prescription, equivalent to the mass renormalisation, \cite{Alday:2020eua}. For 
our discussion it is useful to introduce the following shorthand
\begin{equation}
    f(\lambda) = \log \gcZ - \log \gcZ_0 = \log\frac{\gcZ}{\gcZ_0} \equiv a_1 \lambda_4 + O(\lambda^2)\ .
\end{equation}
Thus, the interacting partition function may be written as
\begin{equation}\label{free-int-relation}
    \gcZ = \gcZ_0 e^{f(\lambda)} = \gcZ_0 e^{a_1 \lambda_4 + O(\lambda^2)} = \gcZ_0 \left(1 + a_1 \lambda_4 + O(\lambda^2) \right)\ .
\end{equation}
On the other hand, we can write character decompositions for the free and interacting partition functions,
\begin{equation}\label{character-decompositions-free-int}
   \gcZ_0 = 1 + \sum_{\mathcal{O}} \chi_{\Delta,\ell}\,, \qquad \gcZ = 1 + \sum_{\mathcal{O}} \chi_{\Delta(\lambda),\ell}\ .
\end{equation}
Implicit in this notation is that both sums run over the same set of free theory operators, which however in the interacting case have shifted scaling dimensions (and unchanged spins). Comparing the equations \eqref{free-int-relation} and \eqref{character-decompositions-free-int}, we learn
\begin{equation}\label{decomposition-derivatives-characters}
    \gcZ_0 a_1 = \sum_{\mathcal{O}} \gamma_\mathcal{O}\, \partial_{\Delta}\chi_{\Delta,\ell} = -\beta \sum_{\mathcal{O}}\gamma_\mathcal{O}\, \chi_{\Delta,\ell}= -\beta \sum_{\Delta,\ell}n_{\Delta,\ell}\,\bar{\gamma}_{\Delta,\ell} \chi_{\Delta,\ell}\,,
\end{equation}
where in the second step we used the property of characters $\partial_\Delta\chi_{\Delta,\ell} = -\beta\chi_{\Delta,\ell}$, while in the last one we have moved from summing over operators to summing over the GFF spectrum. The $n_{\Delta,\ell}$ are the GFF multiplicities, and we have introduced the averaged anomalous dimensions 
\begin{equation}
\bar{\gamma}_{\Delta,\ell}=\left(\sum_{i=1}^{n_{\Delta,\ell}}\gamma_{\mathcal{O}_i}\right)/n_{\Delta,\ell}\,,
\end{equation}
where, for given weight $\Delta$ and spin $\ell$, the summation runs over all primaries $\mathcal{O}_i$ with $\Delta_{\mathcal O_i}=\Delta$
and $\ell_{\mathcal{O}_i} = \ell$. 
By substituting equations \eqref{AdS-propagator}, \eqref{thermal-propagator} and \eqref{chordal-distance-expression} into the expression for the partition function \eqref{interacting-partition-function}, we find the coefficient $a_1$ as a power series in $q$ multiplied by an overall factor of $\beta$,
\begin{equation}\label{eq:lambda_correction_Z}
    a_1 = - 3 \int\limits_{\text{tAdS}} d^4x\sqrt{g}\ G_{\Delta_\phi}^{\beta,\mu}(x,x)^2 = - 6\pi\beta \int \cosh\rho\sinh^2\rho\sin\theta\, d\rho d\theta\ G_{\Delta_\phi}^{\beta,\mu}(x,x)^2\ .
\end{equation}
We used the fact that the term involving the thermal propagator does not depend on $\tau,\varphi$ to perform the integral over these variables. In particular, the factor $\beta$ comes from integration over Euclidean time. It cancels the factor $\beta$ on the right hand side of equation \eqref{decomposition-derivatives-characters} that arises from derivatives of characters. The remaining expressions on both sides expand in powers of $q$. By matching the decompositions order by order, one reads off the (averaged) anomalous dimensions $\bar{\gamma}_{\Delta,\ell}$. 

An interesting family of operators to look at is provided by double-trace operators. Since these do not have multiplicities, no averaging is involved. By matching the power series expansions in $q$ on both sides of the equation \eqref{decomposition-derivatives-characters}, we extract the following expression
\begin{equation}\label{double-trace-anomalous-dimensions}
    \gamma_{2\Delta_\phi+2k,0} = \frac{6\alpha_k}{2\Delta_\phi+2k-\frac32}\lambda_4\,, \quad \alpha_k = \frac{\left(\frac32\right)_k (k+2\Delta_\phi-2)_k (2k+2 \Delta_\phi)_{-\frac12}}{2\pi^{\frac32} k! \left((k+\Delta_\phi)_{-\frac12}\right)^2 \left(k+2\Delta_\phi-\frac32\right)_k}\ . 
\end{equation}
For $\ell>0$, the anomalous dimensions vanish. These 
results agree with previous findings in the existing literature, \cite{Fitzpatrick:2010zm,Kraus:2020nga} (e.g. see equations (1.17) and (1.23) of \cite{Kraus:2020nga}).
\smallskip

In order to compute the averaged anomalous dimensions of multi-twist operators, it is useful to apply the inverse character transform \eqref{inverse-character-transform}, together with equations \eqref{decomposition-derivatives-characters} and \eqref{eq:lambda_correction_Z} to express the anomalous dimension density as
\begin{equation}\label{anomalous-dimensions-density}
    \rho_0(\Delta,\ell)\bar{\gamma}(\Delta,\ell) = -3i \int\limits_{\gamma-i\infty}^{\gamma+i\infty} d\beta \int\limits_{-\pi}^\pi d\mu\, \omega\, \chi_{3-\Delta,\ell}\, \gcZ_0 \int \cosh\rho\sinh^2\rho\sin\theta\, d\rho d\theta\ G_{\Delta_\phi}^{\beta,\mu}(x,x)^2\ .
\end{equation}

\begin{figure}[h]
        \centering
        \includegraphics[width=0.9\linewidth]{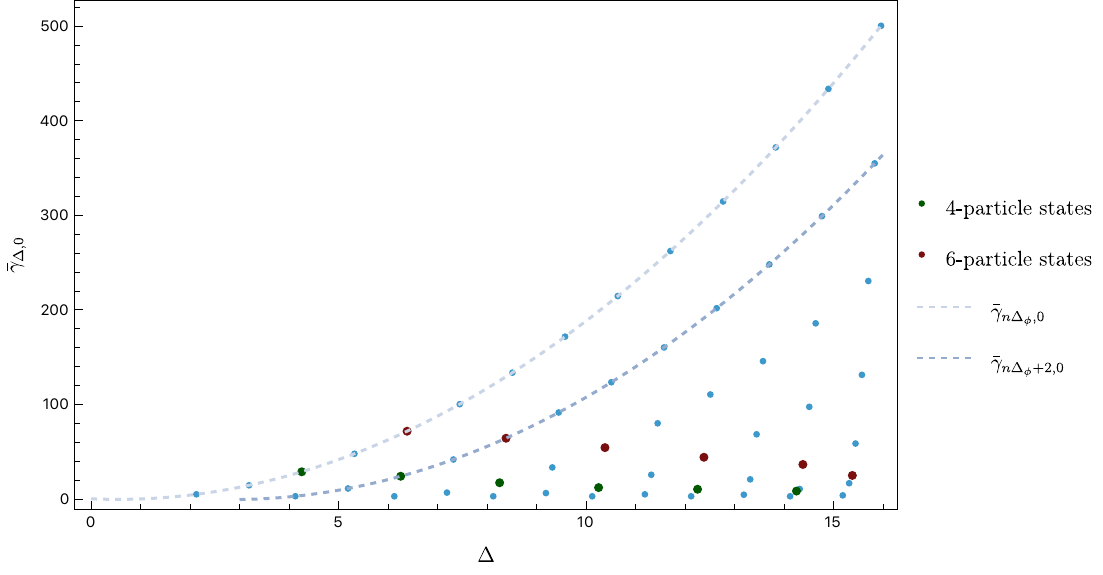}
        \caption{Low-lying anomalous dimensions of scalars, $\Delta_\phi=\frac{33}{31}$. The exact formulas describing the dots on the two dashed curves can be found in \eqref{anom-dim-nDelta} and \eqref{anom-dim-nDelta-plus-2}. The asymptotic formula \eqref{finite-n-anom-dim-asymptotics} captures instead the large $\Delta$ asymptotics of the data when restricted to a fixed particle number, such as the two families of dots we have highlighted in green and dark red.}
        \label{plot-low-lying-anom-dim}
\end{figure}

\paragraph{Multi-particle families.} We now want to use formula 
\eqref{anomalous-dimensions-density} to extract information about the averaged anomalous dimension of multi-particle operators. As in our treatment of the GFF, we can keep track of the particle number by introducing the following generalised thermal propagator,
\begin{equation}
    G_{\Delta_\phi}^{\beta,\mu,g}(x,x) = \sum_{n=-\infty}^\infty g^{|n|}G_{\Delta_\phi}(x,x_n) \,,
\end{equation}
which also depends on the fugacity $g$. By inserting this refined
thermal propagator into equation \eqref{anomalous-dimensions-density}, we can now distinguish the anomalous dimensions from different families of operators. Let us first illustrate this idea on the 
case of double-trace operators. In this case, the inversion formula \eqref{anomalous-dimensions-density} reduces to
\begin{align*}
    \gamma_{2\Delta_\phi+2k+\ell,\ell} &= -3i \int\limits_{\gamma-i\infty}^{\gamma+i\infty} d\beta \int\limits_{-\pi}^\pi d\mu\, \omega\, \chi_{3-\Delta,\ell}\, \left(\gcZ_0 \int \cosh\rho\sinh^2\rho\sin\theta\, d\rho d\theta\ G_{\Delta_\phi}^{\beta,\mu,g}(x,x)^2\right)^{(2)}\\[2mm]
    &= -12i \int\limits_{\gamma-i\infty}^{\gamma+i\infty} d\beta \int\limits_{-\pi}^\pi d\mu\, \omega\, \chi_{3-\Delta,\ell}\, \left( \int \cosh\rho\sinh^2\rho\sin\theta\, d\rho d\theta\ G_{\Delta_\phi}(x,x_1)^2\right)\ .
\end{align*}
The superscript $(2)$ in the first line reminds us to extract the 
coefficient of $g^2$. This then leads to the formula in the second 
line which was indeed used in \cite{Kraus:2020nga} to obtain double-trace anomalous dimensions. For a generic $n-$particle family, the generalisation is straightforward
\begin{equation}\label{anom-dim-density-fixedN}
    \rho_0(\Delta,\ell,n)\bar{\gamma}(\Delta,\ell,n) = -12i \int\limits_{\gamma-i\infty}^{\gamma+i\infty} d\beta \int\limits_{-\pi}^\pi d\mu\, \omega\, \chi_{3-\Delta,\ell}\,\sum_{m=0}^{n-2} \cZ^{(m)}_0\left(\sum_{k=1}^{n-m-1}\mathcal{G}_{\Delta_\phi}^{\,k,n-m-k}\right)\, . 
\end{equation}
Here, $\cZ^{(m)}_0$ is the canonical partition function of the free theory which we defined in eq.~\eqref{eq:canonicalZ} and we have introduced the notation
\begin{equation}\label{integral-of-product-of-prop}
    \mathcal{G}_{\Delta_\phi}^{\,n,m}\equiv\int \cosh\rho\sinh^2\rho\sin\theta\, d\rho d\theta\ G_{\Delta_\phi}(x,x_n)G_{\Delta_\phi}(x,x_m)\ .
\end{equation}
It is straightforward to use the inversion formula \eqref{anomalous-dimensions-density}, or rather expand equations \eqref{decomposition-derivatives-characters} and \eqref{eq:lambda_correction_Z} in powers of $q$ and equate the two sides, to read off anomalous dimensions of any number of low-lying operators. A particular example is illustrated in Figure \ref{plot-low-lying-anom-dim}. To produce the plot we have 
fixed the mass of the scalar field such that $\Delta_\phi=33/31>1$\footnote{We choose the value such that the low-lying operators with different particle numbers have different classical dimensions and are thus distinguished on the plot.}. The dots show the anomalous dimensions of low lying scalar fields. Formula \eqref{anom-dim-density-fixedN} allows to resolve the particle numbers. We have shown this resolution at the example of $4-$ and $6-$particle states which we depicted in a different colour. Of course, 
all other dots also come with a definite particle number even if we did not display that in the plot. The dots on the two dashed curves represent $n-$particle scalar fields on the leading and first subleading trajectory, see also discussion in subsection \ref{sssection:exactAD} below. In the remainder of this subsection, we will derive and discuss various analytic and asymptotic formulas for anomalous dimensions.

\subsubsection{Asymptotic anomalous dimensions}

In order to determine the asymptotics of anomalous dimensions as $\Delta\to\infty$ at finite $\ell$, one may use the inversion formula \eqref{anomalous-dimensions-density} and the $\beta\to0$ behaviour of the integrand (at finite $\Omega$). We will mostly focus on the case of a fixed particle number. On the one hand, this makes contact with the existing literature, and, on the other, allows for a better control of the analytic manipulations we perform in deriving the asymptotics. The relevant inversion formula for this setting is equation \eqref{anom-dim-density-fixedN}.
\smallskip

As usual, in order to extract the large $\Delta$ asymptotics of the integral, we are interested in analysing the $\beta\rightarrow0$ asymptotics of the integrand. While the contribution coming from $\gcZ_0$ can be easily treated, it is less trivial to obtain a high temperature expansion for the integral of the two propagators with shifted arguments. It might be tempting to derive this expansion by simply expanding the integrand around $\beta=0$ and integrating the resulting series term by term. However, exchanging the two operations in this case is not justified, as the high temperature series is obtained by expanding the hypergeometric function entering the propagator at large values of its argument, but for different values of $\rho$ the argument can be either small or large. A way out is to split the domain of $\rho$-integration in different regions and use a different series representation of the hypergeometric in each of them. This cleanly shows how different regions contribute to the overall result with different scalings in $\beta$, but computing the precise coefficients associated to each contribution can become tedious. We shall thus follow a simpler approach that makes use of the Mellin-Barnes representation of $_2F_1$.
\smallskip

When the two propagators in equation \eqref{integral-of-product-of-prop} involve the same thermal image one may use the property (see \cite{Kraus:2020nga})
\begin{equation}\label{product-of-props-identity}
    G_{\Delta_1}(x,y)G_{\Delta_2}(x,y) = \sum_{k=0}^{\infty} \alpha_k  G_{\Delta_1+\Delta_2+2k}(x,y)\,,
\end{equation}
with $\alpha_k$ given in equation \eqref{double-trace-anomalous-dimensions}, together with the identity
\begin{equation}
    \int \cosh\rho\sinh^2\rho\sin\theta\, d\rho d\theta \ G_{\Delta}(x,x_n)=\frac{1}{2\pi(2\Delta-3)}\;\chi_{\Delta,0}(q^n,y^n)\,,
\end{equation}
to rewrite
\begin{equation}
\begin{split}
 &\mathcal{G}_{\Delta_\phi}^{\,n,n}(q,y) = \sum_{k=0}^{\infty} \frac{\alpha_k}{4\pi\left(2\Delta_\phi+2k-\frac32\right)} \;\chi_{2\Delta_\phi+2k,0}(q^n,y^n)\\[2ex]
 & = \frac{(2\Delta_\phi)_{-\frac12}}{4\pi^{\frac52}\,(4\Delta_\phi-3)\,(\Delta\phi)_{-\frac12}^2} \,\chi_{2\Delta_\phi,0}(q^n,y^n)\;{}_4F_3\left(
\begin{matrix}
\frac32,\,\Delta_\phi-1,\,\Delta_\phi,\,2\Delta_\phi-\frac32 \\
\Delta_\phi-\frac12,\,\Delta_\phi+\frac12,\,2\Delta_\phi-2
\end{matrix}\,; \;q^{2n}\right)\ .
 \end{split}
\end{equation}
From this expression one immediately gets the high temperature expansion
\begin{equation}\label{G_nn_high_T}
    \mathcal{G}_{\Delta_\phi}^{\,n,n}=\frac{1}{32n^4\pi^3(1+\Omega^2)\beta^4}+\frac{(4\Delta_\phi-5)(5\Delta_\phi-7)}{192n^3\pi^3(1-\Delta_\phi)(1+\Omega^2)\beta^3}+O\left(\frac{\log(\beta)}{\beta^2}\right)\ .
\end{equation}
The case of two different thermal images is more complicated, as multiple scales are involved coming from the different images. Due to the symmetry of the problem, we may restrict without loss of generality to the case $\mathcal{G}_{\Delta_\phi}^{\,n,m}$ with $n>m$. A convenient tool for extracting the high temperature asymptotics is to move to Mellin space, using the well known representation (derived from the Mellin-Barnes representation of ${}_2F_1$)
\begin{equation*}
    G_{\Delta}(x,y) = \frac{1}{(4\pi)^2} \int\limits_{-i\infty}^{i\infty} \frac{ds}{2\pi i} \frac{\Gamma(s)\Gamma(s-1)\Gamma(\Delta-s)}{\Gamma(s+\Delta-2)} \left(\frac{v}{4}\right)^{-s} \equiv\frac{1}{(4\pi)^2}\int\limits_{-i\infty}^{i\infty} \frac{ds}{2\pi i} \,\mathcal{K}_\Delta(s) \left(\frac{v}{4}\right)^{-s}\ .
\end{equation*}
The last equality defines the function $\mathcal{K}_\Delta(s)$. The integration contour in the complex $s$-plane is taken to separate the two sets of poles (right and left) of the gamma functions in the numerator of $\mathcal{K}_\Delta(s)$. Exchanging the Mellin integrals with those over AdS, and performing the latter, we get
\begin{align}
    \mathcal{G}_{\Delta_\phi}^{\,n,m}&=\frac{1}{2^9 \pi^\frac72}\frac{Q_n}{Q_n-Y_n}\int_{-i\infty}^{i\infty}\frac{ds_1}{2\pi i}\int_{-i\infty}^{i\infty}\frac{ds_2}{2\pi i}\,\left(\frac{Q_n}{4}\right)^{-s_1}\left(\frac{Q_m}{4}\right)^{-s_2}\mathcal{K}_{\Delta_\phi}(s_1)\mathcal{K}_{\Delta_\phi}(s_2)\nonumber\\[1ex]
    &\hskip3cm\frac{\Gamma\left(s_1+s_2-\frac32\right)}{\Gamma(s_1+s_2)}{}_2F_1\left(1,\,s_2,\,s_1+s_2,\,\frac{Y_m/Q_m-Y_n/Q_n}{1-Y_n/Q_n}\right)\,,
\end{align}
where we have introduced the notation $Q_n=q^{-n}(1-q^n)^2,\,Y_n=y^{-n}(1-y^n)^2$. The variables $Q_{n,m}$, which are conjugate to the Mellin space coordinates $s_{1,2}$, behave like $Q_n\sim n^2\beta^2$ at high temperature. Therefore, to find the high temperature asymptotic of $\mathcal{G}_{\Delta_\phi}^{\,n,m}$, we should pick the residues from the poles in $s_1,s_2$ which sit to the \textit{left} of our integration contours. The leading contribution is easy to single out. It comes from the symmetric pole at $(s_1,s_2)=(1,1)$, and gives the scaling
\begin{equation}
    \mathcal{G}_{\Delta_\phi}^{\,n,m}\sim\frac{1}{32n^2m^2\pi^3(1+\Omega^2)\beta^{4}}\ .
\end{equation}
The result generalises equation \eqref{G_nn_high_T} to the case of $n\neq m$. To systematically compute the subleading corrections one may proceed by keeping track of the deformation of the contours in the complex planes of $s_1$ and $s_2$. The first correction for instance, which contributes with a power $\beta^{-3}$, comes from the pole $s_1+s_2=\frac32$.
\smallskip

\begin{figure}[ht]
    \centering
    \begin{minipage}{0.48\textwidth}
        \centering
    \includegraphics[width=\linewidth]{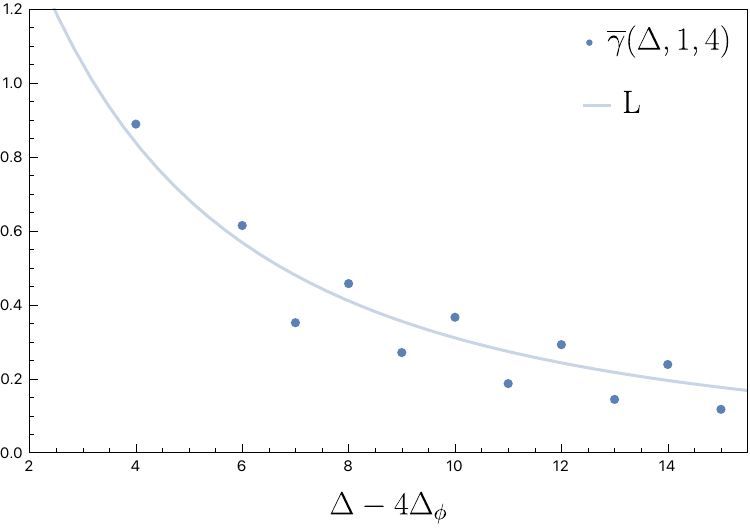}
    \end{minipage}
    \hfill
    \begin{minipage}{0.48\textwidth}
        \centering
    \includegraphics[width=\linewidth]{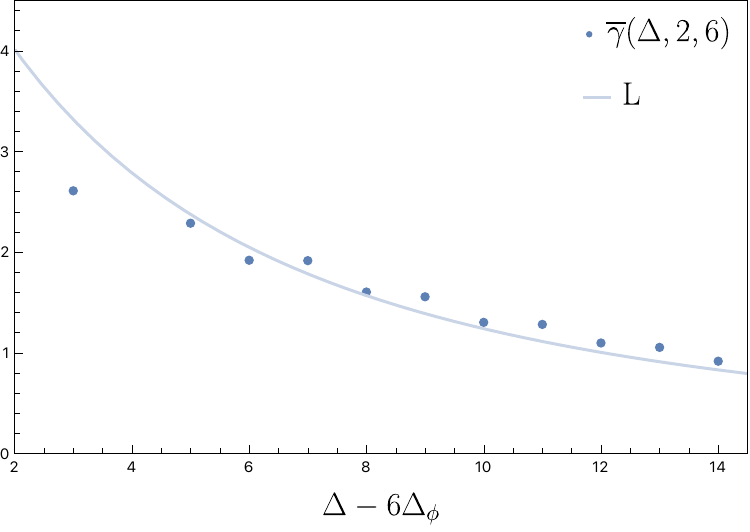}
    \end{minipage}
    \caption{Exact anomalous dimensions of operators with fixed spin $\ell$ and number of particles $n$, compared to the asymptotic formula \eqref{finite-n-anom-dim-asymptotics}. To the left, we display the case $(\ell,n)=(1,4)$, while to the right the case $(\ell,n)=(2,6)$. In both plots we have set $\Delta_\phi=\frac43$.}
    \label{fig::anomal_dim_comparison_fixedN}
\end{figure}

If we combine these results for the integrated propagators $\mathcal{G}_{\Delta_\phi}^{\,n,m}$ with the high temperature expansion of $\cZ^{(n)}_0$, we obtain the following expansion for the quantity entering the inversion formula \eqref{anom-dim-density-fixedN} for the anomalous dimensions of $n$-particle operators
\begin{align}
    &\sum_{m=0}^{n-2}\left(\gcZ_0^{(m)}\right)\left(\sum_{k=1}^{n-m-1}\mathcal{G}_{\Delta_\phi}^{\,k,n-m-k}\right)=\\
    &\frac{1}{32\pi^3(n-2)!(1+\Omega^2)^{n-1}\beta^{3n-2}}+\frac{\Delta_\phi(8\Delta_\phi-23)+3n(\Delta_\phi-1)(2\Delta_\phi-3)+17}{192\pi^3(n-2)!(1-\Delta_\phi)(1+\Omega^2)^{n-1}\beta^{3n-3}}+\dots\ . \nonumber
\end{align}
With this expansion at hand, it is now straightforward to obtain the large-$\Delta$ expansion for the anomalous dimensions, following similar steps to those described in Section \ref{SSS:Asymptotics of OPE_GFF_fixed particles}. We obtain
\begin{equation}\label{finite-n-anom-dim-asymptotics}
    \bar{\gamma}(\Delta,\ell,n)\sim \frac{3(3n-7)(3n-8)n(n-1)(n-2)}{2\pi^2(2n-7)}\Delta^{-2}\,,
\end{equation}
where in performing the integrals of the inversion formula we have assumed $n\geq4$. This is one of the main results of the present subsection. In Figure \ref{fig::anomal_dim_comparison_fixedN} we illustrate how the asymptotic formula \eqref{finite-n-anom-dim-asymptotics} compares to exact anomalous dimensions obtained as explained below equation \eqref{integral-of-product-of-prop}.

\subsubsection{Exact anomalous dimensions}
\label{sssection:exactAD}

As a spin-off of the technology we developed, we briefly discuss some exact expressions for anomalous 
dimensions of families of operators. We will write all anomalous dimensions in units of $\gamma_{\phi^2}$. 
The simplest multi-particle operators that contain no derivatives get anomalous dimensions
\begin{equation}\label{anom-dim-nDelta}
    \frac{\gamma_{\phi^n}}{\gamma_{\phi^2}} = \frac{n(n-1)}{2}\ .
\end{equation}
These operators lie on the leading parabola on Figure \ref{plot-low-lying-anom-dim} (light blue dashed curve). In the limit $n\to\infty$ at finite $k,\ell$ all operators behave as\footnote{This is an observation based on computed anomalous dimensions. Intuitively, when $n\gg k,\ell$, i.e. there are many more $\phi$-s than derivatives in an operator, its behaviour in the leading order is that of $\phi^n$, which is given by eq.~\eqref{anom-dim-nDelta}.}
\begin{equation}\label{large-n-anomalous-dim}
     \frac{\bar{\gamma}_{n\Delta_\phi+k,\ell}}{\gamma_{\phi^2}} \sim \frac{n^2}{2} \qquad\text{as}\qquad n\to\infty\ .
\end{equation}
The formula describes the asymptotic behaviour of the leading and all subleading parabolas in  
Figure~\ref{plot-low-lying-anom-dim}. For instance, the first subleading parabola (dark blue dashed 
curve) corresponds to the family $(n\Delta_\phi+2,0)$, whose anomalous dimensions read
\begin{equation}\label{anom-dim-nDelta-plus-2}
    \frac{\bar\gamma_{n\Delta_\phi+2,0}}{\gamma_{\phi^2}} = \frac12 n^2 - \frac{2\Delta_\phi^2 + 4\Delta_\phi-1}{8\Delta_\phi^2 - 2} n - \frac{\Delta_\phi-2}{8\Delta_\phi^2 - 2}\ .
\end{equation}
Formulas for further subleading parabolas may be written, although the dependence of coefficients on $\Delta_\phi$ quickly becomes complicated, see e.g. equation~\eqref{nDelta+4-anom-dim} for the next family $(n\Delta_\phi+4,0)$. Anomalous dimensions take the simplest form for operators with quantum numbers $(n\Delta_\phi+\ell,\ell)$, $n\geq\ell$. For instance, we have
\begin{equation}
\begin{split}
\frac{\bar\gamma_{n\Delta_\phi+2,2}}{\gamma_{\phi^2}} & = \frac12 n^2 - \frac{4\Delta_\phi+1}{2(2\Delta_\phi+1)} n - \frac{1}{2\Delta_\phi+1}\ , \\[2mm]
\frac{\bar\gamma_{n\Delta_\phi+3,3}}{\gamma_{\phi^2}} & = \frac12 n^2 - \frac{5\Delta_\phi+1}{2(2\Delta_\phi+1)} n - \frac{3}{2\Delta_\phi+1} \ . 
\end{split}
\end{equation}
We have obtained these exact expressions by applying the inversion formula \eqref{anomalous-dimensions-density}
for a large number of low lying operators. The general expression is then easy to reconstruct from this data, at 
least for those families we display here and in Appendix \ref{A:Tables-of-data}.

\subsection{One-point function of \texorpdfstring{$\phi$}{phi} in the interacting theory}

\begin{figure}[h]
        \centering
        \includegraphics[width=0.95\linewidth]{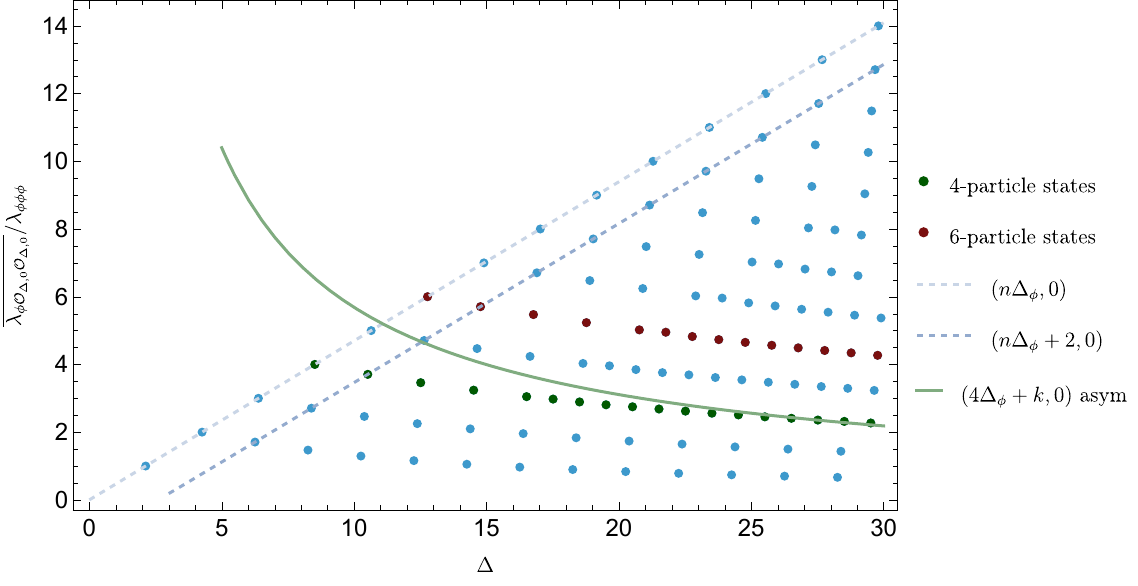}
        \caption{Normalised averaged OPE coefficients $\overline{\lambda_{\phi\mathcal{OO}}}$ for scalar exchange, $\Delta_\phi = \frac{66}{31}$. The two exact formulas describing the dots on the dashed curves are \eqref{leading-OPE-family-cubic} and \eqref{OPE-cubic-nDelta-plus-2}, while the asymptotic green curve is given by \eqref{cubicOPE_asymFixFam}.}
        \label{OPE-exact-plot}
\end{figure}

We turn to one-point functions of the fundamental field $\phi$ on the boundary. From the action in eq.~\eqref{interacting-partition-function}, one can see that, to leading order, the one-point function of $\phi$ can only receive contributions from the cubic interaction, i.e. it only depends on $\lambda_3$. At this order the one-point function of the fundamental field is given by
\begin{equation}
    \langle\phi(\hat x)\rangle_{q,y} = \frac{3\lambda_3}{\sqrt{C_{\Delta_\phi}}} \int\limits_{\text{tAdS}_4} d^4x \sqrt{g}\ G^{q,y}_{\Delta_\phi}(x,x) \Pi^{q,y}_{\Delta_\phi}(x,\hat x) + O(\lambda^2)\ .
\end{equation}
The analogous one-point function in three-dimensional bulk was studied in \cite{Kraus:2017ezw}. Since the GFF one-point function is zero, the $O(\lambda_3)$ term decomposes into thermal blocks themselves, in contrast to the previous decomposition of $O(\lambda_4)$ partition function in derivatives of characters. In the remainder of this section, we will neglect $O(\lambda^2)$ terms and strip off the coupling $\lambda_3$ from the one-point function to write
\begin{equation}\label{cubic_1pt_AdS_representation}
    \langle\upphi(\hat x)\rangle_{q,y} \equiv  \frac{3}{\sqrt{C_{\Delta_\phi}}}\int\limits_{\text{tAdS}_4} d^4x \sqrt{g}\ G^{q,y}_{\Delta_\phi}(x,x) \Pi^{q,y}_{\Delta_\phi}(x,\hat x) \ .
\end{equation}
This integral appears formidable, but we may follow an indirect route and express it in terms of functions related to thermal conformal blocks. First, we expand the sum over images in the bulk-to-bulk propagator
\begin{equation}
    \langle\upphi(\hat x)\rangle_{q,y} = \frac{3}{\sqrt{C_{\Delta_\phi}}} \sum_{n\neq0}\ \int\limits_{\text{tAdS}_4} d^4x \sqrt{g}\ G_{\Delta_\phi}(x,x_n) \Pi^{q,y}_{\Delta_\phi}(x,\hat x) \ .
\end{equation}
Each term in the sum is simply a thermal block \eqref{block-bulk-integral-2}, with shifted argument $q\to q^n$, $y\to y^n$. In the notation of \cite{Buric:2024kxo},
\begin{equation}\label{cubic-one-point-function}
    \gcZ_0\langle\upphi(\hat x)\rangle_{q,y}  = \gcZ_0\, \frac{6C_{\Delta_\phi}^{-1/2}\, \hat r^{-\Delta_\phi}}{A(\Delta_\phi,\Delta_\phi)} \sum_{n=1}^\infty\ g_{\Delta_\phi,0}^{\Delta_\phi,0}\left(q^n,y^n,\hat s\right) \ .
\end{equation}
We have multiplied the one-point function, which is normalised, by the GFF partition function in order to expand in thermal conformal blocks. The right hand side is readily decomposed into blocks corresponding to the GFF spectrum, allowing us to read off the (averaged) OPE coefficients $\overline{\lambda_{\phi\mathcal{OO}}^a}$. An example is given in Figure \ref{OPE-exact-plot}. Different families of operators on this figure are discussed below.

\subsubsection{Asymptotic OPE coefficients}
\label{SSS:Asymptotic OPE coefficients}

As explained in Section \ref{SSS:Asymptotics of OPE coefficients}, the asymptotic $\overline{\lambda_{\phi\mathcal{OO}}^a}$ OPE coefficients can be obtained from the $\beta\to0$ behaviour of the one point function (at fixed $\Omega,s$). From the expression \eqref{cubic-one-point-function} for the one-point function, we see that it suffices to analyse the function
\begin{equation}\label{auxiliary-function-F-cubic}
    F_{\upphi} = \sum_{n=1}^\infty\ g_{\Delta_\phi,0}^{\Delta_\phi,0}\left(n\beta,\Omega,s\right)\ .
\end{equation}
Using the results of \cite{Zagier} it follows that, in order to establish the all-order asymptotic behaviour of the one-point function as $\beta\to0$, it is enough to do the same for the block $g_{\Delta_\phi,0}^{\Delta_\phi,0}\left(\beta,\Omega,s\right)$. As a warmup, we first analyse the case of zero chemical potential, before dealing with generic $\Omega\neq0$.

\paragraph{Zero chemical potential.} At vanishing chemical potential, the scalar exchange blocks are known exactly in terms of the hypergeometric function $_3F_2$, \cite{Gobeil:2018fzy},
\begin{equation}\label{exact-scalar-block}
    g^{\Delta_\phi,0}_{\Delta,0} = \frac{q^\Delta}{(1-q)^{2\Delta}} \, _3F_2\left(\Delta-1,\Delta-\frac{\Delta_\phi}{2},\Delta-\frac{3-\Delta_\phi}{2};\Delta,2 \Delta-2;\frac{-4q}{(1-q)^2}\right)\ .
\end{equation}
Therefore, we may use equations \eqref{cubic-one-point-function} and \eqref{exact-scalar-block} to obtain the $O(\lambda_3)$ one-point function in the expansion around $\beta=0$. The leading behaviour of the block depends on the value of $\Delta_\phi$. Let us first assume that $\Delta_\phi$ is sufficiently large, $\Delta_\phi>2$. Then the leading singular behaviour is
\begin{equation}
    g^{\Delta_\phi,0}_{\Delta,0} = 
    \frac{2^{\Delta_\phi-2\Delta}\,\Gamma\left(2\Delta-1\right)\,
\Gamma\left(\Delta_\phi-\frac32\right)}{\left(\Delta_\phi-2\right)\,
\Gamma\left(\Delta+\frac{\Delta_\phi}{2}-2\right)\,
\Gamma\!\left(\Delta+\frac{\Delta_\phi}{2}-\frac32\right)}
 \beta^{-\Delta_\phi} + \dots\,,
\end{equation}
giving the one-point coefficient
\begin{equation}\label{1-pt-coefficients-holography}
    \oneptcub = \frac{6 C_{\Delta_\phi}^{-1/2}}{A(\Delta_\phi,\Delta_\phi)}\frac{2^{2 \Delta_\phi -5} \Gamma \left(\Delta_\phi -\frac32\right) \Gamma (2 \Delta_\phi -1)}{\sqrt{\pi} (\Delta_\phi -2) \Gamma (3 \Delta_\phi -4)} \zeta(\Delta_\phi)\ .
\end{equation}
If $\Delta_\phi<2$, the leading behaviour of the block changes, namely
\begin{equation}\label{scalings-different-regimes}
   1 < \Delta_\phi < 2: \quad  g^{\Delta_\phi,0}_{\Delta,0} \sim \beta^{-2}\,, \qquad\quad \Delta_\phi<1: \quad g^{\Delta_\phi,0}_{\Delta,0} \sim \beta^{\Delta_\phi-3}\ .
\end{equation}
We have only written the power of the leading singularity, omitting the multiplicative coefficient. When $\Delta_\phi<1$, the AdS integral representation of blocks is not valid and therefore our argument relating the blocks to the one-point function does not apply. From equation \eqref{scalings-different-regimes}, we see that in the case $1<\Delta_\phi<2$ the scaling of blocks gives the leading singularity of the one-point function which is at odds with the limit to the geometry $S^1\times \mathbb{R}^2$. This is discussed in detail in Appendix \ref{A:Flat space limit}. In the rest of this section, we will mostly focus on the case $\Delta_\phi>2$.

\paragraph{Non-zero chemical potential.} In order to obtain the large-$\Delta$ expansion of the OPE coefficients $\overline{\lambda_{\phi\mathcal{OO}}}$ from the inversion formula \eqref{inversion-formula}, it is necessary to consider the one-point function, and hence the blocks, at non-zero chemical potential $\mu$, \cite{Buric:2025uqt}. The appropriate expansion, developed in Appendix \ref{AA:High-temperature expansion of blocks} (specialised here to the case $\Delta=\Delta_\phi$ relevant for the one-point function), takes the form
\begin{align}\label{scalar-block-highT-exp-main-text}
    g^{\Delta_\phi,0}_{\Delta_\phi,0}(\beta,\Omega,s) &= \frac{1}{\beta^{\Delta_\phi}}\left(b_{0,0,0}+b_{0,2,1}\,\Omega^2 s+b_{0,4,2}\,\Omega^4 s^2+\dots\right)\\
    &+\frac{1}{\beta^{\Delta_\phi-2}}\left(b_{2,0,0}+(b_{2,2,0}+b_{2,2,1}\,s)\Omega^2+(b_{2,4,1}\,s+b_{2,4,2}\, s^2)\Omega^4+\dots\right) + \dots\ .\nonumber
\end{align}
The expansion \eqref{scalar-block-highT-exp-main-text} contains two further towers of terms, starting with leading singularities $\beta^{-3+\Delta_\phi}$ and $\beta^{-2}$ (which are multiplied by regular series in $\beta^2$). The asymptotic expansion of the one-point function now follows from equation \eqref{cubic-one-point-function},
\begin{align}\label{cubic-one-point-asymptotic}
    \langle\upphi(s)\rangle_{\beta,\Omega} & = \frac{6 C_{\Delta_\phi}^{-1/2}}{A(\Delta_\phi,\Delta_\phi)}\Bigg( \frac{\zeta(\Delta_\phi)}{\beta^{\Delta_\phi}}\left(b_{0,0,0}+b_{0,2,1}\,\Omega^2 s+b_{0,4,2}\,\Omega^4 s^2+\dots\right)\\
    &+\frac{\zeta(\Delta_\phi-2)}{\beta^{\Delta_\phi-2}}\left(b_{2,0,0}+(b_{2,2,0}+b_{2,2,1}\,s)\Omega^2+(b_{2,4,1}\,s+b_{2,4,2}\, s^2)\Omega^4+\dots\right)\ +\dots \Bigg)\ .\nonumber
\end{align}
We have written the expansion for the case $\Delta_\phi>2$. The expressions for $1<\Delta_\phi<2$ are similar, with the leading singularity being $\beta^{-2}$. The asymptotic series \eqref{cubic-one-point-asymptotic} also involves the term multiplying $\beta^{-1}$, which is computed as explained in the GFF case, see equations \eqref{GFF-highT-1pt-full}-\eqref{betaminusoneGFF} and the discussion around them. Following the same steps as in Section \ref{SSS:Asymptotics of OPE coefficients}, we obtain the asymptotic HHL OPE coefficients. The final result reads
\begin{equation}\label{cubic-OPE-asymptotics>2}
    \overline{\lambda_{\phi\mathcal{OO}}^a}(\Delta,\ell) \sim C_a^{(\Delta_\phi>2)}\, \mathcal{N}_{a,\ell}\, \left(\frac{30\Delta}{\pi^4}\right)^{\frac{\Delta_\phi}{4}}\Delta^{-2a} \quad\text{as}\quad \Delta\to\infty\,,
\end{equation}
for $\Delta_\phi>2$ and
\begin{equation}\label{cubic-OPE-asymptotics<2}
    \overline{\lambda_{\phi\mathcal{OO}}^a}(\Delta,\ell) \sim C_a^{(\Delta_\phi<2)}\, \mathcal{N}_{a,\ell}\, \left(\frac{30\Delta}{\pi^4}\right)^{\frac12}\Delta^{-2a} \quad\text{as}\quad \Delta\to\infty\,,
\end{equation}
for $1<\Delta_\phi<2$. The coefficient $\mathcal{N}_{a,\ell}$ was given in equation \eqref{coefficient_N_a_l}, while the additional constants $C_a$ read
\begin{subequations}
\begin{align}\label{C>}
 C_a^{(\Delta_\phi>2)}& = \sqrt{\frac{\Gamma\left(\Delta_\phi-\frac12\right)}{\Gamma(\Delta_\phi)}} \frac{2^{\Delta_\phi-\frac{9}{2}}(\Delta_\phi-2)\Gamma\left(\frac{\Delta_\phi}{2}-1\right)^3}{\pi^\frac{5}{4}(2\Delta_\phi-3)\Gamma\left(\frac{3\Delta_\phi}{2}-3\right)}\binom{\frac{-\Delta_\phi}{2}}{a}\,\zeta(\Delta_\phi)\,,\\[1.5ex]
  C_a^{(\Delta_\phi<2)} & = \sqrt{\frac{\pi}{\Gamma(2\Delta_\phi-1)}}\frac{(-1)^a \,2^{\Delta_\phi-\frac92}\,\Gamma\left(1-\frac{\Delta_\phi}{2}\right)\Gamma\left(\frac{\Delta_\phi-1}{2}\right)\Gamma\left(\frac{\Delta_\phi}{2}+a\right)^2}{\Gamma\left(1+a\right)^2}\ .\label{C<}
\end{align}
\end{subequations}

A useful way to understand the special role of $\Delta_\phi=2$ is to view the cubic
one-point function as a tadpole diagram in the bulk.  To leading order in $\lambda_3$,
the bulk field $\Phi$ is sourced by the thermal expectation value of the composite operator
$\Phi^2$ in the non-interacting theory, namely $
(\nabla^2-m^2)\langle \Phi\rangle_\beta \sim \lambda_3
\langle \Phi^2\rangle_\beta^{\text{free}}$.
Deep in the AdS$_4$ bulk, where the chordal distance between a point $x$ and its thermal image is small compared to the curvature, the latter has the standard four
dimensional flat space thermal scaling
$
\langle \Phi^2\rangle_\beta \sim \beta^{-2}$. The two additional high temperature contributions to $\langle\phi\rangle_\beta$, with leading term $\beta^{-\Delta_\phi}$ and $\beta^{3-\Delta_\phi}$, are instead dictated by the dimension of the boundary operator. These observations can be made more precise by studying how different regions of the AdS integral \eqref{cubic_1pt_AdS_representation} contribute to the overall result.\\
The transition at $\Delta_\phi=2$ is then simply the point at which two of the high temperature singularities exchange dominance. For $\Delta_\phi>2$, the term $\beta^{-\Delta_\phi}$
is more singular than $\beta^{-2}$ and the high-temperature limit is governed by the
ordinary local scaling of a one-point function of dimension $\Delta_\phi$.
For $1<\Delta_\phi<2$, instead, the tadpole-induced contribution $\beta^{-2}$ is the
dominant one. Possibly logarithmic terms could appear at the crossover.

\smallskip

\begin{figure}[h]
        \centering
        \includegraphics[width=0.9\linewidth]{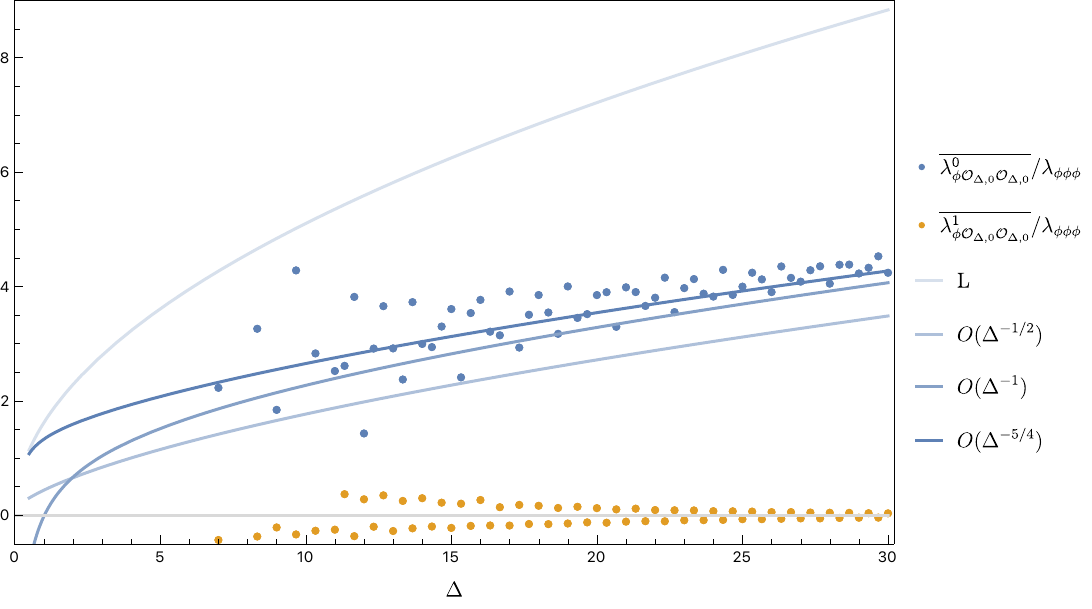}
        \caption{OPE coefficients in the $\lambda_3\Phi^3$ theory for spin one, $\Delta_\phi=\frac43$. We compare normalised exact OPE coefficients with the asymptotic formula \eqref{cubic-OPE-asymptotics<2} for the dominant tensor structure with additional subleading corrections.}
        \label{cubic-OPE-plot-spin-1}
\end{figure}

\begin{figure}[h]
        \centering
        \includegraphics[width=0.9\linewidth]{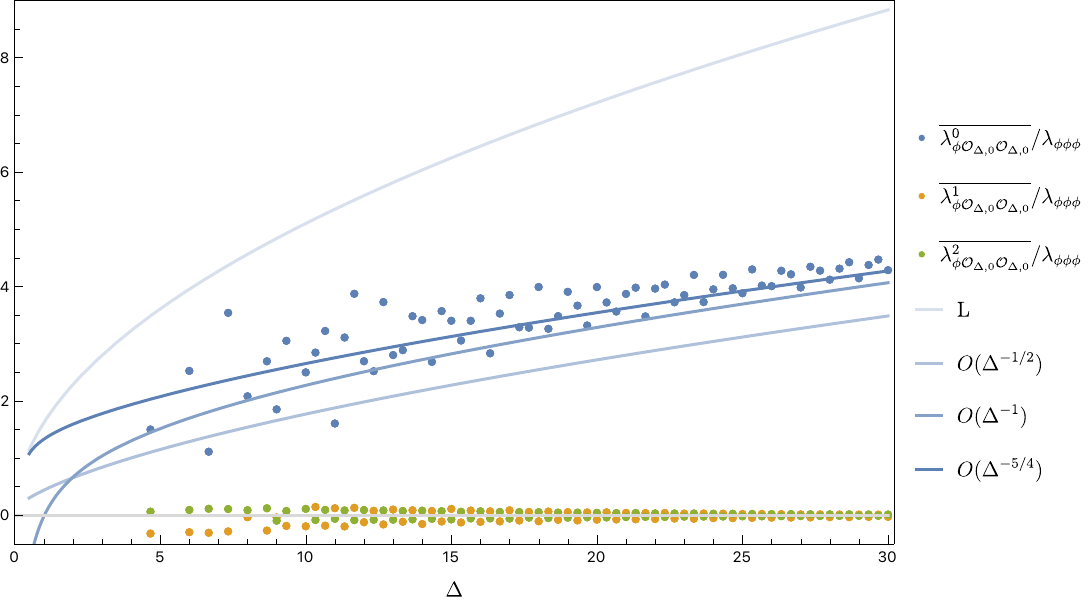}
        \caption{OPE coefficients in the $\lambda_3\Phi^3$ theory for spin two, $\Delta_\phi=\frac43$. We compare normalised exact OPE coefficients with the asymptotic formula \eqref{cubic-OPE-asymptotics<2} for the dominant tensor structure with additional subleading corrections.}
        \label{cubic-OPE-plot-spin-2}
\end{figure}

In \eqref{cubic-OPE-asymptotics>2} and \eqref{cubic-OPE-asymptotics<2} we have only displayed the leading asymptotic result, but a systematic asymptotic expansion is readily available. In Figures \ref{cubic-OPE-plot-spin-1} and \ref{cubic-OPE-plot-spin-2} we show the exact OPE coefficients against asymptotic formulas of increasingly high order. Blue dots correspond to dominant OPE coefficients, $a=0$, and different colours to subdominant tensor structures ($a=1$ orange, $a=2$ green). Upon keeping sufficiently many orders, the asymptotic expansion agrees very well with the exact data.

\subsubsection{OPE coefficients with a fixed particle number}
\label{SSS:OPE coefficients with a fixed particle number}

\begin{figure}[h!]
    \centering
    \begin{minipage}[b]{0.49\textwidth}
        \includegraphics[width=\textwidth]{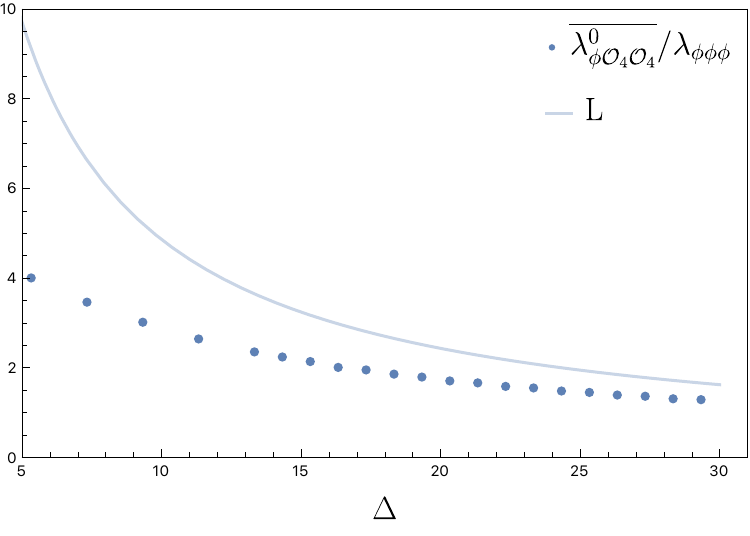}
    \end{minipage}
    \hfill
    \begin{minipage}[b]{0.49\textwidth}
        \includegraphics[width=\textwidth]{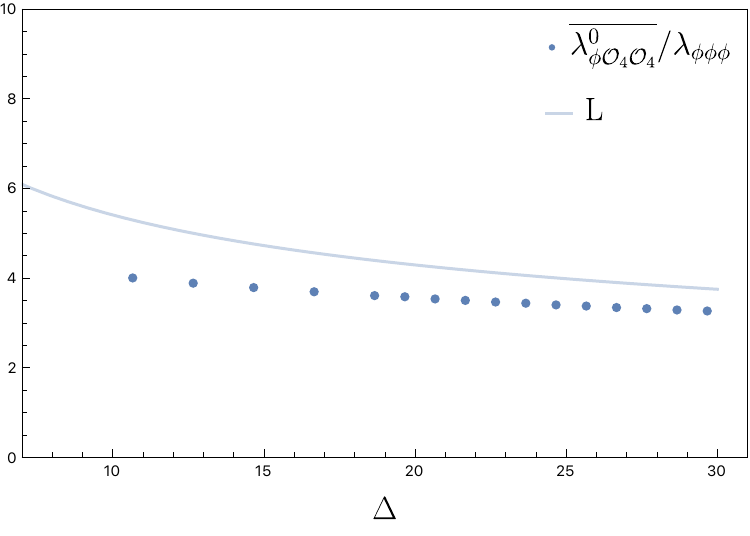}
    \end{minipage}
    \caption{OPE coefficients with four-particle scalar operators for $\Delta_\phi = \frac43$ (left) and $\Delta_\phi = \frac83$ (right). We compare them with the leading formula \eqref{cubicOPE_asymFixFam}.}
    \label{fig:OPE_fixed_family_hol}
\end{figure}

Paralleling the discussion of Section \ref{SSS:Asymptotics of OPE_GFF_fixed particles} for the case of GFF, we can refine the asymptotic formulas \eqref{cubic-OPE-asymptotics>2}-\eqref{cubic-OPE-asymptotics<2} by including the additional constraint on the number of particles. For simplicity, we focus on the case of the dominant tensor structure, $a=0$. The inversion formula \eqref{inversion-formula} now involves the projected one-point function
\begin{equation}
        \left(\gcZ_0\langle\upphi(\hat x)\rangle_{q,y}\right)^{(n)}  = \left(\gcZ_0(\beta,\mu,g) \frac{6C_{\Delta_\phi}^{-1/2}\, \hat r^{-\Delta_\phi}}{A(\Delta_\phi,\Delta_\phi)} \sum_{k=1}^\infty\ g^k g_{\Delta_\phi,0}^{\Delta_\phi,0}\left(q^k,y^k,\hat s\right)\right)^{(n)} \,,
\end{equation}
whose high temperature behaviour directly follows from equations \eqref{eq:Z_U(1)} and \eqref{scalar-block-highT-exp-main-text}. Using the $\beta,\Omega,\hat s$ variables, we obtain the asymptotics of OPE coefficients by simply rescaling $\beta\rightarrow\beta/\Delta$ in the inversion formula, and then integrating term by term the $\Delta\rightarrow\infty$ expansion of the integrand\footnote{As in the GFF case (see section \ref{SSS:Asymptotics of OPE_GFF_fixed particles}), in performing the integrals we assume $n\geq3$.}. As in the unconstrained particle number setting, we find two different leading behaviours depending on the value of $\Delta_\phi$
\begin{equation}\label{cubicOPE_asymFixFam}
    \overline{\lambda_{\phi\mathcal{OO}}^0}(\Delta,\ell,n)\sim
    \begin{cases}
        K_n^{(\Delta_\phi>2)}\Delta^{\Delta_\phi-3}, \quad &\text{if }\Delta_\phi>2\\
        K_n^{(\Delta_\phi<2)}\Delta^{-1}, \quad &\text{if }1<\Delta_\phi<2
    \end{cases}
\end{equation}
where all the dependence in the number of particles $n$ is encoded in the constant prefactors $K_n$, that read
\begin{subequations}
\begin{align}\label{K>}
    K_n^{(\Delta_\phi>2)} & = \frac{1}{\sqrt{C_{\Delta_\phi}}}\frac{2^{\Delta_\phi-4}n(3n-6)!(\Delta_\phi-2)\Gamma\left(\frac{\Delta_\phi}{2}-1\right)^3}{3\pi^2(\Delta_\phi+2n-7)(2\Delta_\phi-3)\Gamma\left(\Delta_\phi+3n-9\right)\Gamma\left(\frac{3\Delta_\phi}{2}-3\right)}\,,\\[1ex]
    K_n^{(\Delta_\phi<2)}& = \frac{1}{\sqrt{C_{\Delta_\phi}}} \frac{3\cdot2^{5-2n-\Delta_\phi}n(n-2)(3n-7)(3n-6)!\,\Gamma\left(\Delta_\phi-1\right)}{\sin\left(\frac{\pi \Delta_\phi}{2}\right)\,\Gamma\left(n-1-\frac{\Delta_\phi}{2}\right)\Gamma\left(n-\frac52+\frac{\Delta_\phi}{2}\right)\Gamma\left(\Delta_\phi-\frac12\right)}\ .\label{K<}
\end{align}
\end{subequations}
In Figure \ref{fig:OPE_fixed_family_hol}, we can see the two different behaviours are compatible with the data. If we had not distinguish the two cases, the data would be off by a factor of three.

\subsubsection{Exact OPE coefficients}

In this brief subsection, we discuss some exact results for OPE coefficients in the cubic theory. For any low-lying operator $\mathcal{O}$, one can readily obtain the coefficients $\overline{\lambda^a_{\phi\mathcal{OO}}}$, see for instance Figure \ref{OPE-exact-plot}. In addition, some infinite families of operators are fairly easily analysed. In the following, we shall write all OPE coefficients in units of $\lambda_{\phi\phi\phi}$. The simplest infinite family consists of operators $\phi^n$, whose OPE coefficients read
\begin{equation}\label{leading-OPE-family-cubic}
    \frac{\lambda_{\phi\phi^n\phi^n}}{\lambda_{\phi\phi\phi}} = n\ .
\end{equation}
These are the largest OPE coefficients for operators of a given dimension and correspond to the leading dashed blue line on Figure \ref{OPE-exact-plot}. Scalar operators with a fixed ‘number of derivatives' $k$, i.e. operators with twist of the form $\Delta-\ell = n\Delta_\phi + k$, are smaller than equation \eqref{leading-OPE-family-cubic} and tend to it in the $n\to\infty$ limit.\footnote{Again, this is an observation based on all computed examples. An intuitive explanation parallels that for anomalous dimensions, see the discussion around equation \eqref{large-n-anomalous-dim}.} For instance, for the family with quantum numbers $(n\Delta_\phi+2,0)$, the OPE coefficients read
{\small \begin{align}\label{OPE-cubic-nDelta-plus-2}
    &\frac{\overline{\lambda_{\phi\mathcal{OO}}}}{\lambda_{\phi\phi\phi}}\\
    &= \frac{2\Delta_\phi(2\Delta_\phi-1) n^2 + \frac16 (\Delta_\phi^4 +2\Delta_\phi^3 - 17\Delta_\phi^2 - 6\Delta_\phi+6) n - \frac16(\Delta_\phi-3)(\Delta_\phi^3 - 3\Delta_\phi^2 + 6\Delta_\phi - 2)}{(2n\Delta_\phi-1)(2\Delta_\phi-1)}\ . \nonumber
\end{align}}

As an additional example, the family with $(\Delta,\ell)=(n\Delta_\phi+4,0)$ can be found in Appendix \ref{AA:OPE coefficients}. The asymptotic linear growth of families with different number of derivatives with the slope $\Delta_\phi^{-1}$ is clearly visible on Figure \ref{OPE-exact-plot}.
\smallskip

\begin{figure}
        \centering
        \includegraphics[width=0.8\linewidth]{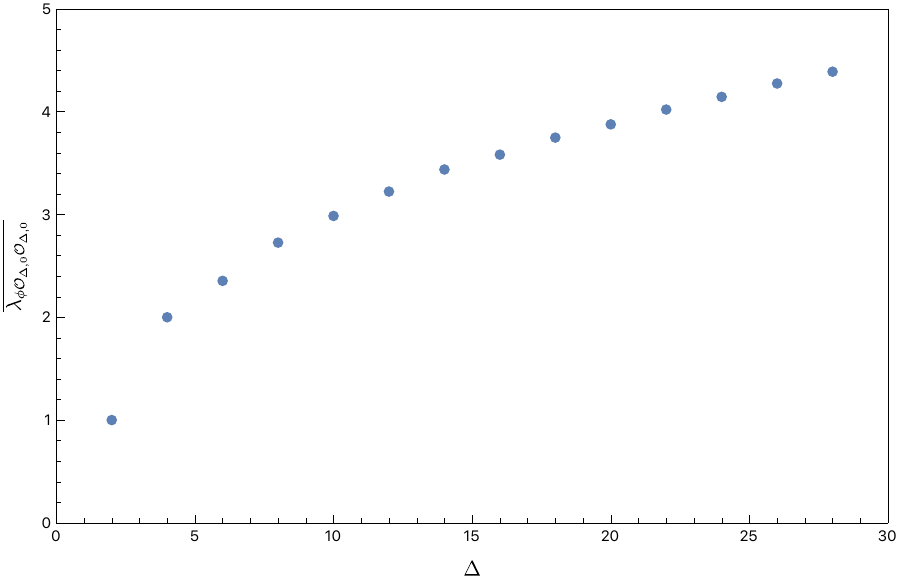}
        \caption{Coarse-grained scalar OPE coefficients taken from Fig. \ref{OPE-exact-plot}.}
        \label{hol-OPE-goarse-grained}
\end{figure}

Finally, let us comment on the emergence of the smooth averaged formulas for OPE coefficients from various families of operators. This process can be exhibited visually by ‘coarse-graining', i.e. averaging OPE coefficients over some intervals of the conformal dimension $\Delta$. In Figure \ref{hol-OPE-goarse-grained}, we show such coarse-graining of the plot from Figure \ref{OPE-exact-plot} over intervals $[2n,2n+2)$ of length two. As advocated, different families of operators average to give a ‘smooth curve', which can be accurately approximated by our asymptotic expansions.

\section{Summary and discussion}
\label{S:Summary and discussion}

In this work, we have extended our previous study of HHL OPE coefficients in three-dimensional CFTs to non-local theories such as the GFF and weakly interacting QFTs in AdS$_4$. We have developed systematic asymptotic expansions around $\Delta=\infty$ and saw that they can be used to approximate exact OPE coefficients very well down to $\Delta\sim15$. A similar study, with similar findings, was performed for the density of primaries. These general asymptotic formulas were refined by the particle number, which is a well-defined quantum number in weakly interacting theories. The resulting expressions, which again accurately reproduce exact CFT data at intermediate $\Delta$, give some insight into multi-particle sectors of CFTs that are difficult to access by traditional methods, and complement recent studies at large spin, \cite{Harris:2024nmr,Kravchuk:2024wmv,Fardelli:2025eun,Fardelli:2025fkn}. In theories with a global $U(1)$ symmetry, multi-particle states correspond to different charge sectors. At large charge, the CFT in these regimes is captured by effective field theory, \cite{Hellerman:2015nra,Cuomo:2017vzg}. It would be interesting to combine the large charge EFT with techniques of this work.
\smallskip

We take the results of this work as an encouraging sign that thermal correlation functions and associated inversion formulas can be used to produce accurate CFT data down to regimes accessible to the numerical bootstrap. As an intermediate step, it would be interesting to connect the asymptotic expansion of the spectral density and/or the OPE coefficients with numerical estimates coming from other methods, such as the fuzzy sphere, \cite{Zhu:2022gjc}. In particular, the recent work \cite{Fardelli:2026zas} used the latter method to produce estimates of $\lambda_{\sigma\mathcal{OO}}$ and $\lambda_{\epsilon\mathcal{OO}}$ OPE coefficients in the 3d Ising model up to $\Delta\sim10$. In order to reproduce these results, several orders of the asymptotic expansion are expected to be required. To this end, it is desirable to use a more economical high-temperature expansion of one-point functions, provided by the thermal EFT.

\paragraph{Thermal effective action.} An interesting future direction is to develop a thermal effective field theory for one-point functions on $S^1\times S^{d-1}$. Generalising the case of the partition function, the thermal EFT is supposed to give a general parametrisation of high-temperature behaviour of thermal one-point functions\footnote{We thank Sridip Pal for several discussions and explanations on the topic.}. For the moment, we will simply observe that one-point functions of this work, up to certain terms described below, take a simple thermal EFT form. Recall that the thermal EFT for a partition function of a local CFT$_3$ takes the form of a derivative expansion
\begin{equation}
    \log\mathcal{Z}_{S^1 \times S^2} = - \int_{S^2} d^2 x \sqrt{\hat g} \left( -f + c_1 \hat R + c_2 F^2 + \dots \right)\ .
\end{equation}
For reader's convenience, we collect the relevant definitions in Appendix \ref{A:Thermal effective field theory}. The one-point functions computed in this paper take a very similar form, namely
\begin{equation}\label{tEFT-1pt}
    \langle\phi(s)\rangle_{\beta,\Omega}^\ast = e^{-\Delta_\phi \Phi} \left(a_0 + a_1 \hat R(s) + a_2 F^2(s) + \dots\right)\ .\\
\end{equation}
Here, $\Phi$ is the value of the dilaton field on $S^1\times S^2$ upon Kaluza-Klein reduction, see equation \eqref{g-A-Phi-evaluated}, and $a_i$ are some theory-dependent constants. Unlike for $\mathcal{Z}$, the geometric invariants $\hat R,F^2\dots$ are not integrated over the sphere, giving the spacetime dependence to $\langle\phi\rangle$. The asterisk on the one-point function in equation \eqref{tEFT-1pt} indicates however that the right hand side does not capture the full asymptotic high-temperature expansion: for $\langle\phi^2\rangle$ in GFF it omits the term multiplying $\beta^{-1}$, while for $\langle\phi\rangle$ in the cubic theory it captures only one of the three asymptotic towers described in Appendix \ref{AA:High-temperature expansion of blocks}. After evaluating the Einstein and Maxwell invariants on $S^2$, see equation  \eqref{Einstein-Maxwell-S1S2}, one gets
\begin{equation}
    \langle\phi(s)\rangle_{\beta,\Omega}^\ast = \frac{\beta^{-\Delta_\phi}}{\left(1+s\Omega^2\right)^{\frac{\Delta_\phi}{2}}} \left(a_0 + a_1 \frac{2\beta^2\left(1 + 5(1-s)\Omega^2-s\Omega^4\right)}{1 + s\Omega^2} + a_2 \frac{8(1-s)\beta^2\Omega^2}{1+s\Omega^2} + \dots\right)\ .
\end{equation}
Refer to Appendix \ref{A:Thermal effective field theory} for the values of coefficients $a_i$ and some of their higher-derivative cousins in GFF and the cubic theory in AdS. The structure of the expansion \eqref{tEFT-1pt} can be understood by introducing a source $J$ for the operator $\phi$ in the thermal path integral. Indeed, upon doing this, the one-point function is obtained in the standard way by differentiating the resulting generating functional with respect to $J$ and setting the source to zero. After integrating out the CFT degrees of freedom and performing the Kaluza-Klein reduction along the thermal circle, $J$ should be treated as an additional background field in the reduced thermal effective action. The local thermal effective action may then contain all terms linear in this source and built from the reduced background fields, such as couplings to $\hat R$, $F^2$, and their higher-derivative generalisations. Differentiating these local terms with respect to the original source directly produces the EFT form in equation \eqref{tEFT-1pt} (the prefactor in \eqref{tEFT-1pt} originating from transformations of $J$ under Weyl rescalings). This also clarifies the meaning of the asterisk there: the expression captures precisely the part of the high-temperature expansion that admits a local derivative expansion in the reduced background fields. Additional contributions, such as non-local  terms, need not be encoded in this local thermal EFT.
\vskip0.05cm

We leave the systematic study of thermal EFTs for one-point functions for future work. More general questions in this direction are to try and relate thermal EFT coefficients of different observables, place positivity bounds on them, or derive sum rules involving both CFT and thermal EFT data. See \cite{Benjamin:2023qsc,Banihashemi:2025qqi} and \cite{Allameh:2024qqp} for related discussions of thermal EFTs for partition and one-point functions, respectively.

\paragraph{Applications in holography and other theories.} While our study of OPE coefficients introduced by cubic interactions in the bulk made use of a specific nature of this interaction, the methods used for anomalous dimensions are rather generic. Thus, they lay some groundwork for more complicated studies of higher loops and/or gauge theories in AdS. The latter are of particular interest in view of their relation to confinement, \cite{Aharony:2012jf,Ciccone:2024guw, Ciccone:2025dqx,Ankur:2026ylr}. Another interesting extension of the present work is to study theories with dynamical gravity. In such theories, thermal one-point functions may encode geometric properties of black hole backgrounds, such as lengths of certain geodesics, \cite{Grinberg:2020fdj,Berenstein:2022nlj,David:2022nfn}. In upcoming work, we will study CFT data in these theories, starting with the pure gravity in the Kerr-AdS black hole phase.\footnote{For a recent study from the gauge theory side at weak coupling, using spin chains, see \cite{Kristjansen:2025xqo}.}

A further analytically tractable example that is ‘in between' holographic and free theories is the $O(N)$ vector model at large $N$. Investigations of this model on $S^1 \times S^2$ have been initiated in \cite{David:2024pir,David:2025tqn}. Deriving high-temperature correlators and the corresponding CFT data in this theory remains an interesting task for the future. 

\paragraph{Further generalisations.} In this paper, we focused on thermal one-point functions of scalar operators. In order to generalise the asymptotic formulas for HHL OPE coefficients to cases where the light operator carries spin (the stress tensor being an important example), the spinning extension of the inversion formula \eqref{inversion-formula} is needed. We will derive this in the upcoming work \cite{Buric:wip}. In particular, it will be explained how the spinning generalisation requires an introduction of an extended space of solutions to Casimir equations on which orthogonality relations may be formulated.
\vskip0.05cm

Complementary to our methods that probe CFT data at finite spin, new techniques have recently been developed to probe an asymptotic regime in which both the spin and the twist become infinite \cite{Anand:2025mfh,Komargodski:2026ain}. These have so far been used to analyse the partition function, but could potentially be generalised to one-point functions. It is worth investigating how the conformal blocks behave in the aforementioned limit and deduce the consequences for the corresponding CFT data.
\vskip0.05cm

Finally, a longer term goal would be to move to two-point functions on $S^1\times S^{d-1}$. The latter have a very rich structure, from possessing non-trivial singularities in holographic theories to satisfying crossing equations, such as the KMS condition. Developing the theory of two-point conformal blocks on $S^1 \times S^{d-1}$ presents an interesting task for the future.

\paragraph{Acknowledgements:} We would like to thank Ivan Gusev, Jeremy Mann, Alessio Miscioscia, Sridip Pal, Marco Serone and Balt van Rees for discussions. IB wishes to thank \' Ecole Polytechnique, where part of this work was done, for hospitality. AV wishes to thank the participants of the workshop ‘‘QFT in AdS 2026" for interesting discussions.
IB is funded by Taighde Éireann – Research Ireland under Grant number SFI-22/FFP-P/11444. FM and VS are funded by the German Research Foundation DFG – SFB 1624 – “Higher structures, moduli spaces and integrability” – 506632645. This project also received funding from the German Research Foundation DFG under Germany’s Excellence Strategy - EXC 2121 Quantum Universe - 390833306. FR is supported by the ERC project number 101087025 “QFTinAdS”. AV is partially supported by the Italian MUR under contract 20223ANFHR (PRIN2022) and FIS-2024-05507 (FIS Advanced Grant).

\appendix

\section{Characters of \texorpdfstring{$SO(3,2)$}{SO32}}

In this Appendix, we collect some background on the harmonic analysis and representation theory of the conformal group $SO(3,2)$. The parabolic Verma modules that appear as state representations in conformal field theory are denoted by $V_{\Delta,\ell}$. Their characters read
\begin{equation}\label{characters}
    \chi_{\Delta,\ell}(q,y) = \text{tr}_{V_{\Delta,\ell}}\left(q^D y^M\right) = 
    \frac{q^\Delta y^{-\ell} \left(1 - y^{2\ell+1}\right)}{(1-q)(1-y)(1-qy)(1-qy^{-1})}\ .
\end{equation}
By a slight abuse of notation, we shall write the characters also as $\chi_{\Delta,\ell}(\beta,\mu)$, bearing in mind the relations \eqref{sets-of-variables}, depending on the set of variables being used. The characters satisfy orthogonality relations
\begin{equation}\label{orthogonality-characters}
   \int\limits_{\gamma-i\infty}^{\gamma+i\infty} d\beta \int\limits_{-\pi}^{\pi} d\mu\ \omega(\beta,\mu)\ 
   \chi_{\Delta_1,\ell_1}(\beta,\mu) \chi_{3-\Delta_2,\ell_2}(\beta,\mu) = 
   2\pi i \delta(\Delta_1-\Delta_2)  \delta_{\ell_1,\ell_2}\,,
\end{equation}
with respect to the measure
\begin{equation}\label{Haar-measure}
    \omega(\beta,\mu) = \left(\frac{8}{\sqrt{\pi}} \sinh\frac{\beta}{2} \sin\frac{\mu}{2} 
    \sinh\frac{\beta+i\mu}{2} \sinh\frac{\beta-i\mu}{2} \right)^2\ .
\end{equation}

\section{Thermal effective field theory}
\label{A:Thermal effective field theory}

In this Appendix, we review the thermal effective action from \cite{Benjamin:2023qsc} (see also \cite{Bhattacharyya:2007vs,Banerjee:2012iz,Jensen:2012jh}), specified to the three-dimensional geometry $S^1 \times S^2$. In the concluding Section \ref{S:Summary and discussion}, a brief discussion of the generalisation to one-point functions is given.
\smallskip

We start with the metric on $S^1 \times S^2$,
\begin{equation}\label{metric-S1xS2}
    G = \beta^2 d\tau^2 + \frac{ds^2}{4s(1-s)} + s d\varphi_1^2\ .
\end{equation}
Here, $s=\sin^2\theta$ and $(\theta,\varphi_1)$ are the usual angular coordinates on the sphere. Next, we put the metric in the Kaluza-Klein form by introducing $\varphi = \varphi_1 - \beta \Omega d\tau$
\begin{equation}\label{metric-KK-form}
    G = \beta^2 \left(1 + s\Omega^2\right) \left(d\tau + \beta^{-1} \frac{s\Omega}{1+s\Omega^2}d\varphi\right)^2 + \frac{ds^2}{4s(1-s)} + \frac{s d\varphi^2}{1+s\Omega^2}\ .
\end{equation}
From the last expression, one reads off the two-dimensional metric, gauge field and the dilaton, respectively
\begin{equation}\label{g-A-Phi-evaluated}
    g = \frac{ds^2}{4s(1-s)} + \frac{s d\varphi^2}{1+s\Omega^2}\,, \qquad A = \beta^{-1} \frac{s\Omega}{1+s\Omega^2}d\varphi\,, \qquad e^{2\Phi} = \beta^2 \left(1 + s\Omega^2\right)\ .
\end{equation}
The CFT partition function at high temperatures, up to exponentially suppressed terms, is given by 
\begin{equation}
    \gcZ\sim e^{-S_{\text{th}}[G]}\,,
\end{equation}
where the thermal effective action
\begin{equation}
    S_{\text{th}}[G] = S_{\text{th}}[g,A,\Phi]\,,
\end{equation}
is a local expression in the KK fields $(g,A,\Phi)$, invariant under two-dimensional coordinate and Weyl transformations, as well as gauge transformations of $A$. In particular, the Weyl invariance
\begin{equation}\label{Weyl-invariance}
    S_{\text{th}}[G] = S_{\text{th}}[e^{2\sigma}G]\,,
\end{equation}
can be used to trivialise the dependence on the dilaton field. Indeed, setting $\sigma = -\Phi$ in equation \eqref{Weyl-invariance}, the metric $\hat G = e^{-2\Phi}G$ has
\begin{equation*}
    \hat g = e^{-2\Phi} g\,, \qquad \hat A = A\,, \qquad \hat\Phi = 0\,,
\end{equation*}
and consequently
\begin{equation}
    S_{\text{th}}[g,A,\Phi] = S_{\text{th}}[e^{-2\Phi}g,A,0] \equiv S[\hat g,A]\ . 
\end{equation}
The simplest invariants are the Ricci scalar associated with $\hat g$ and the squared field strength constructed from $A$, giving the action
\begin{equation}
    S_{\text{th}} = \int d^2 x \sqrt{\hat g} \left( -f + c_1 \hat R + c_2 F^2 + \dots \right)\,,
\end{equation}
where $f$ and $c_{1,2}$ are some theory dependent constants. Although they can be computed in explicit models \cite{Benjamin:2023qsc,David:2024pir,Mauro:2026zus}, they represent arbitrary constants at the level of effective theory.  In the context of conformal field theory, $f$ is the free energy density. On the geometry $S^1\times S^2$, the next two (Einstein and Maxwell) terms evaluate to
\begin{equation}\label{Einstein-Maxwell-S1S2}
    \hat R = \frac{2\beta^2\left(1 + 5(1-s)\Omega^2-s\Omega^4\right)}{1 + s\Omega^2}\,, \qquad F^2 = \hat g^{ik} \hat g^{jl} F_{ij} F_{kl} = \frac{8(1-s)\beta^2\Omega^2}{1+s\Omega^2}\,,
\end{equation}
and contribute
\begin{equation*}
    4\pi \qquad \text{and} \qquad \frac{16\pi\Omega^2}{3\left(1+\Omega^2\right)}\,,
\end{equation*}
to the thermal effective action, respectively. At the fourth derivative order, we have four terms: $\hat R^2$, $(F^2)^2$, $\hat R F^2$ and 
\begin{equation}
    (\nabla F)^2 = \frac{8\beta^4 s \Omega^2 \left(1+\Omega^2\right)^2}{\left(1+s\Omega^2\right)^2}\ .
\end{equation}
While all four terms are linearly independent before integration over $S^2$, after the integration only three independent linear combinations remain, which may be taken to be
\begin{equation}
    \frac{\beta^2}{1+\Omega^2}\,, \qquad \frac{\beta^2\Omega^2}{1+\Omega^2}\,, \qquad \frac{\beta^2\Omega^4}{1+\Omega^2}\ .
\end{equation}
In a similar way one constructs the higher derivative terms, which carry higher powers of $\beta$. For more details, the reader is referred to \cite{Benjamin:2023qsc}.

\subsection{Thermal coefficients in GFF and the interacting theory} 

In this section, we spell out the values of coefficients $a_i$ appearing in equation \eqref{tEFT-1pt} for the thermal one-point functions considered in this paper. For clarity, we will write as indices of EFT coefficients the corresponding invariants, e.g. $a_0$ from equation \eqref{tEFT-1pt} will be denoted by $a_\Lambda$, $a_1$ by $a_{\hat R}$ etc. The coefficients for the one-point function of $\phi^2$ in GFF up to fourth derivative order read
\begin{align}
    & a_\Lambda = \sqrt{2}\, \zeta(2\Delta_\phi)\,, \qquad a_{\hat R} = - \frac{\Delta_\phi\, \zeta(2\Delta_\phi - 2)}{12\sqrt{2}}\,, \qquad a_{F^2} = \frac{5\Delta_\phi\, \zeta(2\Delta_\phi-2)}{48\sqrt{2}}\,, \nonumber\\[2pt]
    & a_{\hat R^2} = \frac{\Delta_\phi(5\Delta_\phi+1) \zeta(2\Delta_\phi-4)}{2880\sqrt{2}}\,, \qquad a_{(F^2)^2} = \frac{5\Delta_\phi(5\Delta_\phi+1) \zeta(2\Delta_\phi-4)}{9216\sqrt{2}}\,,\\[2pt]
    & a_{\hat R F^2} = -\frac{\Delta_\phi(5\Delta_\phi+1) \zeta(2\Delta_\phi-4)}{1152 \sqrt{2}}\,, \qquad a_{(\nabla F)^2} = -\frac{\Delta_\phi\zeta(2 \Delta_\phi-4)}{1440 \sqrt{2}}\ . \nonumber
\end{align}
The coefficients for the one-point function of $\phi$ in the interacting model up to second derivative order read
\begin{align}
    & a_\Lambda = \oneptcub\,, \qquad a_{\hat R} = \frac{\Delta_\phi\left(11\Delta_\phi^2 - 49 \Delta_\phi +56\right) \zeta(\Delta_\phi -2)}{48 (\Delta_\phi-4) (2\Delta_\phi-5) \zeta(\Delta_\phi )} \oneptcub\,,\\[2pt] 
    &a_{F^2} = -\frac{\Delta_\phi \left(55\Delta_\phi^2-263 \Delta_\phi+316\right) \zeta(\Delta_\phi-2)}{192(\Delta_\phi -4)(2\Delta_\phi-5)\zeta(\Delta_\phi)} \oneptcub\,,
\end{align}
where $\oneptcub$ is given in \eqref{1-pt-coefficients-holography}.
 
\section{New results on blocks with external scalars}
\label{A:New results on blocks with external scalars}

In this Appendix, we derive several exact formulas for thermal conformal blocks, akin to equation \eqref{exact-scalar-block}, by resumming their low-temperature expansions. In particular, these results allow to study the high-temperature behaviour of conformal blocks.
\smallskip

We focus on the case of a scalar external field. For completeness, we recall the Laplace-Casimir operator entering the Casimir equation \eqref{Casimir-eqn} (see \cite{Gobeil:2018fzy,Buric:2024kxo,Buric:2025uqt} for its derivation)
\begin{align}
    &\mathcal{C} _{\Delta_\phi} = -2q^2 \partial_q^2-2y^2\partial_y^2 +\frac{4q^2}{1-q}\partial_q +
    \frac{4y^2}{1-y}\partial_y-2\frac{q+y}{q-y}(q\partial_q-y\partial_y)-2\frac{qy+1}{qy-1}
    (q\partial_q+y\partial_y)\nonumber\\[2pt]
    &+ \frac{q}{(q-1)^2}\mathcal{D}^{(1)}_{\Delta_\phi}(s) + \frac{y}{(y-1)^2}
    \mathcal{D}^{(2)}_{\Delta_\phi}(s) + \left( \frac{qy}{(q-y)^2} + 
    \frac{qy}{(qy-1)^2}\right) \mathcal{D}^{(3)}_{\Delta_\phi}(s)\,, \label{Laplace-Casimir-op}
\end{align}
where
\begin{align}\label{D1-s-variable}
    \mathcal{D}^{(1)}_{\Delta_\phi} & = 8s^2(1-s)\partial_s^2 +4\big(1+2\Delta_\phi 
    -2s(\Delta_\phi+1)\big) s\partial_s - 2\Delta_\phi \big(1 + (s-1)\Delta_\phi \big) 
    \, ,\\[2pt]
    \mathcal{D}^{(2)}_{\Delta_\phi} & = 8s(s-1)\partial_s^2 + 4(3s-2)
    \partial_s\,,\quad \mathcal{D}^{(3)}_{\Delta_\phi} = - 
    \frac{\mathcal{D}^{(1)}_{\Delta_\phi} + \mathcal{D}^{(2)}_{\Delta_\phi}}{2} + \Delta_\phi 
    (\Delta_\phi-3)\ .\label{D2-s-variable}
\end{align}
The low-temperature expansion simplifies greatly by passing to a new variable $\bb\equiv\frac{q}{(1-q)^2}$, which is invariant under the symmetry $q\rightarrow q^{-1}$ of the Casimir equation. We expand the blocks as
\begin{align}
    g^{\Delta_\phi,a}_{\Delta ,\ell }(q,u,s) & = 
    \bb^{\Delta }\sum_{n_1=0}^{\infty} \, f_{n_1}(u,s) \,\bb^{n_1}, \\
    f_{n_1}(u,s)=\sum_{n_2,n_3} B_{(n_1,n_2,n_3)} u^{n_2} s^{n_3}\,, \qquad & \text{with} \qquad  
    0\leq n_2\leq n_1+ \ell  \,, \ \ 0\leq n_3 \leq n_2\,,        
\end{align}
This choice of variable allows for a simple derivation of several new results:
\medskip

\textbf{1) Blocks at vanishing chemical potential}: Given a fixed spin $\ell$ and a fixed tensor structure label $a$, it is possible to write down a closed formula for the block at $\mu=0$. Recall that these blocks do not depend on $s$ due to extended Ward identities, \cite{Buric:2024kxo}. The simplest case occurs for $a=\ell$, in which we find
\begin{align}\label{exact-blocks-subdominant}
    g^{\Delta_\phi,a=\ell}_{\Delta,\ell} & = \,c_{\ell,\ell}\; 
    \frac{16^{\ell}\,\left(\tfrac12\right)_{\ell}\,\left(\tfrac32-\tfrac{\Delta_{\phi}}{2}\right)_{\ell}^{2}
    }{\left(2-\Delta\right)_\ell\,\left(-2+2\Delta\right)_{\ell}\,\left(\Delta+\ell\right)_\ell}
    \\[1ex]
    & \bb^{\Delta+\ell} \;
    _3F_2\left(\Delta+\ell-1,\Delta+\ell-\frac{\Delta_\phi}{2},\Delta+\ell+\frac{\Delta_\phi-3}{2};\Delta+2\ell,2 \Delta+\ell-2;\;\bb\right)\,,\nonumber
\end{align}
where $c_{\ell,\ell}$ is the normalisation constant specified in \cite{Buric:2025uqt}, see equation (2.36) of that work. Closed form expressions for the other tensor structures are more complicated, but can be derived at least on a case-by-case basis. For example, in the case of spin one exchange, the remaining tensor structure (the dominant one) can be written as the sum of two terms
\begin{align}
    g^{\Delta_\phi,a=0}_{\Delta,1} &= \bb^\Delta \, \Bigg[3\;_3F_2\left(\Delta-1,\Delta-\frac{\Delta_\phi}{2},\Delta+\frac{\Delta_\phi-3}{2};\Delta,2 \Delta-2;\;\bb\right)\\
    &+\frac{(2\Delta-3)(\Delta_\phi-3)\Delta_\phi}{(\Delta-2)\Delta(\Delta^2-1)}\,\bb\;_3F_2\left(\Delta,1+\Delta-\frac{\Delta_\phi}{2},\Delta+\frac{\Delta_\phi-1}{2};\Delta+2,2 \Delta-1;\;\bb\right)\Bigg]\ . \nonumber
\end{align}
It is suggestive to note that the second term essentially coincides with the block for the other tensor structure $a=1$ present for spin $\ell=1$, while the first term has the same functional form as the scalar exchange block \eqref{exact-scalar-block}. Similar expressions can be obtained for $\ell\geq2$.
\medskip

\textbf{2) Scalar exchange blocks at $\mu\neq0$}: We find the following closed-form of the expansion
\begin{align}\label{exact-full-scalar-block}
    &g^{\Delta_\phi,0}_{\Delta,0}=\\
    & \bb^\Delta\hspace{-4pt}\sum_{k,m,n\geq0}
    \frac{(1)_{m+n}(\Delta-1)_k\left(\Delta+\frac{\Delta_\phi-3}{2}\right)_k\left(\Delta-\frac{\Delta_\phi}{2}\right)_{k+m}\left(\frac{\Delta_\phi}{2}\right)_n^{2}}
    {(2\Delta-2)_k\,(\Delta)_{k+m+n}}\;
    \frac{(-4\bb)^k}{k!}\,\frac{(\bb u)^m}{m!}\,\frac{(\bb us)^n}{(n!)^2}\ . \nonumber
\end{align}
This representation falls into the broad class of three-variable hypergeometric series, but doesn't appear to be one of the most studied cases. The regions of convergence of this type of series are typically known in the literature \cite{Srivastava1985MultipleGH}.  For some applications it may be useful to rewrite one of the sums in terms of a generalised hypergeometric function. For example, performing the sum over $k$ yields
\begin{align}
    g^{\Delta_\phi,0}_{\Delta,0}=\,\bb^\Delta\hspace{-4pt}\sum_{m,n\geq0}&
    \;_3F_2\left(\Delta-1,m+\Delta-\frac{\Delta_\phi}{2},\Delta+\frac{\Delta_\phi-3}{2};m+n+\Delta,2 \Delta-2;\;\bb\right)\nonumber\\
   &\frac{(m+n)!\left(\Delta-\frac{\Delta_\phi}{2}\right)_m\left(\frac{\Delta_\phi}{2}\right)_n^2}
    {(\Delta)_{m+n}}\,\frac{(\bb u)^m}{m!}\,\frac{(\bb us)^n}{(n!)^2}\ .\label{exact-scalar-block-resummed}
\end{align}
From the last expression, it is manifest that setting $m=n=0$ recovers the known result \eqref{exact-scalar-block} for the block at zero chemical potential.
\medskip

To conclude this section, let us mention that in the recent work \cite{Alkalaev:2026sha}, the scalar exchange thermal blocks at $\mu=0$ have been linked both to a particular limit of flat space four-point blocks and to a particular configuration of defect two-point blocks. It would be interesting to understand if the closed formulas for the more general type of thermal blocks presented here can be also obtained by using a similar representation. It is suggestive to notice that the relation between the coordinate $\bb$ and $q$ used in our thermal setting is very reminiscent of that defining radial coordinates for four-point flat space blocks \cite{Hogervorst:2013sma} and two-point defect blocks \cite{Lauria:2017wav}.

\subsection{High-temperature expansion of blocks}
\label{AA:High-temperature expansion of blocks}

We shall now use the above results to derive an expansion of scalar exchange blocks around $\beta=0$ with finite $\Omega$. These blocks are particularly relevant for the $\lambda_3\Phi^3$ theory studied in the main text.
\smallskip

Rewriting $\bb,u$ in terms of $\beta,\Omega$ and expanding the double sum in equation \eqref{exact-scalar-block-resummed} term by term around $\beta=0$, we obtain
\begin{equation}\label{scalar-block-highT-exp}
\begin{split}
    g^{\Delta_\phi,0}_{\Delta,0} & = \frac{1}{\beta^{\Delta_\phi}}\left(b_{0,0,0}+b_{0,2,1}\,\Omega^2 s+b_{0,4,2}\,\Omega^4 s^2+\dots\right)\\
    &+\frac{1}{\beta^{\Delta_\phi-2}}\left(b_{2,0,0}+(b_{2,2,0}+b_{2,2,1}\,s)\Omega^2+(b_{2,4,1}\,s+b_{2,4,2}\, s^2)\Omega^4+\dots\right) + \dots  \\
    & +\frac{1}{\beta^{3-\Delta_\phi}}\left(b'_{0,0,0}+\left(b'_{0,2,0}+b'_{0,2,1}\,s\right)\Omega^2 +\dots\right)+O\left(\beta^{1-\Delta_\phi}\right)\\
    &+\frac{1}{\beta^2}\left(b_{0,0,0}''+\left(b_{0,2,0}'' + b_{0,2,1}''\,s\right)\Omega^2 +\dots\right)+O\left(\beta^0\right)\,,
\end{split}
\end{equation}
for some coefficients $b_{2i,2j,k}$, $ b_{2i,2j,k}'$ and $b_{2i,2j,k}''$. There are three towers of terms in the asymptotic expansion, with leading powers $\beta^{-\Delta_\phi}$, $\beta^{-3+\Delta_\phi}$ and $\beta^{-2}$. Each of these powers is multiplied by a series in $\beta^2$. We have displayed the first two terms of the tower beginning with $\beta^{-\Delta_\phi}$ in the first two lines of \eqref{scalar-block-highT-exp} and the leading term of each of the remaining towers in the third and fourth line. Further, the powers of $\Omega,s$ respect the general structure of a thermal one-point function conjectured in \cite{Buric:2025uqt}. 
\smallskip

From the series \eqref{exact-scalar-block-resummed} one may compute all the coefficients appearing in the expansion \eqref{scalar-block-highT-exp} one by one. An alternative, and simpler, way to obtain them is to substitute the expansion \eqref{scalar-block-highT-exp} into the Casimir equation. In the latter method, a subset of coefficients is needed as an ‘initial condition' that fixes the remaining ones. In the rest of this appendix, we shall assume that $\Delta_\phi$ is sufficiently large so that all terms considered come from the $\beta^{-\Delta_\phi}$-tower. In this case, the coefficients needed as the initial conditions for the Casimir equation are $b_{2i,0,0}$. In turn, these coefficients are extracted from the block at $\mu=0$, equation \eqref{exact-scalar-block}, and are readily available. We learn, for example, that the leading high temperature behaviour, as a function of $\Omega,s$ is simply
\begin{equation}\label{scalar-block-exact-highT}
    g^{\Delta_\phi,0}_{\Delta,0} = \frac{b_{0,0,0}}{\beta^{\Delta_\phi}
   \left(1+s\Omega^2\right)^{\frac{\Delta\phi}{2}}} + O\left(\beta^{-\Delta_\phi+2}\right)\,,
\end{equation}
where
\begin{equation}\label{coefficient-b000}
b_{0,0,0} = \frac{2^{\Delta_\phi-2\Delta}\,\Gamma\left(2\Delta-1\right)\,
\Gamma\left(\Delta_\phi-\frac32\right)}{\left(\Delta_\phi-2\right)\,
\Gamma\left(\Delta+\frac{\Delta_\phi}{2}-2\right)\,
\Gamma\!\left(\Delta+\frac{\Delta_\phi}{2}-\frac32\right)}\ .
\end{equation}
It is interesting to note that all the dependence on the dimension $\Delta$ of the exchanged field is encoded in the overall prefactor. Apart from this prefactor, the function is the same as the one describing the high temperature behaviour of the  $\phi^2$ one-point function in GFF (upon identifying $\Delta_\phi$ appearing here with the dimension of the field $\phi^2$). In Section \ref{S:Summary and discussion}, we recognise this function in the thermal EFT language, where it is related to the value of the dilaton field on $S^1\times S^2$. It is straightforward to extend equation \eqref{scalar-block-exact-highT} to higher orders in $\beta$, e.g.
\begin{equation}
    g^{\Delta_\phi,0}_{\Delta,0} = \frac{b_{0,0,0}}{\beta^{\Delta_\phi}
   \left(1+s\Omega^2\right)^{\frac{\Delta\phi}{2}}} \left(1+\beta^2\,\frac{c_1(1-s \Omega^4)+c_2\,\Omega^2(1-s)}{(1+s\Omega^{2})}+O(\beta^4)\right)\,,
\end{equation}
where
\begin{equation}\label{coefficients-c1-c2}
\begin{split}
    c_1=&\frac{b_{2,0,0}}{b_{0,0,0}}=\frac{72\Delta-24\Delta^2-16\Delta_\phi-36\Delta\Delta_\phi
+12\Delta^2\Delta_\phi+11\Delta_\phi^2-\Delta_\phi^3}{24\,(\Delta_\phi-4)\,(2\Delta_\phi-5)}\,,\\[2ex]
c_2=&\frac{b_{2,2,0}}{b_{0,0,0}}=\frac{6\left(2\Delta-\Delta_\phi\right)\left(2\Delta+\Delta_\phi-6\right)}{24\,(\Delta_\phi-4)\,(2\Delta_\phi-5)}\ .
\end{split}
\end{equation}

\section{One-point function on \texorpdfstring{$S^1_\beta \times \mathbb{R}^2$}{S1R2}}
\label{A:Flat space limit}

\subsection{Flat space limit of thermal AdS}

If we are interested in the boundary geometry $S^1_\beta \times \mathbb{R}^2$, the bulk space to consider is the Poincaré patch of Euclidean AdS$_4$. Poincar\'e coordinates $(z,x^\mu)$, $\mu=1,2,3$, on EAdS$_4$ are defined in terms of the embedding coordinates by
\begin{equation}
    \left(X^0+X^4,X^0-X^4,X^\mu\right) = \left(\frac{1}{z},\frac{z^2+x^2}{z},\frac{x^\mu}{z}\right)\,,
\end{equation}
so that the metric reads
\begin{equation}\label{bdy-metric}
    ds^2 = \frac{dz^2 + \delta_{\mu\nu} dx^\mu dx^\nu}{z^2}\ .
\end{equation}
For ordinary EAdS$_4$, the boundary is located at $\{z=0\}$ and $x^\mu$ serve as the usual Cartesian coordinates. In particular, the dilation generator is given by $D = z\partial_z+x^\mu\partial_\mu$. In the flat space limit, we instead write (following conventions of \cite{Alday:2020eua}) $x^\mu=(\mathcal{T},\mathbf{x})$, with the same metric \eqref{bdy-metric}, but now viewed as coordinates on $S^1_\beta \times \mathbb{R}^2$, i.e. $\mathcal{T}\sim\mathcal{T}+\beta$. In particular, the dilation generator is now simply $\partial_{\mathcal{T}}$. The bulk-to-boundary propagator reads
\begin{equation}
    \Pi^\beta_{\Delta_\phi}(x,\hat x)=C^{1/2}_{\Delta_\phi}\sum_{n=-\infty}^{\infty}\left(\frac{z}{(\mathcal{\hat T}-\mathcal{T}+n\beta)^2+(\hat{\mathbf{x}}-\mathbf{x})^2+z^2}\right)^{\Delta_\phi}\,,
\end{equation}
while the  bulk-to-bulk propagator is given by
\begin{equation}
    G^\beta_{\Delta_\phi}(x,x) = C_{\Delta_\phi}\sum_{n\neq0} \left(\frac{z}{\beta n}\right)^{2\Delta_\phi}\ _2F_1\left(\Delta_\phi,\Delta_\phi-1,2\Delta_\phi-2,-\frac{4z^2}{\beta^2 n^2}\right)\ .
\end{equation}
We have dropped the divergent zeroth image, which corresponds to mass renormalisation, \cite{Alday:2020eua}.

\subsection{One-point function}

In the cubic theory, the one-point function of the fundamental field $\phi$ is given now by (at leading order in $\lambda)$
\begin{equation}
    \langle\phi(\hat x)\rangle_\beta =  3\lambda_3 \int_0^\infty dz\int_0^\beta d\mathcal{T}\int d^2x  \sqrt{g}\ \Pi^{\beta}_{\Delta_\phi}(x,\hat x)G^{\beta}_{\Delta_\phi}(x,x)\ .
\end{equation}
Using translational invariance of the theory, we may safely specialize $\hat{x}=0$, and trade the sum over thermal images in the bulk-to-boundary propagator for the extension of the $\mathcal{T}$ integral over the whole real line. As the bulk-to-bulk propagator only depends on $z$, we may readily integrate over the other coordinates, using the standard result
\begin{equation}
    \int d\mathcal{T} dx^2\left(\frac{z}{\mathcal{T}^2+\mathbf{x}^2+z^2}\right)^{\Delta_\phi}=z^{3-\Delta_\phi}\frac{\pi^{3/2}\Gamma(\Delta_\phi-\frac32)}{\Gamma(\Delta_\phi)}, \qquad \text{for }\Delta_\phi>\frac32 \ .
\end{equation}
Using $\sqrt{g}=z^{-4}$ and switching the sum over images in the bulk-to-bulk propagator with the $z$ integral, we are thus left with
\begin{align}
    &\langle\phi(\hat x)\rangle_{\beta} =  \frac{6\lambda_3\, (\pi\,C_{\Delta_\phi})^{3/2}\,\Gamma(\Delta_\phi-\frac32)}{\Gamma(\Delta_\phi)} \\
    &\hskip3cm\sum_{n=1}^{\infty}\int_0^\infty\frac{dz}{z^{1+\Delta_\phi}} \left(\frac{z}{\beta n}\right)^{2\Delta_\phi}\!\!\!\ _2F_1\left(\Delta_\phi,\Delta_\phi-1,2\Delta_\phi-2,-\frac{4z^2}{\beta^2 n^2}\right)\ . \nonumber
\end{align}
Rescaling $z\rightarrow n\beta z$ we can then at the same time factor out the $\beta$ dependence and decouple the sum over $n$ from the integral. The former gives us a factor $\zeta(\Delta_\phi)$, while the latter evaluates as 
\begin{equation*}
    \int_0^\infty dz\, z^{\Delta_\phi-1} \ _2F_1\left(\Delta_\phi,\Delta_\phi-1,2\Delta_\phi-2,-4z^2\right)=\frac{\pi  2^{2-3 \Delta_\phi } \Gamma \left(\frac{\Delta_\phi }{2}-1\right) \Gamma (2 \Delta_\phi -2)}{\Gamma \left(\frac{\Delta_\phi -1}{2}\right) \Gamma \left(\frac{\Delta_\phi +1}{2}\right) \Gamma \left(\frac{3 \Delta_\phi }{2}-2\right)}\,,
\end{equation*}
for $\Delta_\phi>2$. Crucially, the last integral converges only for $\Delta_\phi>2$. This is the way in which different scalings as $\beta\to0$ that we observed in the main text manifest themselves in the flat space limit. Collecting all factors, we find thus $\langle\phi(\hat x)\rangle_{\beta} = b_\phi \beta^{-\Delta_\phi}$, with
\begin{equation}
b_\phi= \frac{\pi^{\frac14}\,3\cdot 2^{\frac52-3 \Delta_\phi } \Gamma \left(\frac{\Delta_\phi }{2}-1\right) \sqrt{\frac{\Gamma (\Delta_\phi )}{\Gamma \left(\Delta_\phi -\frac12\right)}} \Gamma (2 \Delta_\phi -3) \zeta (\Delta_\phi )}{\Gamma \left(\frac{\Delta_\phi -1}{2}\right) \Gamma \left(\frac{\Delta_\phi +1}{2}\right) \Gamma \left(\frac{3 \Delta_\phi}{2}-2\right)}\,,
\end{equation}
provided that $\Delta_\phi>2$. This precisely matches the one-point function \eqref{1-pt-coefficients-holography} obtained in the main text.

\section{An alternative view on multiplicities in GFF}
\label{app:deMelloKoch}

In the main text we described our approach to multiplicities of 
finite twist primary fields and their large $k$ asymptotics that 
was based on the inversion formula \eqref{density-inversion-U1}. 
 An alternative perspective on the results we obtained, and in 
 particular the asymptotic density \eqref{GFF-U1-asymptotics-spectrum}, 
 is provided by the analysis of \cite{deMelloKoch:2018klm}, that we 
 briefly discuss. The starting point of this analysis is a 
 representation-theoretic description of the multiplicities 
\begin{equation}\label{counting-muliplicities}
N(n,\ell,k)=\operatorname{Mult\Big(Sym^k}( (1)\otimes[n-1,1])\,;\,(\ell)\otimes [n]\Big)\ .
\end{equation}
In this equation, representations appearing are those of $U(\mathfrak{so}(3))\otimes\mathbb{C}[S_n]$ and Mult denotes the multiplicity of the second representation inside the first. We label representations of $\mathfrak{so}(3)$ and $S_n$ by Gelfand-Tsetlin labels, i.e. lengths of rows of the corresponding Young diagram. For clarity, we use ordinary brackets for representations of $\mathfrak{so}(3)$ and angle brackets for those of $S_n$. For instance, $(1)$ is the vector representation of $\mathfrak{so}(3)$, while $[n-1,1]$ and $[n]$ are the $(n-1)$-dimensional hook representation and the trivial representation of $S_n$, respectively.
\smallskip

In some cases, the description \eqref{counting-muliplicities} can be used to derive explicit exact expressions for the multiplicities, typically by realising the representations involved on spaces of polynomials ($n$-particle GFF primary states are put in 1-1 correspondence with polynomials in $nd$ variables, subject to a system of linear differential equations and permutation invariance, see \cite{deMelloKoch:2018klm} for more details). On the other hand, equation \eqref{counting-muliplicities} is also useful for analysis of large-$k$ asymptotics. Indeed, we can write a Molien-type representation for the generating function of multiplicities as
\begin{equation}
    F(s)\equiv\sum_{k=0}^{\infty}N(n,\ell,k)s^k = \frac{1}{|S_n|}\sum_{\sigma\in S_n}\int\limits_{SO(3)} dg \,\chi_\ell(g)\frac{1}{\det(\mathbf{1}-s\,g\,\otimes\rho_{[n-1,1]}(\sigma))}\ .
\end{equation}
As is standard practice in the use of generating functions, one can read off the large $k$ asymptotics of $N(n,\ell,k)$ by studying the expansion of $F(s)$ around $s=1$. In the Molien representation of $F(s)$, the leading contribution in this regime comes from the identity permutation $\sigma=e$ in the sum over $S_n$ and the elements of $SO(3)$ close to the identity element $g=\mathbf{1}$ in the integral. After restricting the sum over $S_n$ to the identity, and integrating over two of the Euler angles of $SO(3)$, we get
\begin{equation}
    F(s)\sim \frac{1}{n!} \frac{2}{\pi}
    \int_0^\pi dw\sin^2\frac{w}{2}\, \chi_\ell(w)\frac{1}{(1-s)^{n-1}(1-s e^{iw})^{n-1}(1-s e^{-iw})^{n-1}}\,,
\end{equation}
where $\chi_\ell$ is the $SO(3)$ character
\begin{equation}
    \chi_\ell(w)=\frac{\sin\frac{(2\ell+1)w}{2}}{\sin\frac{w}{2}}\ .
\end{equation}
To capture the leading contribution to the integral, which is coming from $w\sim0$, we may use the change of variable $u=w/(1-s)$ and expand the integrand around $s=1$ to get
\begin{equation}
    F(s)\sim \frac{2\ell+1}{2\pi\, n!}(1-s)^{6-3n}\int_0^\infty du \frac{u^2}{(1+u^2)^{n-1}}=\frac{(2\ell+1)}{8\sqrt{\pi}\, n!}\frac{\Gamma(n-\frac52)}{(n-2)!}(1-s)^{6-3n}\ .
\end{equation}
We have assumed $n\geq4$ to perform the last integral. The behaviour of $F(s)$ near $s=1$ is precisely that required to recover the asymptotics \eqref{GFF-U1-asymptotics-spectrum}, as we have
\begin{equation}
    N(n,\ell,k)\underset{k\rightarrow\infty}{\sim}\operatorname{Res}_{s=1} \left( \frac{F(s)}{s^{k+1}}\right) = \frac{2\ell+1}{8\sqrt{\pi} \,n!}\frac{\Gamma(n-\frac52)}{(n-2)!}\frac{k^{3n-7}}{(3n-7)!} + O\left(k^{3n-8}\right)\ .  
\end{equation}

\section{Tables of anomalous dimensions and OPE coefficients}
\label{A:Tables-of-data}
In this Appendix, we list some exact $O(\lambda)$ anomalous dimensions and OPE coefficients induced by quartic and cubic couplings in the bulk, respectively.

\subsection{Anomalous dimensions}
\label{AA:Anomalous dimensions}

Anomalous dimensions for the family of operators with quantum numbers $(n\Delta_\phi+4,0)$, corresponding to the second subleading parabola on Figure \ref{plot-low-lying-anom-dim} (not drawn), are given by
\begin{align}\label{nDelta+4-anom-dim}
    \frac{\bar\gamma_{n\Delta_\phi+4,0}}{\gamma_{\phi^2}} = \frac12 n^2 &- \frac{258\Delta_\phi^5 + 563\Delta_\phi^4 + 48\Delta_\phi^3-269 \Delta_\phi^2 + 18}{12 \left(2 \Delta_\phi+3\right) \left(4 \Delta_\phi^2-1\right)^2} n \\
    &+ \frac{116\Delta_\phi^5 - 314\Delta_\phi^4 - 243 \Delta_\phi^3 + 491\Delta_\phi^2 + 76\Delta_\phi - 120}{12 \left(2\Delta_\phi+3\right) \left(4\Delta_\phi^2-1\right)^2}\,, \nonumber
\end{align}
for $n\geq 4$. The cases of $n=2,3$ need to be considered separately, and are given by
\begin{center}
    \begin{tabular}{ | c | c | }
    \hline
    $n$ & $\bar\gamma_{\Delta,\ell}/\gamma_{\phi^2}$\\
    \hline
      2 & $\frac{15 \Delta_\phi^2 \left(16 \Delta_\phi^3-13 \Delta_\phi+3\right)}{4 (2 \Delta_\phi+3) \left(1-4 \Delta_\phi^2\right)^2}$\rule[-1.5ex]{0pt}{5ex}\\
    \hline
      3 &  $\frac{750 \Delta_\phi^5 + 461 \Delta_\phi^4 - 739 \Delta_\phi^3 - 110 \Delta_\phi^2 + 76 \Delta_\phi + 24}{8 (2 \Delta_\phi + 3) \left(1-4 \Delta_\phi^2\right)^2}$\rule[-1.5ex]{0pt}{5ex}\\
      \hline
    \end{tabular}
\end{center}
Some anomalous dimensions for families of the form $(n\Delta_\phi+\ell,\ell)$ read
\begin{center}
    \begin{tabular}{ | c | c | }
    \hline
    $(\Delta,\ell)$ & $\bar\gamma_{\Delta,\ell}/\gamma_{\phi^2}$\\
    \hline
         $(n\Delta_\phi+4,4)$ & $\frac12 n^2 - \frac{23\Delta_\phi^2+39\Delta_\phi+6}{4(2\Delta_\phi+1)(2\Delta_\phi+3)} n + \frac{3\Delta_\phi^2-9\Delta_\phi-21}{2(2\Delta_\phi+1)(2\Delta_\phi+3)}$\rule[-1.5ex]{0pt}{5ex} \\
    \hline
         $(n\Delta_\phi+5,5)$ & $\frac12 n^2 - \frac{6(17\Delta_\phi^2 + 29\Delta_\phi + 4)}{16(2\Delta_\phi+1)(2\Delta_\phi+3)} n + \frac{3\Delta_\phi^2 - 19\Delta_\phi - 39}{2(2\Delta_\phi+1)(2\Delta_\phi+3)}$\rule[-1.5ex]{0pt}{5ex}\\
    \hline
    $(n\Delta_\phi+6,6)$  &  $\frac12 n^2 - \frac{235 \Delta_\phi^3 + 982 \Delta_\phi^2+1035 \Delta_\phi +120}{16 (2\Delta_\phi+1)(2\Delta_\phi+3)(2\Delta_\phi+5)} n + \frac{63\Delta_\phi^3 + 72\Delta_\phi^2 - 513\Delta_\phi - 750}{8 (2 \Delta_\phi +1) (2 \Delta_\phi+3) (2 \Delta_\phi +5)} $\rule[-1.5ex]{0pt}{5ex}\\
    \hline
    $(n\Delta_\phi+7,7)$ & $\frac12 n^2 - \frac{499 \Delta_\phi^3 + 2095 \Delta_\phi^2 + 2220 \Delta_\phi + 240}{32 (2\Delta_\phi+1)(2\Delta_\phi+3)(2\Delta_\phi+5)} n + \frac{63\Delta_\phi^3 - 15\Delta_\phi^2 - 906\Delta_\phi - 1200}{8 (2 \Delta_\phi +1) (2 \Delta_\phi+3) (2 \Delta_\phi +5)} $\rule[-1.5ex]{0pt}{5ex}\\
    \hline
    \end{tabular}
\end{center}
Further anomalous dimensions are available upon request.

\subsection{OPE coefficients}
\label{AA:OPE coefficients}

Here we spell out the averaged OPE coefficients $\bar \lambda_{\phi\mathcal{O}\mathcal{O}}$  for the family of operators $\mathcal{O}$ with quantum numbers $(\Delta,\ell)=(n\Delta_\phi+4,0)$. For $n\geq4$, these are captured by the formula
\begin{equation*}
    \frac{\bar \lambda_{\phi\mathcal{O}\mathcal{O}}}{\lambda_{\phi\phi\phi}}=\frac{\sum_{k=0}^5 C_k(\Delta_\phi)\,n^k}{480 n (\Delta_\phi+1)(2\Delta_\phi-1)(2\Delta_\phi+1)(n\Delta_\phi+1)(2n\Delta_\phi+1)(2n\Delta_\phi+3)}\,,
\end{equation*}
where the coefficients of the polynomial in the numerator read
\begin{equation*}
\begin{aligned}
    C_0(\Delta_\phi)= & 1440 + 2712 \Delta_\phi - 6680 \Delta_\phi^2 - 11694 \Delta_\phi^3 + 3952 \Delta_\phi^4 + 3436 \Delta_\phi^5- 1112 \Delta_\phi^6\\
    &- 206 \Delta_\phi^7+ 96 \Delta_\phi^8 - 8 \Delta_\phi^9\,,\\[1ex]
C_1(\Delta_\phi)=&5760 + 11832 \Delta_\phi - 20916 \Delta_\phi^2 - 58270 \Delta_\phi^3 - 33263 \Delta_\phi^4 + 22100 \Delta_\phi^5 + 7398 \Delta_\phi^6\\
&- 3278 \Delta_\phi^7 - 247 \Delta_\phi^8+ 160 \Delta_\phi^9 - 12 \Delta_\phi^{10}\,,\\[1ex]
C_2(\Delta_\phi)=&-4320 + 14160 \Delta_\phi + 43556 \Delta_\phi^2 - 72108 \Delta_\phi^3 - 133753 \Delta_\phi^4 - 21100 \Delta_\phi^5 + 36286 \Delta_\phi^6\\
&+ 2452 \Delta_\phi^7- 2561 \Delta_\phi^8 + 84 \Delta_\phi^9 + 24 \Delta_\phi^{10}\,,\\[1ex]
C_3(\Delta_\phi)=&-15840 \Delta_\phi + 1440 \Delta_\phi^2 + 82512 \Delta_\phi^3 - 29520 \Delta_\phi^4 - 100892 \Delta_\phi^5 + 8240 \Delta_\phi^6\\
&+ 16648 \Delta_\phi^7- 976 \Delta_\phi^8 - 188 \Delta_\phi^9 - 16 \Delta_\phi^{10}\,,\\[1ex]
C_4(\Delta_\phi)=&-17280 \Delta_\phi^2 - 12480 \Delta_\phi^3 + 73520 \Delta_\phi^4 + 41392 \Delta_\phi^5 - 24220 \Delta_\phi^6\\
&+ 5952 \Delta_\phi^7 + 1384 \Delta_\phi^8 + 80 \Delta_\phi^9 + 4 \Delta_\phi^{10}\,,\\[1ex]
C_5(\Delta_\phi)=&-5760 \Delta_\phi^3 - 5760 \Delta_\phi^4 + 23040 \Delta_\phi^5 + 23040 \Delta_\phi^6\ .
\end{aligned}
\end{equation*}
The cases of $n=2,3$ need to be considered separately, and are given by
\begin{center}
    \begin{tabular}{ | c | c | }
    \hline
    $n$ & $\bar\lambda_{\phi\mathcal{OO}}/\lambda_{\phi\phi\phi}$\rule{0pt}{2.5ex}\\
    \hline
      2 & $\frac{(\Delta_\phi+2)(2 \Delta_\phi^8+101 \Delta_\phi^7+1693 \Delta_\phi^6+7452 \Delta_\phi^5+21992 \Delta_\phi^4+9985 \Delta_\phi^3-7799 \Delta_\phi^2-5826 \Delta_\phi-720)}{480 (\Delta_\phi+1) (2\Delta_\phi-1) (2\Delta_\phi+1) (4\Delta_\phi+1) (4 \Delta_\phi+3)}$\rule[-1.5ex]{0pt}{5ex} \\
    \hline
      3 &  $\frac{18\Delta_\phi^9 + 649\Delta_\phi^8 + 15190\Delta_\phi^7 + 124634\Delta_\phi^6 + 626562\Delta_\phi^5 + 793045\Delta_\phi^4 + 71294\Delta_\phi^3 - 233320\Delta_\phi^2 - 87816\Delta_\phi - 7920}{4320 (\Delta_\phi +1) (2\Delta_\phi - 1) (2\Delta_\phi +1) (3\Delta_\phi+1) (6\Delta_\phi+1)}$\rule[-1.5ex]{0pt}{5ex}\\
      \hline
    \end{tabular}
\end{center}
Further OPE coefficients are available upon request.

\bibliographystyle{JHEP}
\bibliography{bibliography}

\end{document}